\documentclass[aps,prb,twocolumn,groupedaddress,showpacs,floatfix]{revtex4}
\usepackage{graphicx}
\usepackage{braket}

\begin{document}

\title{Melting of Lennard-Jones  rare gas clusters doped with a single impurity atom}
\author{Nicol\'as Quesada}
\affiliation{Instituto de F\'isica, Universidad de Antioquia, AA 1226 Medell\'in, Colombia}
\author{Gloria E. Moyano}
\email[Corresponding author: ]{gloria.moyano@exactas.udea.edu.co}
\affiliation{Instituto de Qu\'imica, Universidad de Antioquia, AA 1226 Medell\'in, Colombia}

\date{July 13, 2010}
\begin{abstract}
Single impurity effect on the melting process of magic number Lennard-Jones,
rare gas, clusters of up to 309 atoms is studied on the basis of Parallel
Tempering Monte Carlo simulations in the canonical ensemble. A decrease on the
melting temperature range is prevalent, although such effect is dependent on the size of the impurity atom relative to the cluster size. Additionally, the difference between the atomic sizes of the impurity and the main component of the cluster should be considered. We demonstrate that solid-solid transitions due to migrations of the impurity become apparent and are clearly differentiated from the melting up to cluster sizes of 147 atoms.
\end{abstract}
\pacs{36.40.Ei, 61.46.+w}

\maketitle

\section{Introduction}

Alloying effects in atomic nanoclusters cover a domain of property behavior
wider and more complex than those corresponding to individual atoms and bulk
matter, with strong particle size specificities which combine with composition
and finite-size effects. 
Even for pure substances the structure of their atomic nanoclusters is very
dependent upon the number of atoms per particle. There are ``magic'' numbers,
corresponding to cluster structures characterized by their conspicuous
energetic stabilities relative to size, but for a given finite cluster
structure, stability results from a trade-off between packing and surface
effects. General non-monotonic property trends as a function of size
characterize finite clusters, so complex structural transitions may occur
during the growth from finite sizes to the bulk. The addition of dopant atoms
to a pure atomic cluster can alter its structure and growth patterns depending
upon the nature of both the impurity and the cluster, the cluster size, and
the concentration of dopant atoms. The possibility to manipulate nanoparticle
structures and so, tune their physico-chemical properties (\textit{e.g.}
catalytic, electronic, thermodynamic) has motivated a lot of recent research
on alloy nanoclusters \cite{Ferrando:2008chrev, RevModPhys.77.371}. \\
Regarding the phase changes, the melting process of pure and alloy clusters
has attracted considerable attention in experimental as well as in theoretical
studies. A number of specific features have been recognized in the melting
mechanisms of finite particles such as solid-solid structure changes prior to
melting\cite{doye:8143}, \emph{premelting}\cite{RevModPhys.77.371} effects of
surface loosening (formation of ``liquid-like'' surface layers
\cite{calvo:2888, Chen-nano}), coexistence of different atom-packing
schemes\cite{noya:104503}, oscillations between the liquid and solid phases
\cite{Duan200757}, etc.\\     
The melting temperature as a function of the cluster size has been studied on
the basis of several models which agree on predicting that the melting
temperature decreases linearly or quasi-linearly with the inverse of the
radius of the particle\cite{RevModPhys.77.371,0022-3727-24-3-017,
  0957-4484-12-1-312}. The pioneering work by Pawlow is summarized in the
formula:  
\[T_M (N)=T_M (\infty)(1-CN^ {-1/3}),\] 
in which $T_M (N)$ and $T_M (\infty)$ represent the melting temperatures of a
$N$-sized spherical cluster and the bulk, respectively; and $C$ is a constant
(see Ref [\onlinecite{RevModPhys.77.371}] for a derivation of this law and further
correction terms). Pawlow's law is consistent with several experimental
results and, although deviations occur for the smaller clusters, whose shapes
are far from spherical, the melting point of nanoclusters is usually
depressed. Nevertheless, there is experimental evidence of exceptions to this
trend for cases like the ionic tin clusters with 10-30 atoms, whose melting
points are at least 50K above that of the bulk\cite{PhysRevLett.85.2530}. In
addition to the size effects, the melting temperatures of alloy clusters can
be increased\cite{Mottet:2005} or decreased\cite{Hock:2008} with respect to
those of the pure components \cite{Ferrando:2008chrev}. The amount and
direction of the shiftings of the melting point in finite doped atomic
clusters can be attributed to several factors: alterations of the cluster
structure, whether or not the impurity is soluble in the cluster, many-body
energetic effects, and/or other complex energetic-entropic effects\cite{Hock:2008}.\\
The phenomenology seen in the melting mechanisms of pure clusters is also
apparent for binary and multiple-component clusters, but having the
composition as an additional variable enormously increases the complexity of
structural behaviors\cite{PhysRevLett.95.063401,
  Calvo-Yurtsever:2004}. Alloying effects in mixed atomic clusters depend upon
the differences between the atomic sizes, cluster surface energies, overall
structure strain, number and strength of the interactions between unlike
atoms. \cite{Calvo-Yurtsever:2004}. Further contributing aspects may be
kinetic factors, specific electronic/magnetic effects, and environmental
conditions \cite{Ferrando:2008chrev}. Alloying effects can be significant even
when a single impurity is introduced into a cluster of the order of a hundred
atoms.\cite{Mottet:2005}.\\ 
An efficient scheme to model the melting of doped atomic clusters has to
address the issues associated with the increased complexity of the energy
landscapes to explore during the simulations of mixed clusters, the
occurrence of homotop structures, as well as convergence difficulties related
to quasi-ergodicity that have been described elsewhere \cite{review,
  PhysRevE.72.037102}. Methods such as replica exchange Molecular Dynamics and
Parallel Tempering Monte Carlo (PTMC) have been developed to address the
quasi-ergodicity by improved sampling. PTMC is a  powerful method to sample
rugged energy  surfaces which takes advantage of the fact that replicas
running at high temperature are able to sample most of the relevant
configuration space. At the same time, through configuration exchange PTMC
connects high temperature replicas, which can visit most of the configuration
space, with replicas at low temperatures so that the latter do not get trapped in local
minima \cite{Earl-Deem}. \\
The paper has been written as follows: In section II we present the
methodology for optimal structure search, sampling and observable calculations
to monitor the cluster  melting process. Then, in Section III we discuss the
features that differentiate the melting of doped clusters from that of the
pure ones, taking into account their composition and cluster size. Special
detail is given to the study of the low temperature solid-solid
transitions. Finally we present some general conclusions.

\section{METHODOLOGY}
In this work we used the scaled Lennard-Jones (LJ) parameters $\sigma_i$ and
$\epsilon_i$ for the rare gas interactions reported in [\onlinecite{Calvo-Yurtsever:2004}].

\subsection{Optimal Structures}
To obtain the (putative) global minima presented in Table \ref{table:minima} 
(excepting for the cases of the pure LJ clusters, the 13 atom clusters,
Ar$_{54}$Xe and ArXe$_{54}$ which had already been reported in
[\onlinecite{CCD, Calvo-Yurtsever:2004}]) we perfomed three types of calculations:
\begin{enumerate}
 \item Local optimizations using the Fletcher-Reeves conjugate gradient
algorithm (FRCGA) were performed starting from the structures of the global
minima of each pure cluster, in which one atom of the pure cluster was
substituted by the dopant atom. This way we obtained a set of icosahedral low
energy structures.  
 \item In a complementary, ampler search, we used the Basin-Hopping
   method(BH) \cite{bh}. To sample the energy surfaces two types
of random moves were perfomed: Moving all the atoms at the same time and
swapping the dopant atom with an atom of the matrix.
We performed at least 20000 steps (=swaps+moves) in which, after each move, we
performed a local optimization using the FRCGA. For all the compositions the BH
method arrived to the same result of the first procedure.
\item Additionally, after the finite temperature simulations described in section
{\ref{sec:sampling}} we quenched samples saved at different
temperatures  for each composition. For the smallest clusters we performed
around 25000 local minimizations, and for larger clusters about
55000 local optimizations. 
\end{enumerate}
The results were equivalent for all procedures in the above list. We note that
the first strategy was computationally much cheaper than the other two. 
The minima in Table \ref{table:minima} were used to initialize the finite temperature
simulations.\\
Considering the lowest energy structures the dopant atom takes the central
position of the cluster when the impurity is Ar or Kr, while it remains in
one of the two most external shells when the impurity is Xe.

 \begin{table} 
 \begin{ruledtabular}
 \begin{tabular}{lccccc}
 Cluster & $E_0/\epsilon_{Ar}$ & $E_0/\epsilon_{i}$ & Dopant   & Point & $T_M/K$\\
         &                     &                    & Position & Group & \\
 \hline
 LJ$_{13}$ & -44.3268 & -44.3268 & - & $I_h$ & -\\
 Ar$_{12}$Xe & -47.6981 & -47.6981 & 1/1 & $C_{5v}$ & 30.22\\
 ArXe$_{12}$ & -78.6977 & -42.4934 & 0/1 & $I_h$ & 59.96\\
 Kr$_{12}$Xe & -62.5139 & -45.5132 & 1/1 & $C_{5v}$ & 41.51\\
 KrXe$_{12}$ & -81.0895 & -43.7848 & 0/1 & $I_h$ & 62.74\\
\hline 
LJ$_{55}$ & -279.248 & -279.248 & - & $I_h$ & -\\
 Ar$_{54}$Xe & -284.276 & -284.276 & 2/2 & $C_{2v}$ & 31.25\\
 ArXe$_{54}$ & -516.170 & -278.709 & 0/2 & $I_h$ & 63.78\\
 Kr$_{54}$Xe & -386.018 & -281.040 & 2/2 & $C_{2v}$ & 42.93\\
 KrXe$_{54}$ & -517.631 & -279.498 & 0/2 & $I_h$ & 65.33\\
\hline
 LJ$_{147}$ & -876.461 & -876.461 & - & $I_h$ & -\\
 Ar$_{146}$Xe & -882.335 & -882.335 & 3/3 & $C_{3v}$ & 42.33\\
 ArXe$_{146}$ & -1625.44 & -877.667 & 0/3 & $I_h$ & 79.86\\
 Kr$_{146}$Xe & -1206.77 & -878.584 & 3/3 & $C_{3v}$ & 58.14\\
 KrXe$_{146}$ & -1625.44 & -877.666 & 0/3 & $I_h$ & 79.59\\
\hline
 LJ$_{309}$ & -2007.22 & -2007.22 & -& $I_h$ & -\\
 Ar$_{308}$Xe & -2013.39 & -2013.39 & 3/4 & $C_{3v}$& 50.18 \\
 ArXe$_{308}$ & -3722.59 & -2010.04 & 0/4 & $I_h$ & 94.13\\
 Kr$_{308}$Xe & -2760.16 & -2009.53 & 3/4 & $C_{3v}$ & 68.92\\
 KrXe$_{308}$ & -3721.20 & -2009.28 & 0/4 & $I_h$ & 93.06\\
 \end{tabular}
 \end{ruledtabular}
 \caption{\label{table:minima} Global minima for the different
compositions considered, corresponding energies in absolut units $E^0/\epsilon_{Ar}$, in
units relative to their matrix composition $E^0/\epsilon_{i}$, shell position of
the dopant atom in the structure (number of shell containing the
impurity/total number of shells in cluster, the 0$^{th}$ shell is the geometric
center of the icosahedron), point group, and the melting temperature of each
cluster in Kelvin.}
 \end{table}

\subsection{Sampling Strategy \label{sec:sampling}}
To sample the complex energy surfaces of our systems in the Canonical Ensemble
we used the PTMC method \cite{review}. 
For each replica we have used two types of moves. On the one hand, single
particle moves (SPM) have been implemented using an adapstive step that assures
that half of the time the new configuration will be accepted. On the other hand,
since we have two different atomic species in each cluster we have also
implemented particle exchange moves. This sampling strategy consists in exchanging
the position of two different atoms in the clusters. 
The simulation temperatures were chosen according to the geometric progression $T_i= T_{0}
\lambda^i$. The number of temperatures for the simulations as well as their maximum and
minimum values are summarized in Table \ref{table:params}. 

For each system the number of equilibration steps was always equal to the
number of Monte Carlo steps ($N_{MC}$). To prevent the evaporation of the
clusters we implemented hard sphere constraining potentials for the constraining radii
listed in Table \ref{table:params}.\\
Finally, the swapping acceptance ratios between replicas in all the systems
simulated remained around 60-70\% and never went below 35\%.

 \begin{table} 
 \begin{ruledtabular}
 \begin{tabular}{lcccccc}
 $N$ & $n$ &$k_B T_0/\epsilon_i$ & $k_B T_f/\epsilon_i$ & $R_c/\sigma_i$ & $N_{MC}$ & $N_{swap}$\\
 \hline
13 & 31 &0.01 & 0.4 & 2.5 & 4 $\times$ 10$^8$ & 100 \\
 \hline
55 & 71 & 0.01 & 0.4 & 3.5 & 8 $\times$ 10$^8$ & 100 \\
 \hline
147 & 71 &0.01 \& 0.2 & 0.4 \& 0.5 & 4.5 & 1.6 $\times$ 10$^9$ & 250 \\
\hline
309 & 71 & 0.2 & 0.5 & 5.5 & 2 $\times$ 10$^9$ & 500 \\
 \end{tabular}
 \end{ruledtabular}
 \caption{\label{table:params} Cluster sizes ($N$), number of
temperatures simulated ($n$), their minimum ($T_0$) and maximum ($T_f$) values,
constraining radii ($R_c$), number of Monte Carlo steps ($N_{MC}$) and
 frequencies at which swaps between adjacent replicas were attempted
($N_{swap}$). The constraining radii $R_c$ and the temperatures $T_0$ and $T_f$
are given in units of the LJ parameters of the atoms of the matrix.}
 \end{table}

\subsection{Observables}
We analyse the melting process by monitoring various
observables. Firstly, the heat capacity $C_V$, which
is calculated according to the formula:
\[
 C_V(T)=\frac{1}{k_B T^2} \left( \braket{E^2}_T - \braket{E}^2_T\right).
\]
To interpolate the points obtained with the PTMC simulation and have a smooth
dependence in the $C_V(T)$ curve we used the multihistogram method \cite{multi1,multi2}.
Note that the formula given above depends on the volume in which the system
is constrained to move. In figure \ref{fig:volume} we compare the $C_V(T)$ curves for
two constraining volumes. Notice that although the second volume is twice the
first ($V(R_C=4.5 \sigma_{Ar})/V(R_C=3.5 \sigma_{Ar})\approx2.13$) the main peak is
not strongly affected and the  features of the curve below the main peak
basically do not change (As one expects from a ``solid'' phase).

To monitor the effects of the dopant atom in each cluster we calculated
the radial distributions functions (RDF) $g(r)$ of the dopant atom and of the rest of the atoms in the
matrix, for these calculations all the distances $r$ have been taken with respect to the
\emph{geometric} center of the cluster $\vec r_{geom}$, where
$ \vec r_{geom}=\frac{1}{N}\sum_i^ N \vec r_i$,  $N$ represents the \emph{total} number of atoms
in the cluster.\\  
To further quantify the delocalization of the atom we calculated the standard
deviation of the RDF of the dopant atom ($\xi$) according to:
\[
\xi=\sqrt{\braket{r_{dopant}^2}-\braket{r_{dopant}}^2}.
\]

\subsection{Harmonic Superposition Method}
To understand the solid-solid transitions that occur in a doped cluster
between homotops of the same stoichiometry we have used the Harmonic
Superposition Method (HSM)  \cite{berry,calvo3}.

This method assumes that there is a number $m$ of well defined states that
make most of the contribution to the partition function in a certain range of
temperatures. Then, one approximates the contribution of each state to the
partition function ($Z(T)$) as the contribution of its harmonic part. Such partition function is
obtained from the normal modes and frequencies by expanding the
potential around the corresponding minimum in a power series up to quadratic order:
\[
 V(\vec R)=V(\vec R_0)+\frac{1}{2} \vec R^{T} \hat H \vec R+\mathcal{O}(R^3),
\]
where $\vec R= \left(\vec r_1,\ldots, \vec r_N \right)$,  $\vec R_0$ is the
equilibrium position and $\hat H$ is the Hessian Matrix of that minimum.
To obtain the partition function (and the thermodynamics of the system) one
adds the Simple Harmonic Oscillator partition functions of each
state:
\[
%\label{eq:superpos}
 Z(T)=\sum_{\alpha}^{m} n_{\alpha} \frac{\exp(-\beta E_{\alpha})}{(\beta h
\bar{\nu_{\alpha}})^{3N-6}}=\sum_{\alpha}^{m} n_{\alpha} Z_{\alpha}(T),
\]
where $\beta=1/k_B T$, $E_{\alpha}$ is the energy of each state, $n_{\alpha}$
is its degeneracy due to symmetry ($n_{\alpha}=2 p! (N-p)!/h_{\alpha} $, $N$
the number of atoms, $p$ the number of impurities, $p=1$ and $h_{\alpha}$ the
order of the point group of the state $\alpha$). $\bar{\nu_{\alpha}}$ is the
geometric \emph{mean} vibrational frequency of each state (which is proportional to
geometric mean of the square roots of the eigenvalues of the matrix $\hat H$)
and $N$ is the number of atoms considered.

\section{RESULTS AND DISCUSSION}

\subsection{Size dependence of the melting temperature}
In figure \ref{fig:melting} we present the results of our calculations
regarding the variation of the melting temperatures as a function of
$N^{-\frac{1}{3}}$. It is seen that as the size of the cluster is increased the
melting temperature of the clusters also increases, this behavior has been
verified for Ar and Xe clusters \cite{calvo2}. 
The case of $N=13$ is certainly out of any linear  tendency for all the
compositions, yet, for other larger clusters $N=55,147,309$ where surface
effects are less marked the dependence of the melting temperature as a
function of $N^{1/3}$  can be well described by a line, and in all cases
increases with the size of the cluster. From figure \ref{fig:melting} is
clearly seen that doping effects are very strong for small doped  clusters
($N=13, 55$), whose atoms have the highest differences between their LJ
parameters, $\epsilon$ and $\sigma$ (In this case Argon and Xenon). It is also
seen that for the largest cluster sizes studied here their melting
temperatures are almost equal for the doped and pure clusters. \\

\subsection{Doping effects: Composition and Size Dependence} 

So far we have discussed the doping effects solely in terms of the position,
on the temperature scale, of the peak associated with the melting of the
cluster. Yet as is seen in figures \ref{fig:13y55cv} and \ref{fig:147y309cv}
the peak changes, not only its position but also its height and width, for some
compositions. For instance, for Ar$_{146}$Xe, the change in height with respect
to Ar$_{147}$ is noticeable although the displacement of the maximum is just
around 1\%. Other characteristic of the $C_V(T)$ that is modified by
the presence of the dopant atom is the occurence of a small peak or bump in
the low temperature region. As we will demonstrate  in section \ref{lowT}, for
the clusters with sizes ($N=13,55,147$), this is due to a solid-solid
transition. Some general trends for compositions ArXe$_{N-1}$ and KrXe$_{N-1}$
($N=\{13,55,147,309\}$) are: Regarding their lowest energy configurations,
each pure Xe$_N$ cluster and the doped ArXe$_{N-1}$ and KrXe$_{N-1}$ clusters
have the same symmetry group ($I_h$), and are also very close in
geometry. After a small temperature increase, the dopant atom in both
ArXe$_{N-1}$ and KrXe$_{N-1}$ behaves the same way. It starts to move from the
the center of the cluster in the lowest energy configuration, to the second
most energetically favorable position, in the outer shell of the cluster, as
seen in the first two rows of figures \ref{fig:gs13}, \ref{fig:gs55},
\ref{fig:gs147} and \ref{fig:gs309}. A pictorical representation of the
process is given in figure \ref{bolitas}. Nevertheless, excepting for the
smallest cluster size, $N=13$, the dopant atom never relocalizes completely in
a stable configuration different from the global minimum, this occurs because
for larger structures $N>13$ there is more than one icosahedral stable
structure in which  the dopant atom is located in the outer shell of the
clusters. To support this, see in figure \ref{fig:var} that, excepting for
the cases ArXe$_{12}$ and KrXe$_{12}$, the standard
deviation of the position of the dopant atom ($\xi_{Xe}$) is always an increasing function of the
temperature, until the cluster melts. The bottom rows in figures
\ref{fig:gs13}, \ref{fig:gs55}, \ref{fig:gs147} and \ref{fig:gs309} show that
upon melting, the RDFs of the matrix and the dopant
show the same structure. This indicates that, in the liquid phase, and for the
compositions studied, Ar and Kr are not segregated by the Xe matrix. 
Finally, as one would expect based on the similarities of their LJ parameters,
the Xe-Kr doped clusters show more resemblance to the pure cluster in their
$C_V$ curves.\\ 
The clusters Kr$_{N-1}$Xe are the ones that show a more similar behavior to
the pure clusters LJ$_N$, considering their $C_V$ curves. For these
compositions the standard deviation of the position of the dopant atom
($\xi_{Xe}$) is always an increasing function of the temperature (see figure
\ref{fig:var}). This implies that the Xe atom does not leave completely its
external shell location, as in the lowest energy configuration (see the fourth
column on Table \ref{table:minima}). Such configuration plays a significant
role in the thermodynamics of the system until the phase change. This can be
seen on the RDFs of Xe in the Kr$_{N-1}$Xe clusters,
as plotted in the fourth column of figures \ref{fig:gs13}, \ref{fig:gs55},
\ref{fig:gs147} and \ref{fig:gs309}. Also, on the spectra of quenched energies
of Kr$_{12}$Xe and Kr$_{54}$Xe, in figures \ref{fig:13quench} and
\ref{fig:55quench}. The shape of $\xi_{Xe}$  for Kr$_{N-1}$Xe  is
qualitatively different depending on the cluster size $N$, indicating that the
temperature ranges for the migration of the dopant atom and the melting of the
cluster overlap for the smaller sizes. For $N=13$, $\xi_{Xe}$ simply increases
once the cluster starts to melt. For $N=55$, the dopant atom starts to
delocalize smoothly  between the second and first shells, until the migration
is met by the melting of the cluster (see the last column on figure
\ref{fig:gs55}). For the largest structures Kr$_{146}$Xe and Kr$_{308}$Xe the
dopant atom migrates to several positions in different shells of the
structure, as seen in the last column of figures \ref{fig:gs147} and
\ref{fig:gs309}. Upon melting, these compositions show the same behavior
observed in ArXe$_{N-1}$ and KrXe$_{N-1}$ , \emph{i.e.} there is no
segregation between the Kr atoms  and the Xe atom of the cluster. The
composition that shows more features during the heating process is
Ar$_{N-1}$Xe. The largest doping effect is seen in the cluster Ar$_{12}$Xe.
For this cluster we see that the melting temperature (taken as the position of
the maximum in the $C_V(T)$ curve) drops by around $\Delta T=0.037
\epsilon_{Ar}/k_B$, which is around $13\%$ of the melting temperature of the 
pure cluster. 
A comparable change in the melting point occurs for Ar$_{54}$Xe with respect
to Ar$_{55}$. This is not the only feature that changes drastically when
replacing one atom, with respect to the melting of Ar$_{55}$. From figure
\ref{fig:13y55cv} it is also seen that the melting peak in the $C_V(T)$ curve
for the doped cluster is smaller, by almost a factor of 2, as compared 
with the pure cluster, in other words the latent heat associated with the melting
is smaller in the doped cluster.
The reduction in the latent heat is a feature present in all the Argon clusters,
doped with Xenon. For the case of Ar$_{54}$Xe two different transitions are
seen in the RDF, $g(r)$, of the dopant atom (see the
third column of figure \ref{fig:gs55}). These transitions are seen in the
non-monotonous behavior of the standard deviation of the position of the Xe
atom in figure \ref{fig:var}b. In the first transition the Xe atom 
migrates
from the outer shell to the inner shell, and remains there. As it was mentioned in the last
section, this causes a small bump in the $C_V$ curve. Then, as the temperature is
further increased, the atom starts to migrate between the center of the cluster, the first
shell and the outer shell. This occurs near the temperature range for the
phase change. Finally, when the cluster reaches the liquid-like phase
an interesting effect occurs, namely the Xe atom is segregated from the Ar atoms. This is
clearly seen in the last row of figures \ref{fig:gs13}, \ref{fig:gs55},
\ref{fig:gs147} and \ref{fig:gs309}. We note that, for all the cases studied, the segregation is related to a maximum size contrast 
between the impurity and other atoms in the cluster.
 \begin{table} 
 \begin{ruledtabular}
 \begin{tabular}{lcccccc}
Cluster & $\Delta/\epsilon_{Ar}$ & $\Delta/\epsilon_{i}$ & P. Group &
$\sigma_i^2 \bar{\nu_0}/ \epsilon_i$ & $\sigma_i^2\bar{\nu_1}/\epsilon_i$ & $k_B T_h /\epsilon_i$ \\
        &                        &                       & 2$^{nd}$ state &         &         &                        \\

\hline
 Kr$_{12}$Xe & 2.02627 & 1.47523 &  $I_h$    & 11.622 & 9.8675 & 0.50598  \\
 ArXe$_{12}$ & 0.04162 & 0.02247 &  $C_{5v}$ & 12.777 & 11.826 & 0.00446 \\
 Ar$_{12}$Xe & 3.46358 & 3.46358 &  $I_h$    & 11.349 & 6.4721 & 0.21580 \\
 KrXe$_{12}$ & 0.87796 & 0.47406 &  $C_{5v}$ & 13.136 & 12.208 & 0.09668 \\
 Kr$_{54}$Xe & 0.63580 & 0.46289 &  $C_{5v}$ & 13.032 & 12.844 & 0.15423 \\
 ArXe$_{54}$ & 2.79753 & 1.51055 &  $C_{5v}$ & 14.374 & 13.954 & 0.19544 \\
 Ar$_{54}$Xe & 1.15348 & 1.15348 &  $C_{5v}$ & 12.544 & 11.919 & 0.14918 \\
 KrXe$_{54}$ & 2.53307 & 1.36775 &  $C_{5v}$ & 14.307 & 14.019 & 0.21457 \\
 Ar$_{146}$Xe & 0.36404 & 0.36404 &  $C_{2v}$ & 13.519 & 13.381 & 0.08944

 \end{tabular}
 \end{ruledtabular}
 \caption{\label{table:hsm} Parameters used for the HSM. $\Delta/\epsilon_{Ar}$
is the energy difference between the lowest energy structure and the second
stable structure considered. $\Delta/\epsilon_{i}$ is the scaled difference of
the two levels considered in terms of the $\epsilon$ parameter of the matrix.
The fourth column is the point group of the low energy minimum. $\nu_i$ is the square root of the
geometric mean of the non zero eigenvalues of the Hessian Matrix $\hat H$
for each stable structure and $k_B T_h=\frac{\Delta}{(3N-6) \ln \bar{\nu_0}/\bar{\nu_1}-\ln h_0/h_1}$ is the temperature at which the transition
between the two solid structures is expected, \emph{i.e.} at which the partition functions associated with each minimum are equal \cite{calvo3}.} 
 \end{table}

\subsection{Low $T$ behavior \label{lowT}}
One of the most interesting features of the $C_V$ calculations presented in
figure \ref{fig:13y55cv} and \ref{fig:147y309cv} is the occurence of a second
small peak, not seen in the pure clusters, for some of the doped structures. The most
noticeable case being that of KrXe$_{12}$. Such peak has been associated with
a solid-solid transition, and studied in detail for rare gas clusters of
6\cite{white}  and 13
\cite{frantz:10030,frantz:1992,sabo:847,sabo:856,Munro20021} atoms. It has
been suggested 
\cite{frantz:10030,frantz:1992,sabo:847,sabo:856,Munro20021,white} that this 
bump is due to structural transitions between isomers of the same
composition. We reach the same conclusion via an analysis of around 1000
structures, which we sampled, for each replica and each composition in
clusters with up to 147 atoms. We later quenched those structures. 
The energies and relative sampling frequencies of the set of minima obtained
for each composition are presented in figures \ref{fig:13quench} and
\ref{fig:55quench}. From these figures we note that the extra peak correlates
extremely well with the appearance of a second stable structure that becomes
increasingly important until the cluster melts. This second structure
corresponds to an icosahedron in which the dopant atom swaps positions with an 
atom in a different shell.\\
To further support our conclusion we have performed Harmonic Superposition
Method (HSM) calculations for some compositions. The input values used in the
HSM calculations  are shown in the first six columns of Table
\ref{table:hsm}. 
In the last column of the same table, we show the results (\emph{i.e.} the
predicted temperature for the solid-solid change asociated with the transition
between the two minima ). The predicted temperatures agree well with those
obtained from the PTMC simulations in figures \ref{fig:13y55cv}, and
\ref{fig:147y309cv}. The Table \ref{table:hsm} also shows why the extra peak
is not present in all clusters. For Kr$_{12}$Xe, Ar$_{12}$Xe and Kr$_{54}$Xe,
as can be seen in the insets of figure \ref{fig:13y55cv}, the temperature of
the solid-solid transition is so close to the melting peak, that when the
structure can change to a different minima it has started to sample ``liquid
like'' configurations.

\section{CONCLUSIONS}
PTMC simulations for rare gases (LJ) doped clusters with up to 309 atoms
showed that a single atom impurity can cause doping effects such as the depletion of the
melting range (with respect to the pure cluster), and the occurence of a
solid-solid transition in the low temperature range \footnote{The $C_V$ curve
  of LJ$_{309}$ shows considerably large peaks in the temperature range
  $0.39<k_B T/\epsilon<0.43$. These have been studied in detail by Noya and
  Doye\cite{noya:104503}. They point out that prior to the melting there is complex
  structural transformation involving several processes such as surface
  roughening and the formation of structures with diverse symmetries.}. The
shifting of the melting range due to the presence of the single atom impurity
decreases with increasing cluster size. In terms of absolute temperature it is
noticeable for clusters with less than a 100 atoms, for instance for
Ar$_{54}$Xe it represents 3.4 K. Several criteria (\emph{i.e.} $C_V$ curves,
radial distribution functions, spectra of quenched energies, and HSM) have
been used to support that a solid-solid transition peak may arise for doped
clusters with up to 147 atoms.

\begin{acknowledgments}
We thank Fundaci\'on para la Promoci\'on de la Investigaci\'on y la
Tecnolog\'ia del Banco de la Rep\'ublica and Universidad de Antioquia for
Financial support, CRESCA-SIU for computational resources. We also thank
F. Calvo for helpful discussions and for providing the multihistogram method
computer code. 

\end{acknowledgments}

\bibliography{BS11199}

\begin{thebibliography}{30}
\expandafter\ifx\csname natexlab\endcsname\relax\def\natexlab#1{#1}\fi
\expandafter\ifx\csname bibnamefont\endcsname\relax
  \def\bibnamefont#1{#1}\fi
\expandafter\ifx\csname bibfnamefont\endcsname\relax
  \def\bibfnamefont#1{#1}\fi
\expandafter\ifx\csname citenamefont\endcsname\relax
  \def\citenamefont#1{#1}\fi
\expandafter\ifx\csname url\endcsname\relax
  \def\url#1{\texttt{#1}}\fi
\expandafter\ifx\csname urlprefix\endcsname\relax\def\urlprefix{URL }\fi
\providecommand{\bibinfo}[2]{#2}
\providecommand{\eprint}[2][]{\url{#2}}

\bibitem[{\citenamefont{Ferrando et~al.}(2008)\citenamefont{Ferrando, Jellinek,
  and Johnston}}]{Ferrando:2008chrev}
\bibinfo{author}{\bibfnamefont{R.}~\bibnamefont{Ferrando}},
  \bibinfo{author}{\bibfnamefont{J.}~\bibnamefont{Jellinek}}, \bibnamefont{and}
  \bibinfo{author}{\bibfnamefont{R.~L.} \bibnamefont{Johnston}},
  \bibinfo{journal}{Chemi. Rev.} \textbf{\bibinfo{volume}{108}},
  \bibinfo{pages}{845} (\bibinfo{year}{2008}).

\bibitem[{\citenamefont{Baletto and Ferrando}(2005)}]{RevModPhys.77.371}
\bibinfo{author}{\bibfnamefont{F.}~\bibnamefont{Baletto}} \bibnamefont{and}
  \bibinfo{author}{\bibfnamefont{R.}~\bibnamefont{Ferrando}},
  \bibinfo{journal}{Rev. Mod. Phys.} \textbf{\bibinfo{volume}{77}},
  \bibinfo{pages}{371} (\bibinfo{year}{2005}).

\bibitem[{\citenamefont{Doye et~al.}(1998)\citenamefont{Doye, Wales, and
  Miller}}]{doye:8143}
\bibinfo{author}{\bibfnamefont{J.~P.~K.} \bibnamefont{Doye}},
  \bibinfo{author}{\bibfnamefont{D.~J.} \bibnamefont{Wales}}, \bibnamefont{and}
  \bibinfo{author}{\bibfnamefont{M.~A.} \bibnamefont{Miller}},
  \bibinfo{journal}{J. Chem. Phys.} \textbf{\bibinfo{volume}{109}},
  \bibinfo{pages}{8143} (\bibinfo{year}{1998}).

\bibitem[{\citenamefont{Calvo and Spiegelmann}(2000)}]{calvo:2888}
\bibinfo{author}{\bibfnamefont{F.}~\bibnamefont{Calvo}} \bibnamefont{and}
  \bibinfo{author}{\bibfnamefont{F.}~\bibnamefont{Spiegelmann}},
  \bibinfo{journal}{J. Chem. Phys.} \textbf{\bibinfo{volume}{112}},
  \bibinfo{pages}{2888} (\bibinfo{year}{2000}).

\bibitem[{\citenamefont{Chen and Johnston}(2007)}]{Chen-nano}
\bibinfo{author}{\bibfnamefont{F.}~\bibnamefont{Chen}} \bibnamefont{and}
  \bibinfo{author}{\bibfnamefont{R.~L.} \bibnamefont{Johnston}},
  \bibinfo{journal}{ACS Nano} \textbf{\bibinfo{volume}{2}},
  \bibinfo{pages}{165} (\bibinfo{year}{2007}).

\bibitem[{\citenamefont{Noya and Doye}(2006)}]{noya:104503}
\bibinfo{author}{\bibfnamefont{E.~G.} \bibnamefont{Noya}} \bibnamefont{and}
  \bibinfo{author}{\bibfnamefont{J.~P.~K.} \bibnamefont{Doye}},
  \bibinfo{journal}{J. Chem. Phys.} \textbf{\bibinfo{volume}{124}},
  \bibinfo{eid}{104503} (\bibinfo{year}{2006}).

\bibitem[{\citenamefont{Duan et~al.}(2007)\citenamefont{Duan, Ding, Rosén,
  Harutyunyan, Curtarolo, and Bolton}}]{Duan200757}
\bibinfo{author}{\bibfnamefont{H.}~\bibnamefont{Duan}},
  \bibinfo{author}{\bibfnamefont{F.}~\bibnamefont{Ding}},
  \bibinfo{author}{\bibfnamefont{A.}~\bibnamefont{Rosén}},
  \bibinfo{author}{\bibfnamefont{A.~R.} \bibnamefont{Harutyunyan}},
  \bibinfo{author}{\bibfnamefont{S.}~\bibnamefont{Curtarolo}},
  \bibnamefont{and} \bibinfo{author}{\bibfnamefont{K.}~\bibnamefont{Bolton}},
  \bibinfo{journal}{Chemical Physics} \textbf{\bibinfo{volume}{333}},
  \bibinfo{pages}{57 } (\bibinfo{year}{2007}).

\bibitem[{\citenamefont{Wautelet}(1991)}]{0022-3727-24-3-017}
\bibinfo{author}{\bibfnamefont{M.}~\bibnamefont{Wautelet}},
  \bibinfo{journal}{J. Phys. D: Appl. Phys.} \textbf{\bibinfo{volume}{24}},
  \bibinfo{pages}{343} (\bibinfo{year}{1991}).

\bibitem[{\citenamefont{Vall\'ee et~al.}(2001)\citenamefont{Vall\'ee, Wautelet,
  Dauchot, and Hecq}}]{0957-4484-12-1-312}
\bibinfo{author}{\bibfnamefont{R.}~\bibnamefont{Vall\'ee}},
  \bibinfo{author}{\bibfnamefont{M.}~\bibnamefont{Wautelet}},
  \bibinfo{author}{\bibfnamefont{J.~P.} \bibnamefont{Dauchot}},
  \bibnamefont{and} \bibinfo{author}{\bibfnamefont{M.}~\bibnamefont{Hecq}},
  \bibinfo{journal}{Nanotechnology} \textbf{\bibinfo{volume}{12}},
  \bibinfo{pages}{68} (\bibinfo{year}{2001}).

\bibitem[{\citenamefont{Shvartsburg and Jarrold}(2000)}]{PhysRevLett.85.2530}
\bibinfo{author}{\bibfnamefont{A.~A.} \bibnamefont{Shvartsburg}}
  \bibnamefont{and} \bibinfo{author}{\bibfnamefont{M.~F.}
  \bibnamefont{Jarrold}}, \bibinfo{journal}{Phys. Rev. Lett.}
  \textbf{\bibinfo{volume}{85}}, \bibinfo{pages}{2530} (\bibinfo{year}{2000}).

\bibitem[{\citenamefont{Mottet et~al.}(2005)\citenamefont{Mottet, Rossi,
  Baletto, and Ferrando}}]{Mottet:2005}
\bibinfo{author}{\bibfnamefont{C.}~\bibnamefont{Mottet}},
  \bibinfo{author}{\bibfnamefont{G.}~\bibnamefont{Rossi}},
  \bibinfo{author}{\bibfnamefont{F.}~\bibnamefont{Baletto}}, \bibnamefont{and}
  \bibinfo{author}{\bibfnamefont{R.}~\bibnamefont{Ferrando}},
  \bibinfo{journal}{Phys. Rev. Lett.} \textbf{\bibinfo{volume}{95}},
  \bibinfo{pages}{035501} (\bibinfo{year}{2005}).

\bibitem[{\citenamefont{Hock et~al.}(2008)\citenamefont{Hock, Stra\ss{}burg,
  Haberland, v.~Issendorff, Aguado, and Schmidt}}]{Hock:2008}
\bibinfo{author}{\bibfnamefont{C.}~\bibnamefont{Hock}},
  \bibinfo{author}{\bibfnamefont{S.}~\bibnamefont{Stra\ss{}burg}},
  \bibinfo{author}{\bibfnamefont{H.}~\bibnamefont{Haberland}},
  \bibinfo{author}{\bibfnamefont{B.}~\bibnamefont{v.~Issendorff}},
  \bibinfo{author}{\bibfnamefont{A.}~\bibnamefont{Aguado}}, \bibnamefont{and}
  \bibinfo{author}{\bibfnamefont{M.}~\bibnamefont{Schmidt}},
  \bibinfo{journal}{Phys. Rev. Lett.} \textbf{\bibinfo{volume}{101}},
  \bibinfo{pages}{023401} (\bibinfo{year}{2008}).

\bibitem[{\citenamefont{Doye and Meyer}(2005)}]{PhysRevLett.95.063401}
\bibinfo{author}{\bibfnamefont{J.~P.~K.} \bibnamefont{Doye}} \bibnamefont{and}
  \bibinfo{author}{\bibfnamefont{L.}~\bibnamefont{Meyer}},
  \bibinfo{journal}{Phys. Rev. Lett.} \textbf{\bibinfo{volume}{95}},
  \bibinfo{pages}{063401} (\bibinfo{year}{2005}).

\bibitem[{\citenamefont{Calvo and Yurtsever}(2004)}]{Calvo-Yurtsever:2004}
\bibinfo{author}{\bibfnamefont{F.}~\bibnamefont{Calvo}} \bibnamefont{and}
  \bibinfo{author}{\bibfnamefont{E.}~\bibnamefont{Yurtsever}},
  \bibinfo{journal}{Phys. Rev. B} \textbf{\bibinfo{volume}{70}},
  \bibinfo{pages}{045423} (\bibinfo{year}{2004}).

\bibitem[{\citenamefont{Topper et~al.}(2003)\citenamefont{Topper, Freeman,
  Bergin, and LaMarche}}]{review}
\bibinfo{author}{\bibfnamefont{R.}~\bibnamefont{Topper}},
  \bibinfo{author}{\bibfnamefont{D.}~\bibnamefont{Freeman}},
  \bibinfo{author}{\bibfnamefont{D.}~\bibnamefont{Bergin}}, \bibnamefont{and}
  \bibinfo{author}{\bibfnamefont{K.}~\bibnamefont{LaMarche}},
  \bibinfo{journal}{Rev. Comput. Chem.} \textbf{\bibinfo{volume}{19}},
  \bibinfo{pages}{1} (\bibinfo{year}{2003}).

\bibitem[{\citenamefont{Frantsuzov and Mandelshtam}(2005)}]{PhysRevE.72.037102}
\bibinfo{author}{\bibfnamefont{P.~A.} \bibnamefont{Frantsuzov}}
  \bibnamefont{and} \bibinfo{author}{\bibfnamefont{V.~A.}
  \bibnamefont{Mandelshtam}}, \bibinfo{journal}{Phys. Rev. E}
  \textbf{\bibinfo{volume}{72}}, \bibinfo{pages}{037102}
  (\bibinfo{year}{2005}).

\bibitem[{\citenamefont{Earl and Deem}(2005)}]{Earl-Deem}
\bibinfo{author}{\bibfnamefont{D.~J.} \bibnamefont{Earl}} \bibnamefont{and}
  \bibinfo{author}{\bibfnamefont{M.~W.} \bibnamefont{Deem}},
  \bibinfo{journal}{Phys. Chem. Chem. Phys.} \textbf{\bibinfo{volume}{7}},
  \bibinfo{pages}{3910} (\bibinfo{year}{2005}).

\bibitem[{\citenamefont{Wales et~al.}()\citenamefont{Wales, Doye, Dullweber,
  Hodges, Naumkin, Calvo, Hern\'andez-Rojas, and Middleton}}]{CCD}
\bibinfo{author}{\bibfnamefont{D.}~\bibnamefont{Wales}},
  \bibinfo{author}{\bibfnamefont{J.}~\bibnamefont{Doye}},
  \bibinfo{author}{\bibfnamefont{A.}~\bibnamefont{Dullweber}},
  \bibinfo{author}{\bibfnamefont{M.}~\bibnamefont{Hodges}},
  \bibinfo{author}{\bibfnamefont{F.}~\bibnamefont{Naumkin}},
  \bibinfo{author}{\bibfnamefont{F.}~\bibnamefont{Calvo}},
  \bibinfo{author}{\bibfnamefont{J.}~\bibnamefont{Hern\'andez-Rojas}},
  \bibnamefont{and}
  \bibinfo{author}{\bibfnamefont{T.}~\bibnamefont{Middleton}},
  \emph{\bibinfo{title}{The cambridge cluster database}},
  \urlprefix\url{http://www-wales.ch.cam.ac.uk/CCD.html}.

\bibitem[{\citenamefont{Wales and Doye}(1997)}]{bh}
\bibinfo{author}{\bibfnamefont{D.~J.} \bibnamefont{Wales}} \bibnamefont{and}
  \bibinfo{author}{\bibfnamefont{J.~P.~K.} \bibnamefont{Doye}},
  \bibinfo{journal}{J. Phys. Chem. A} \textbf{\bibinfo{volume}{101}},
  \bibinfo{pages}{5111} (\bibinfo{year}{1997}).

\bibitem[{\citenamefont{Ferrenberg and Swendsen}(1989)}]{multi1}
\bibinfo{author}{\bibfnamefont{A.~M.} \bibnamefont{Ferrenberg}}
  \bibnamefont{and} \bibinfo{author}{\bibfnamefont{R.~H.}
  \bibnamefont{Swendsen}}, \bibinfo{journal}{Phys. Rev. Lett.}
  \textbf{\bibinfo{volume}{63}}, \bibinfo{pages}{1195} (\bibinfo{year}{1989}).

\bibitem[{\citenamefont{Labastie and Whetten}(1990)}]{multi2}
\bibinfo{author}{\bibfnamefont{P.}~\bibnamefont{Labastie}} \bibnamefont{and}
  \bibinfo{author}{\bibfnamefont{R.~L.} \bibnamefont{Whetten}},
  \bibinfo{journal}{Phys. Rev. Lett.} \textbf{\bibinfo{volume}{65}},
  \bibinfo{pages}{1567} (\bibinfo{year}{1990}).

\bibitem[{\citenamefont{Amar and Berry}(1986)}]{berry}
\bibinfo{author}{\bibfnamefont{F.~G.} \bibnamefont{Amar}} \bibnamefont{and}
  \bibinfo{author}{\bibfnamefont{R.~S.} \bibnamefont{Berry}},
  \bibinfo{journal}{J. Chem. Phys.} \textbf{\bibinfo{volume}{85}},
  \bibinfo{pages}{5943} (\bibinfo{year}{1986}).

\bibitem[{\citenamefont{Doye and Calvo}(2001)}]{calvo3}
\bibinfo{author}{\bibfnamefont{J.~P.~K.} \bibnamefont{Doye}} \bibnamefont{and}
  \bibinfo{author}{\bibfnamefont{F.}~\bibnamefont{Calvo}},
  \bibinfo{journal}{Phys. Rev. Lett.} \textbf{\bibinfo{volume}{86}},
  \bibinfo{pages}{3570} (\bibinfo{year}{2001}).

\bibitem[{\citenamefont{Pahl et~al.}(2008)\citenamefont{Pahl, Calvo, Koccaroni,
  and Schwerdtfeger}}]{calvo2}
\bibinfo{author}{\bibfnamefont{E.}~\bibnamefont{Pahl}},
  \bibinfo{author}{\bibfnamefont{F.}~\bibnamefont{Calvo}},
  \bibinfo{author}{\bibfnamefont{L.}~\bibnamefont{Koccaroni}},
  \bibnamefont{and}
  \bibinfo{author}{\bibfnamefont{P.}~\bibnamefont{Schwerdtfeger}},
  \bibinfo{journal}{Angew. Chem. Int. Ed.} \textbf{\bibinfo{volume}{47}},
  \bibinfo{pages}{8207} (\bibinfo{year}{2008}).

\bibitem[{\citenamefont{White et~al.}(2005)\citenamefont{White, Cleary, and
  Mayne}}]{white}
\bibinfo{author}{\bibfnamefont{R.~P.} \bibnamefont{White}},
  \bibinfo{author}{\bibfnamefont{S.~M.} \bibnamefont{Cleary}},
  \bibnamefont{and} \bibinfo{author}{\bibfnamefont{H.~R.} \bibnamefont{Mayne}},
  \bibinfo{journal}{J. Chem. Phys.} \textbf{\bibinfo{volume}{123}},
  \bibinfo{eid}{094505} (\bibinfo{year}{2005}).

\bibitem[{\citenamefont{Frantz}(1996)}]{frantz:10030}
\bibinfo{author}{\bibfnamefont{D.~D.} \bibnamefont{Frantz}},
  \bibinfo{journal}{J. Chem. Phys.} \textbf{\bibinfo{volume}{105}},
  \bibinfo{pages}{10030} (\bibinfo{year}{1996}).

\bibitem[{\citenamefont{Frantz}(1997)}]{frantz:1992}
\bibinfo{author}{\bibfnamefont{D.~D.} \bibnamefont{Frantz}},
  \bibinfo{journal}{J. Chem. Phys.} \textbf{\bibinfo{volume}{107}},
  \bibinfo{pages}{1992} (\bibinfo{year}{1997}).

\bibitem[{\citenamefont{Sabo et~al.}(2004{\natexlab{a}})\citenamefont{Sabo,
  Doll, and Freeman}}]{sabo:847}
\bibinfo{author}{\bibfnamefont{D.}~\bibnamefont{Sabo}},
  \bibinfo{author}{\bibfnamefont{J.~D.} \bibnamefont{Doll}}, \bibnamefont{and}
  \bibinfo{author}{\bibfnamefont{D.~L.} \bibnamefont{Freeman}},
  \bibinfo{journal}{J. Chem. Phys.} \textbf{\bibinfo{volume}{121}},
  \bibinfo{pages}{847} (\bibinfo{year}{2004}{\natexlab{a}}).

\bibitem[{\citenamefont{Sabo et~al.}(2004{\natexlab{b}})\citenamefont{Sabo,
  Predescu, Doll, and Freeman}}]{sabo:856}
\bibinfo{author}{\bibfnamefont{D.}~\bibnamefont{Sabo}},
  \bibinfo{author}{\bibfnamefont{C.}~\bibnamefont{Predescu}},
  \bibinfo{author}{\bibfnamefont{J.~D.} \bibnamefont{Doll}}, \bibnamefont{and}
  \bibinfo{author}{\bibfnamefont{D.~L.} \bibnamefont{Freeman}},
  \bibinfo{journal}{J. Chem. Phys.} \textbf{\bibinfo{volume}{121}},
  \bibinfo{pages}{856} (\bibinfo{year}{2004}{\natexlab{b}}).

\bibitem[{\citenamefont{Munro et~al.}(2002)\citenamefont{Munro, Tharrington,
  and Jordan}}]{Munro20021}
\bibinfo{author}{\bibfnamefont{L.~J.} \bibnamefont{Munro}},
  \bibinfo{author}{\bibfnamefont{A.}~\bibnamefont{Tharrington}},
  \bibnamefont{and} \bibinfo{author}{\bibfnamefont{K.~D.}
  \bibnamefont{Jordan}}, \bibinfo{journal}{Comput. Phys. Commun.}
  \textbf{\bibinfo{volume}{145}}, \bibinfo{pages}{1 } (\bibinfo{year}{2002}).

\end{thebibliography}

\begin{widetext}

\begin{figure}
\includegraphics[width=0.55\textwidth,angle=-90]{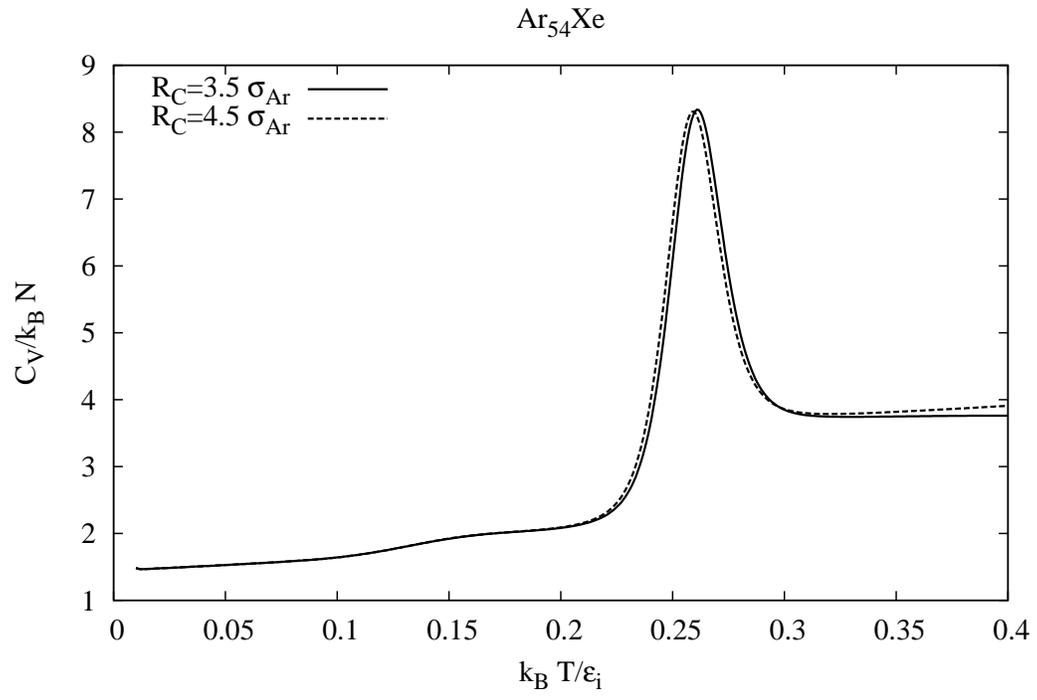}%
\caption{\label{fig:volume} Volume dependence of the $C_V$ curve for Ar$_{54}$Xe.}
\end{figure}

\begin{figure}
\includegraphics[width=0.95\textwidth]{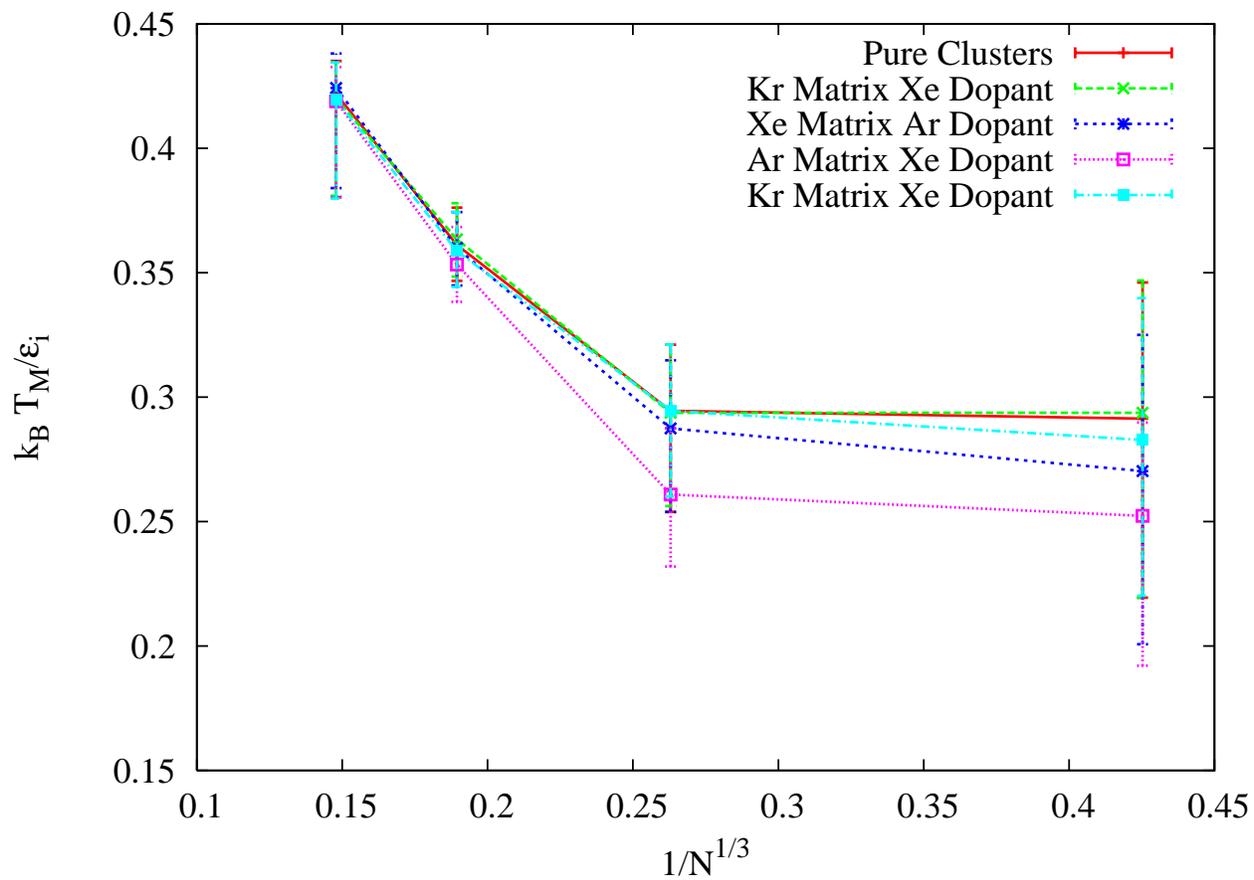}%
\caption{\label{fig:melting} Melting temperatures ($T_M$) as a function of
$N^{-\frac{1}{3}}$ for each type of composition studied. The bars represent the width of the peak associated with the melting of the cluster.}
\end{figure}

\begin{figure}[!ht]
    \centering a)
  \begin{minipage}{\textwidth}
    \includegraphics[width=0.8\textwidth]{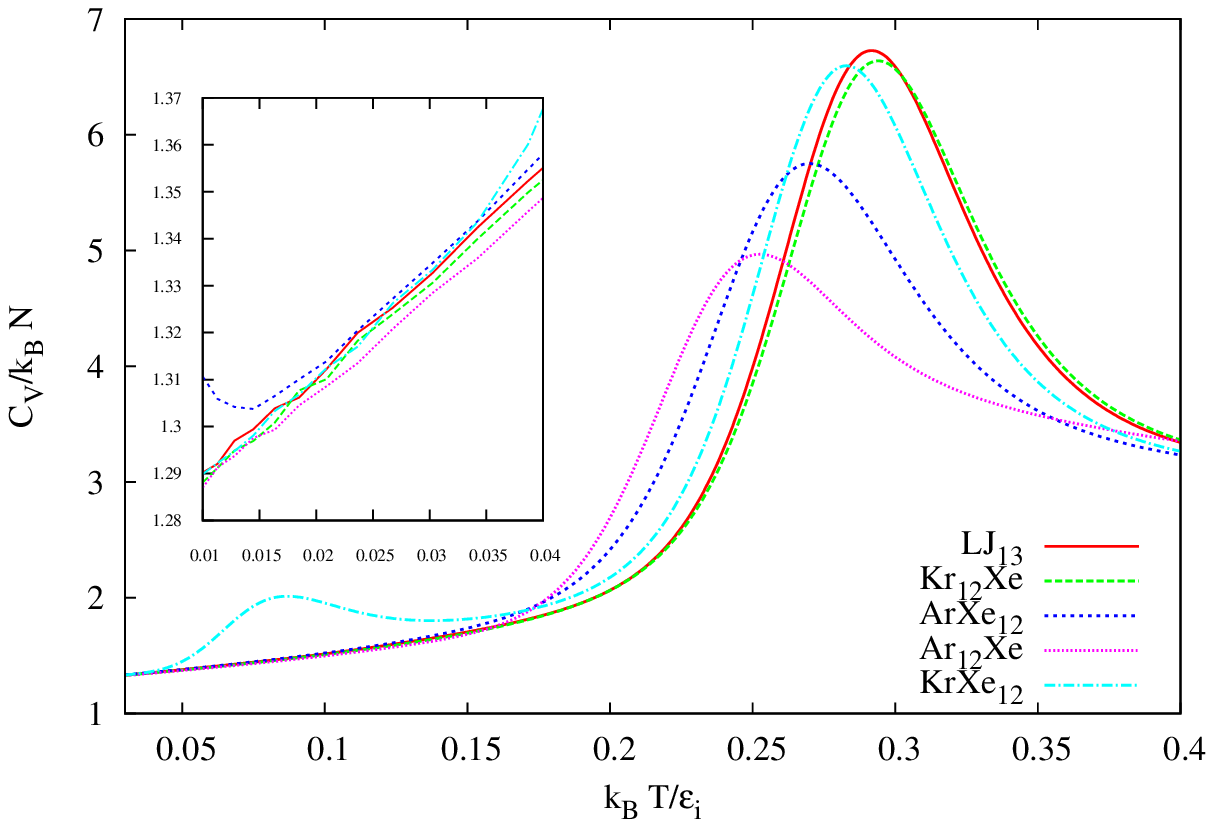}
  \end{minipage}
    \centering b)
  \begin{minipage}{\textwidth}
    \includegraphics[width=0.8\textwidth]{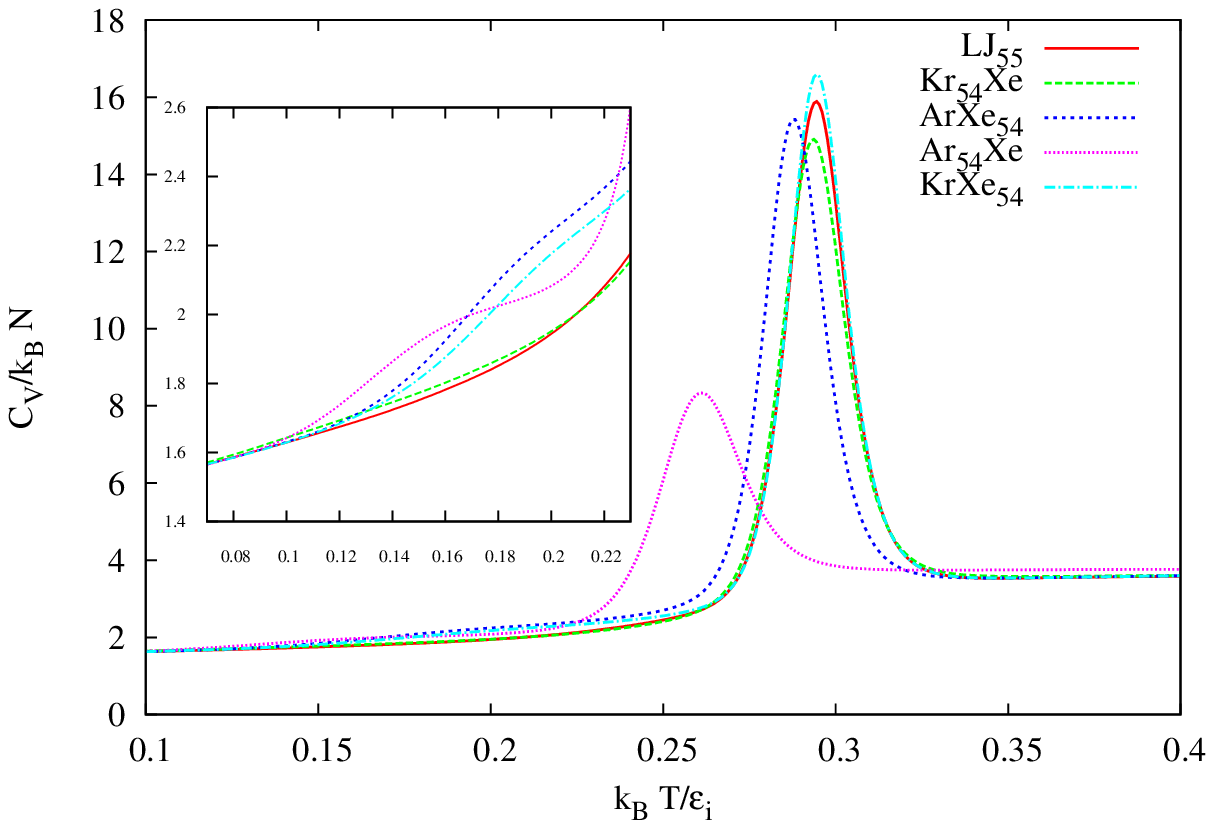}
  \end{minipage}
\caption{\label{fig:13y55cv}Constant Volume Heat Capacities $C_V$ as a function of
temperature for cluster sizes 13 and 55.}
\end{figure}

\begin{figure}[!ht]
    \centering a)
 \begin{minipage}{\textwidth}
    \includegraphics[width=0.8\textwidth]{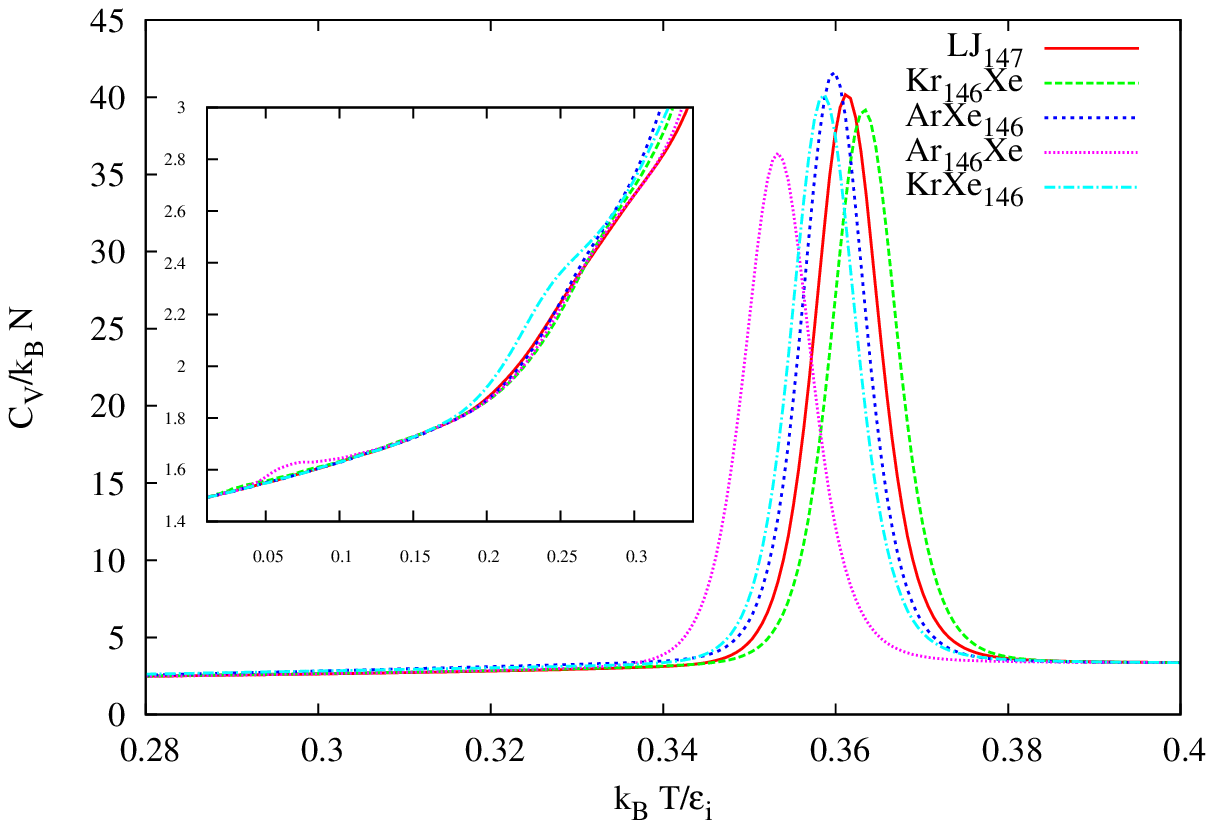}
  \end{minipage}
\centering b)
  \begin{minipage}{\textwidth}
    
    \includegraphics[width=0.8\textwidth]{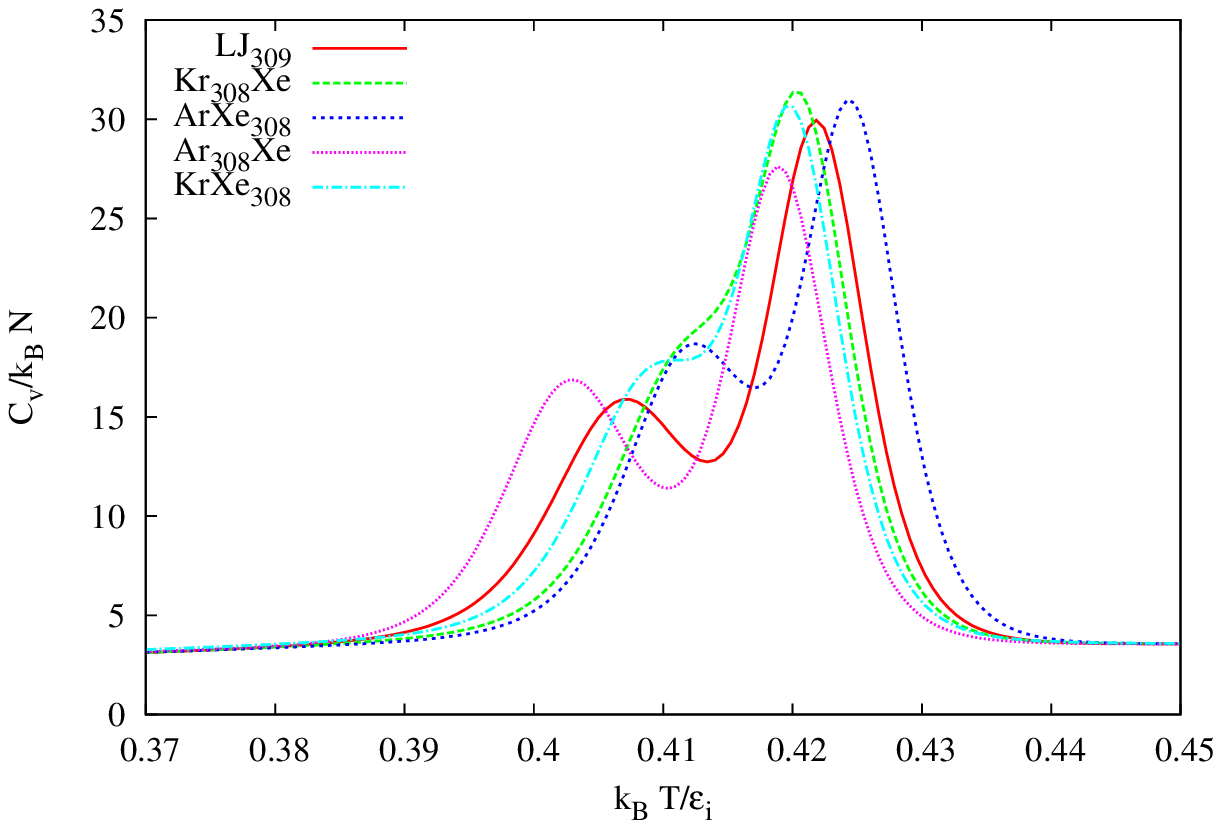}
  \end{minipage}
\caption{\label{fig:147y309cv}Constant Volume Heat Capacities $C_V$ as a function of
temperature for cluster sizes 147 and 309.}
\end{figure}

\begin{figure}[!ht]
  \begin{minipage}{0.23\textwidth}
    \centering ArXe$_{12}$
  \end{minipage}
  \ \hfill 
  \begin{minipage}{0.23\textwidth}
    \centering KrXe$_{12}$
  \end{minipage}
  \ \hfill
  \begin{minipage}{0.23\textwidth}
    \centering Ar$_{12}$Xe
  \end{minipage}
  \begin{minipage}{0.23\textwidth}
    \centering Kr$_{12}$Xe
  \end{minipage}
  \\
\bigskip
\bigskip
 \begin{minipage}{0.23\textwidth}
    \includegraphics[width=\textwidth]{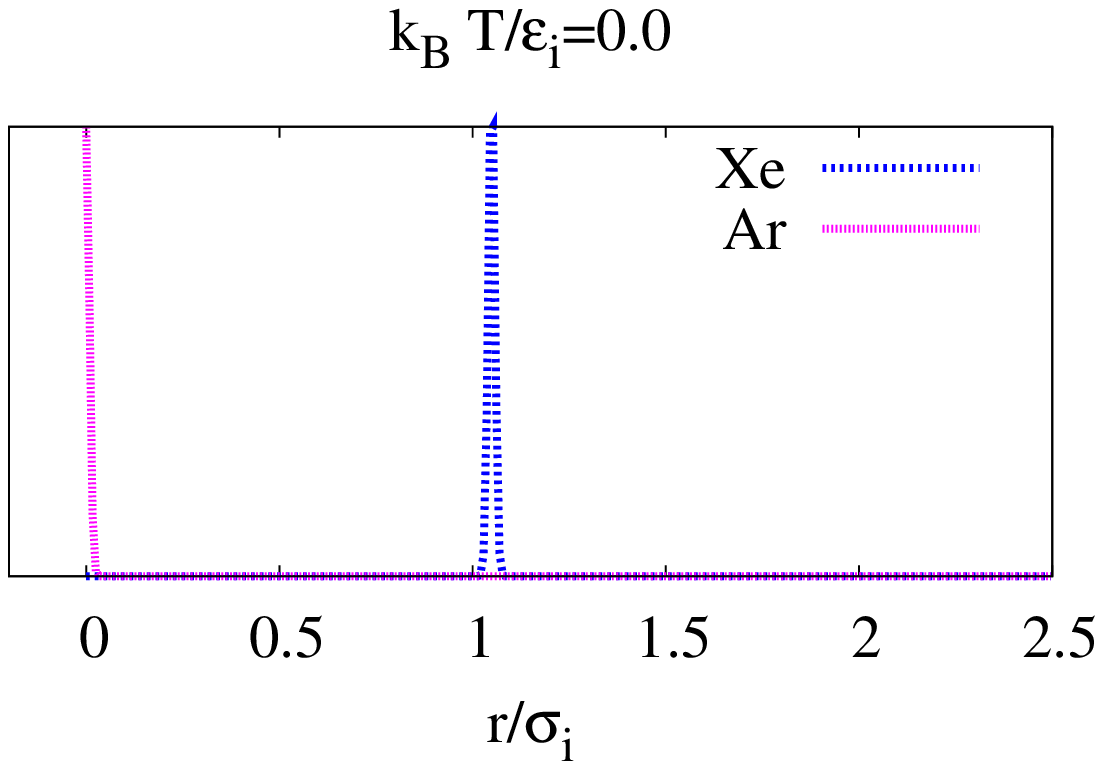}
  \end{minipage}
  \ \hfill 
 \begin{minipage}{0.23\textwidth}
    \includegraphics[width=\textwidth]{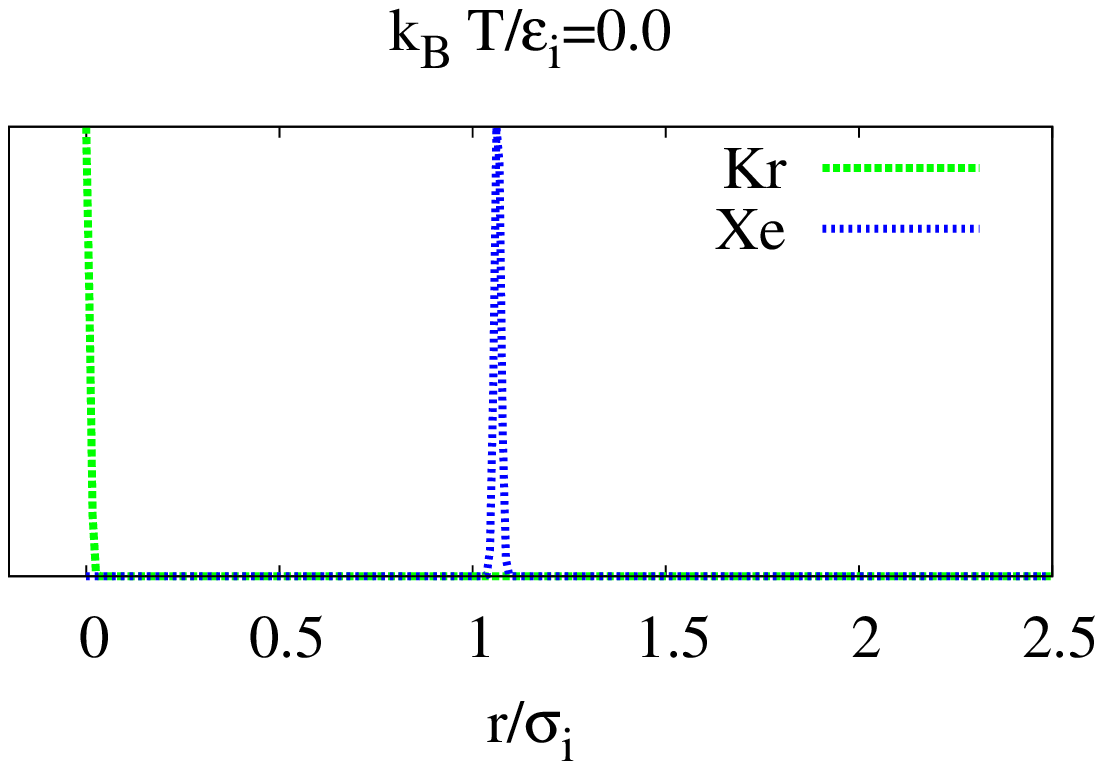}
  \end{minipage}
  \ \hfill 
  \begin{minipage}{0.23\textwidth}
    \includegraphics[width=\textwidth]{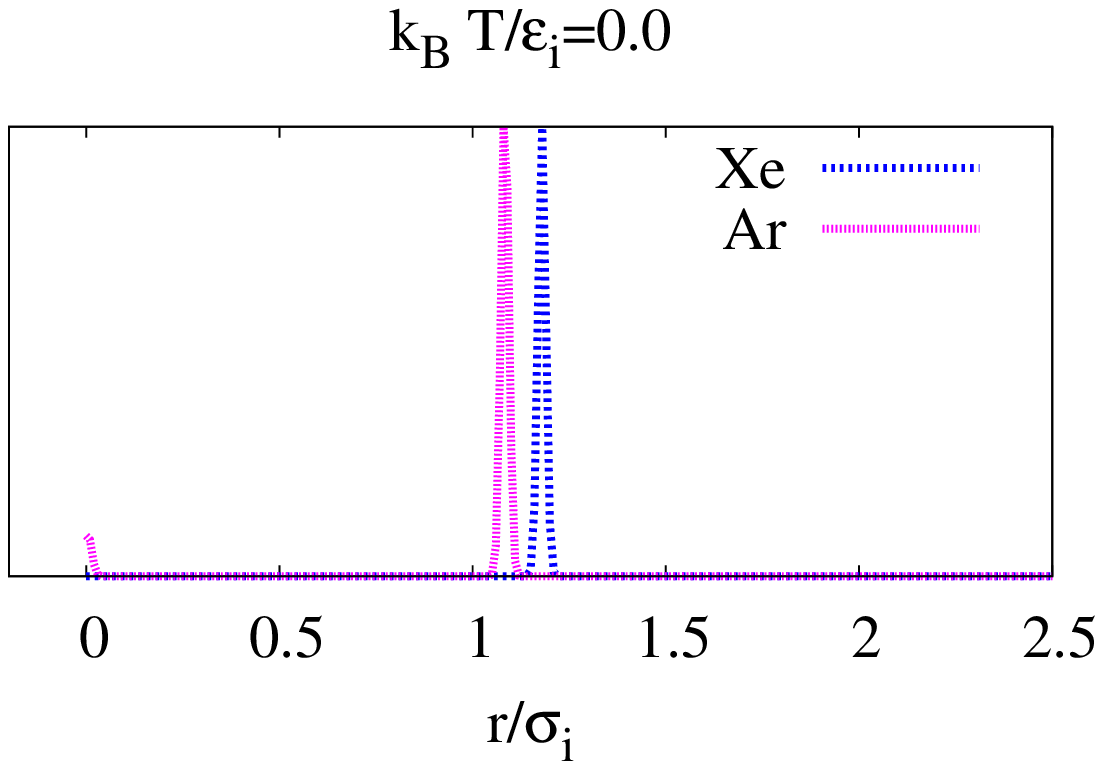}
  \end{minipage}
  \ \hfill
  \begin{minipage}{0.23\textwidth}
    \includegraphics[width=\textwidth]{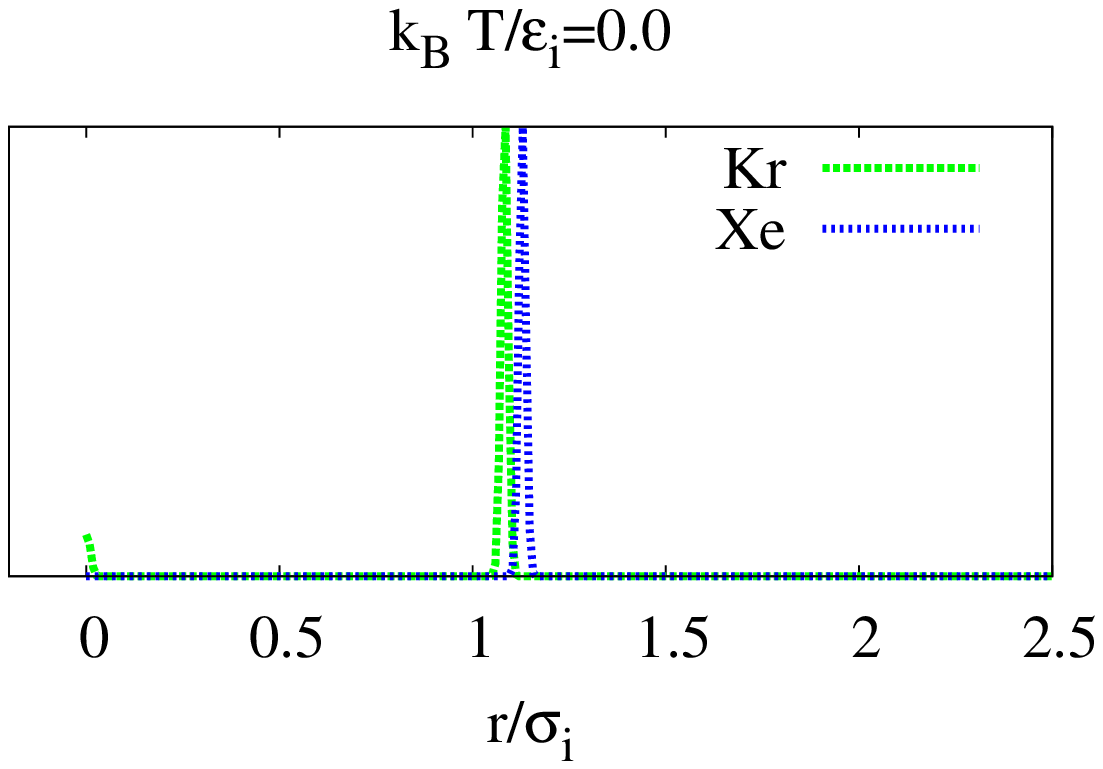}
  \end{minipage}
  \\
\bigskip
\bigskip
 \begin{minipage}{0.23\textwidth}
    \includegraphics[width=\textwidth]{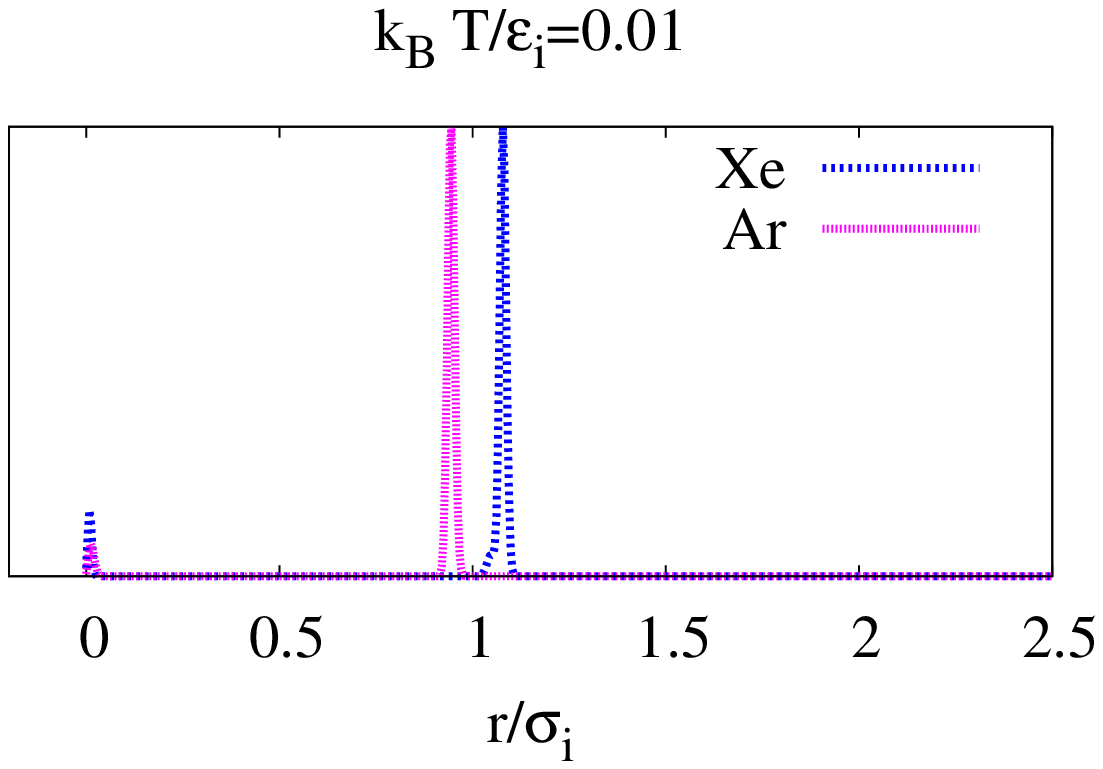}
  \end{minipage}
  \ \hfill 
 \begin{minipage}{0.23\textwidth}
    \includegraphics[width=\textwidth]{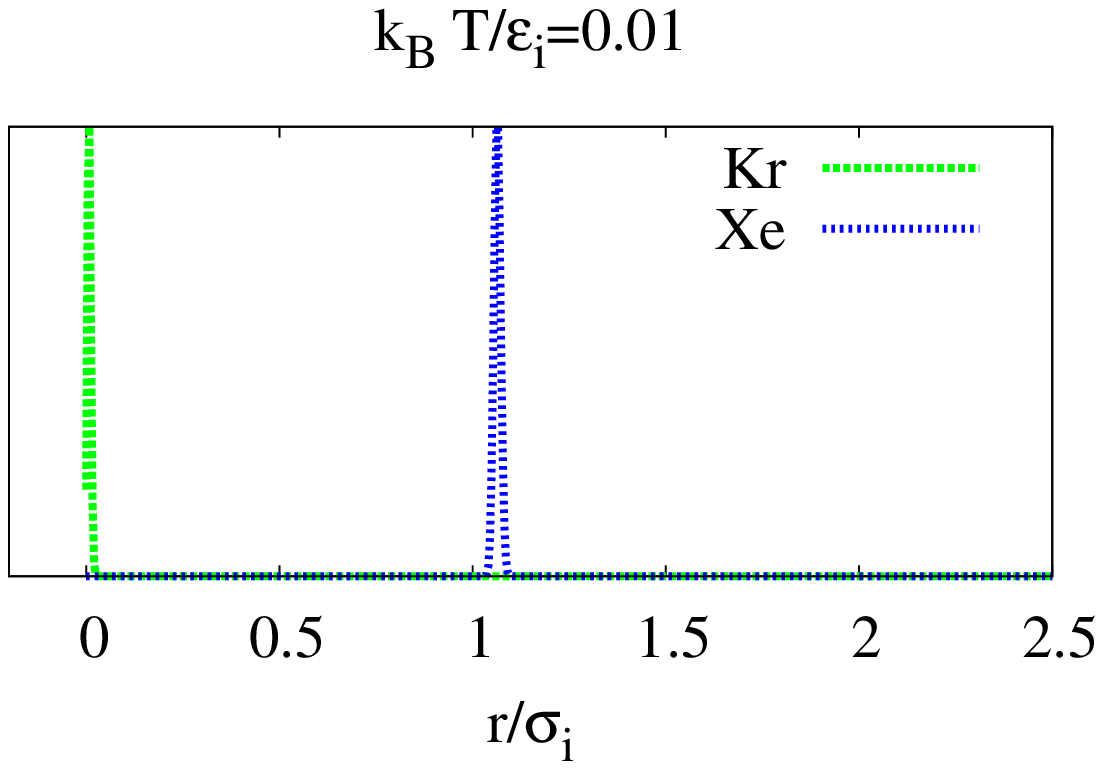}
  \end{minipage}
  \ \hfill 
  \begin{minipage}{0.23\textwidth}
    \includegraphics[width=\textwidth]{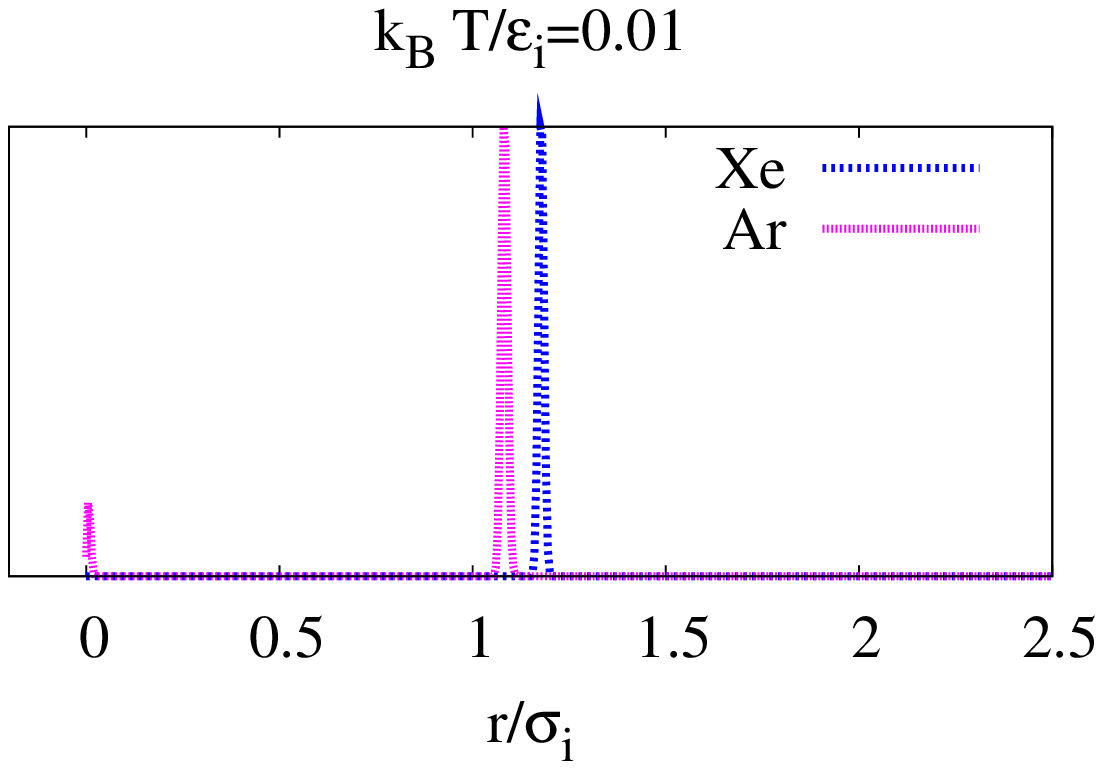}
  \end{minipage}
  \ \hfill
  \begin{minipage}{0.23\textwidth}
    \includegraphics[width=\textwidth]{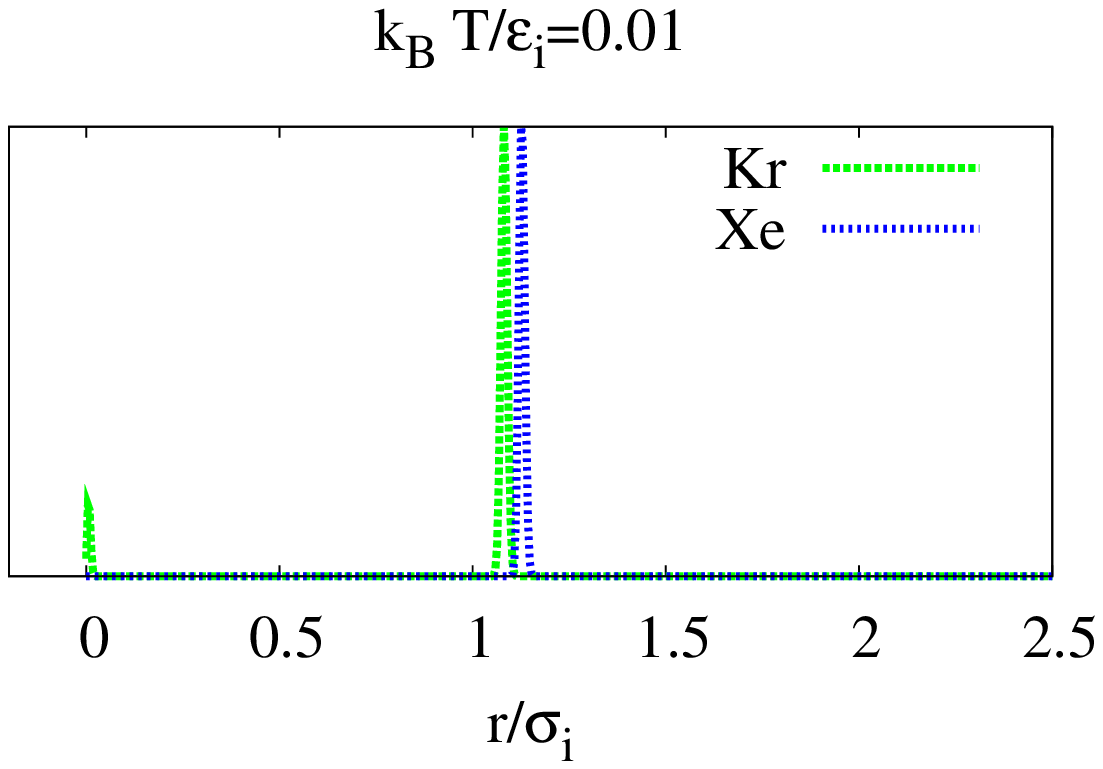}
  \end{minipage}
  \\
\bigskip
\bigskip
 \begin{minipage}{0.23\textwidth}
    \includegraphics[width=\textwidth]{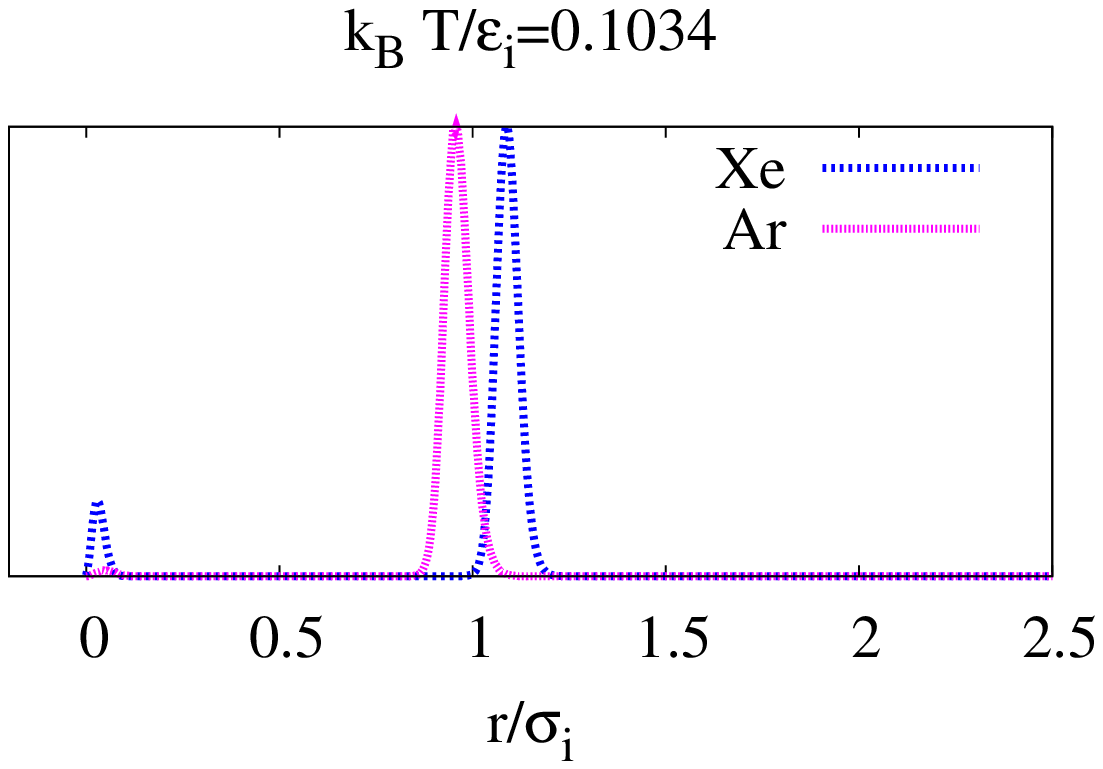}
  \end{minipage}
  \ \hfill 
 \begin{minipage}{0.23\textwidth}
    \includegraphics[width=\textwidth]{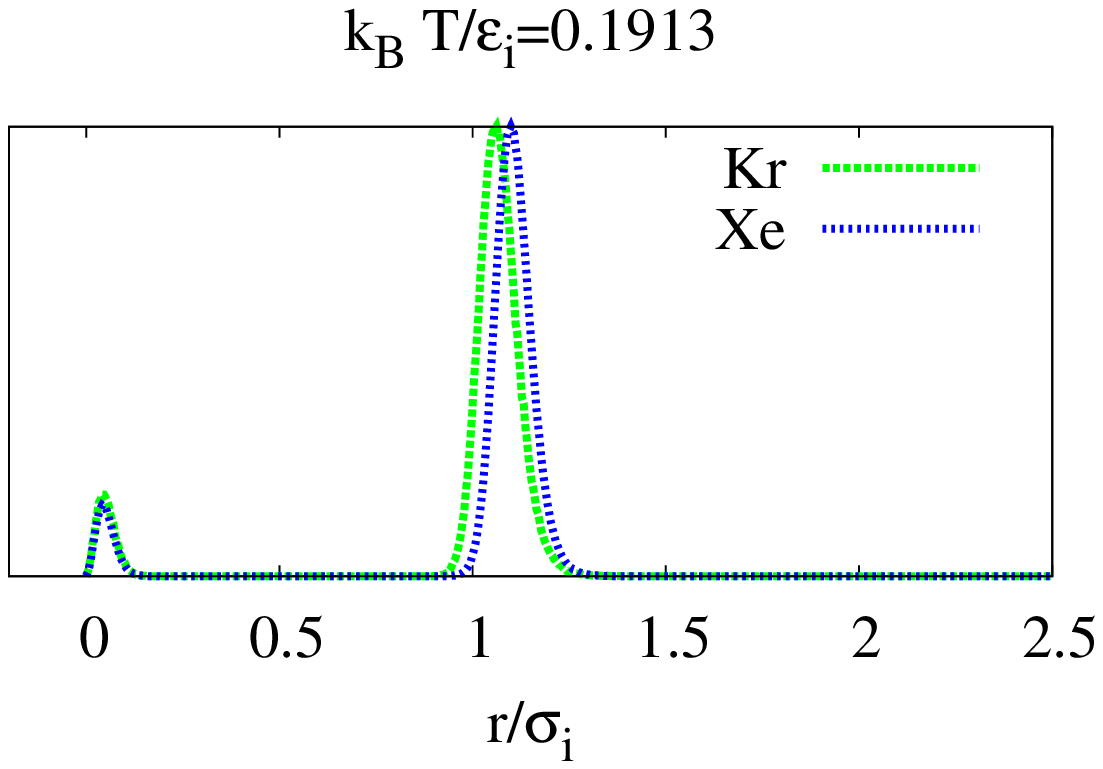}
  \end{minipage}
  \ \hfill 
  \begin{minipage}{0.23\textwidth}
    \includegraphics[width=\textwidth]{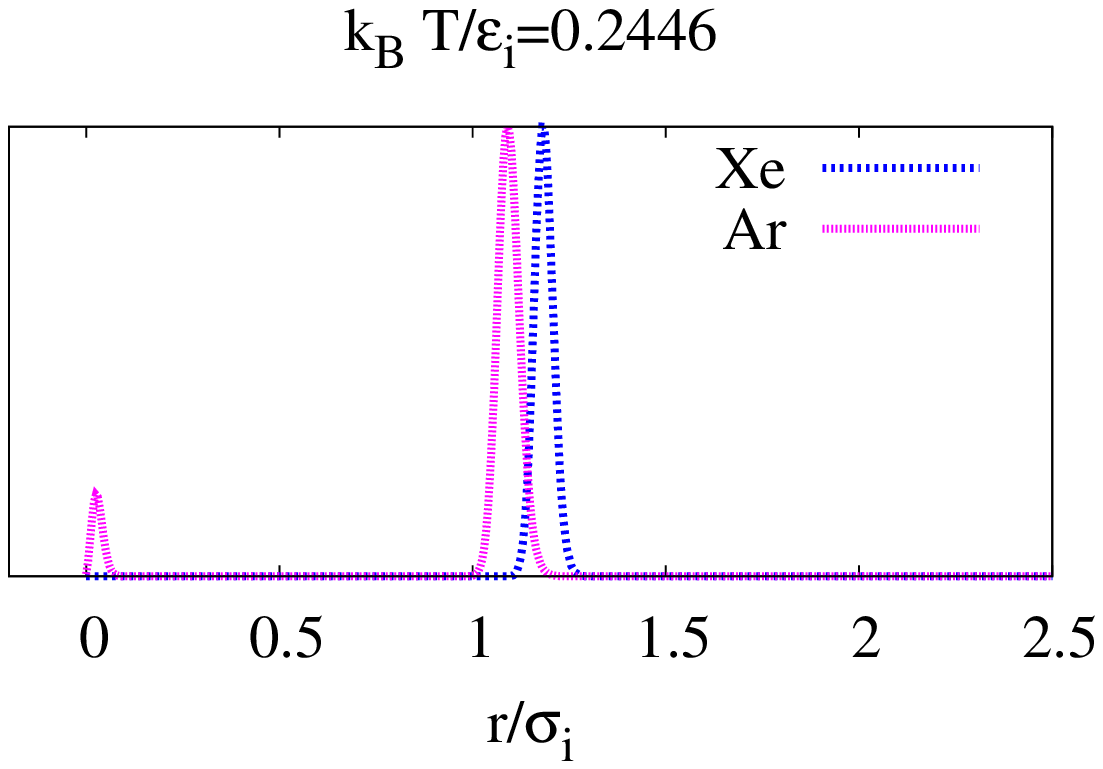}
  \end{minipage}
  \ \hfill
  \begin{minipage}{0.23\textwidth}
    \includegraphics[width=\textwidth]{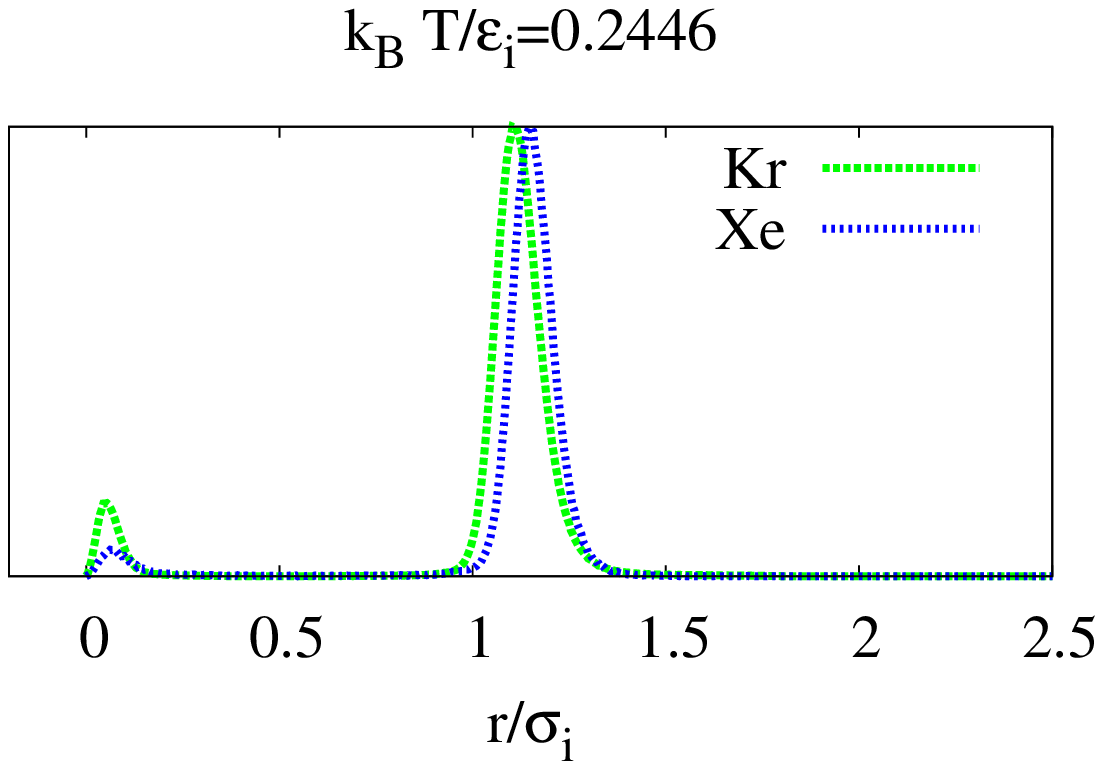}
  \end{minipage}
  \\
\bigskip
\bigskip
 \begin{minipage}{0.23\textwidth}
    \includegraphics[width=\textwidth]{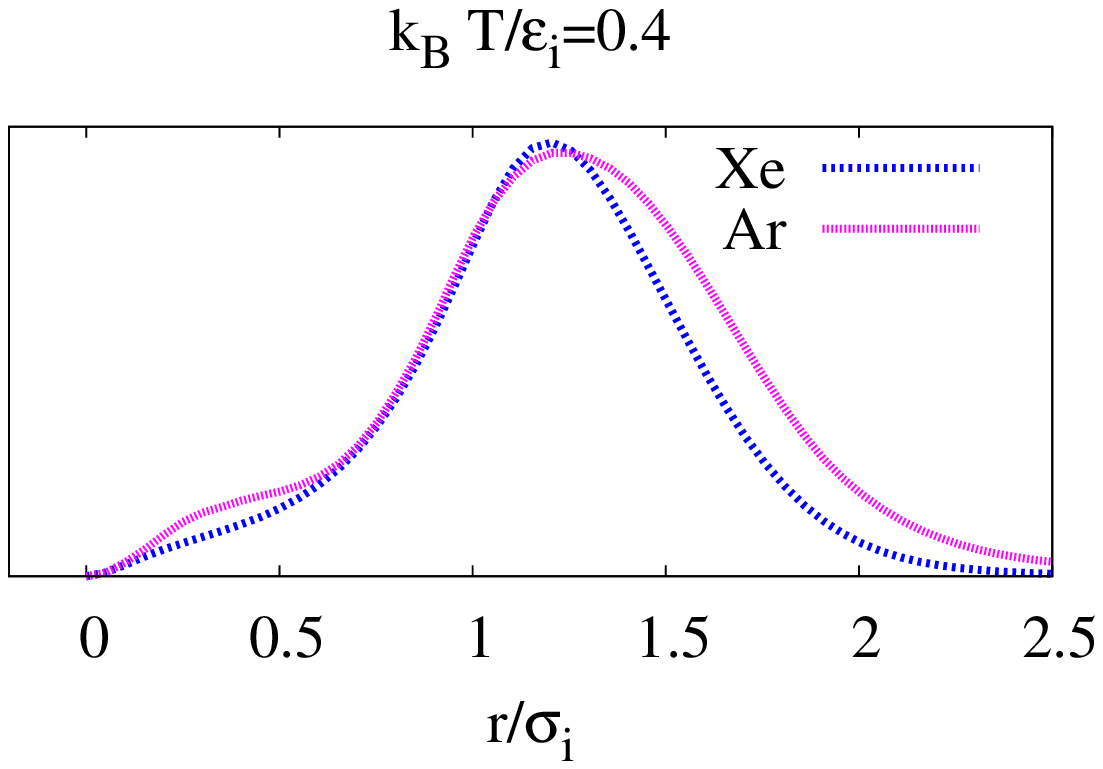}
  \end{minipage}
  \ \hfill  \begin{minipage}{0.23\textwidth}
    \includegraphics[width=\textwidth]{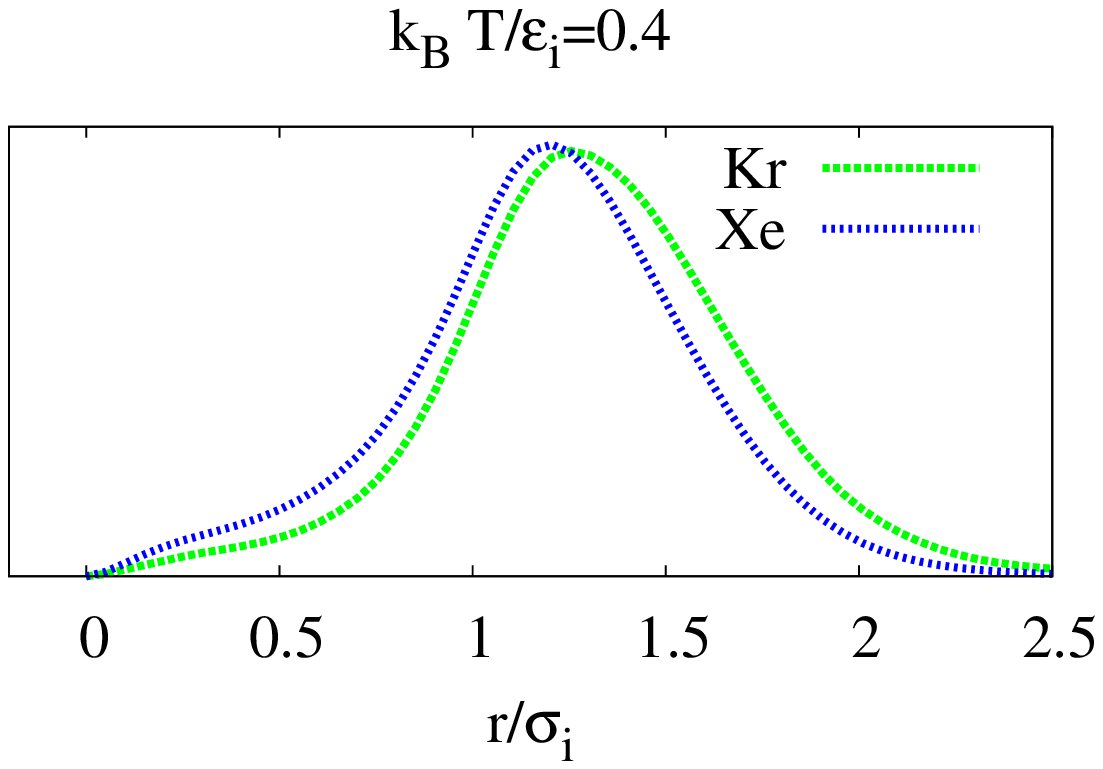}
  \end{minipage}
  \ \hfill 
  \begin{minipage}{0.23\textwidth}
    \includegraphics[width=\textwidth]{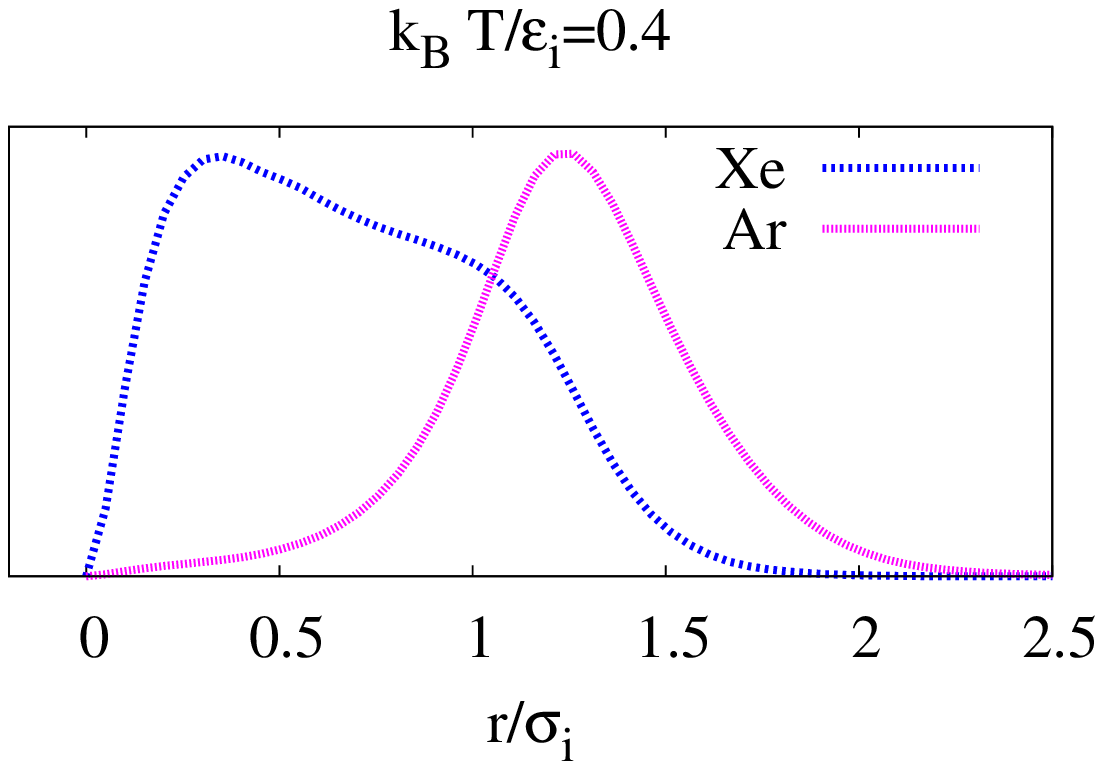}
  \end{minipage}
  \ \hfill
  \begin{minipage}{0.23\textwidth}
    \includegraphics[width=\textwidth]{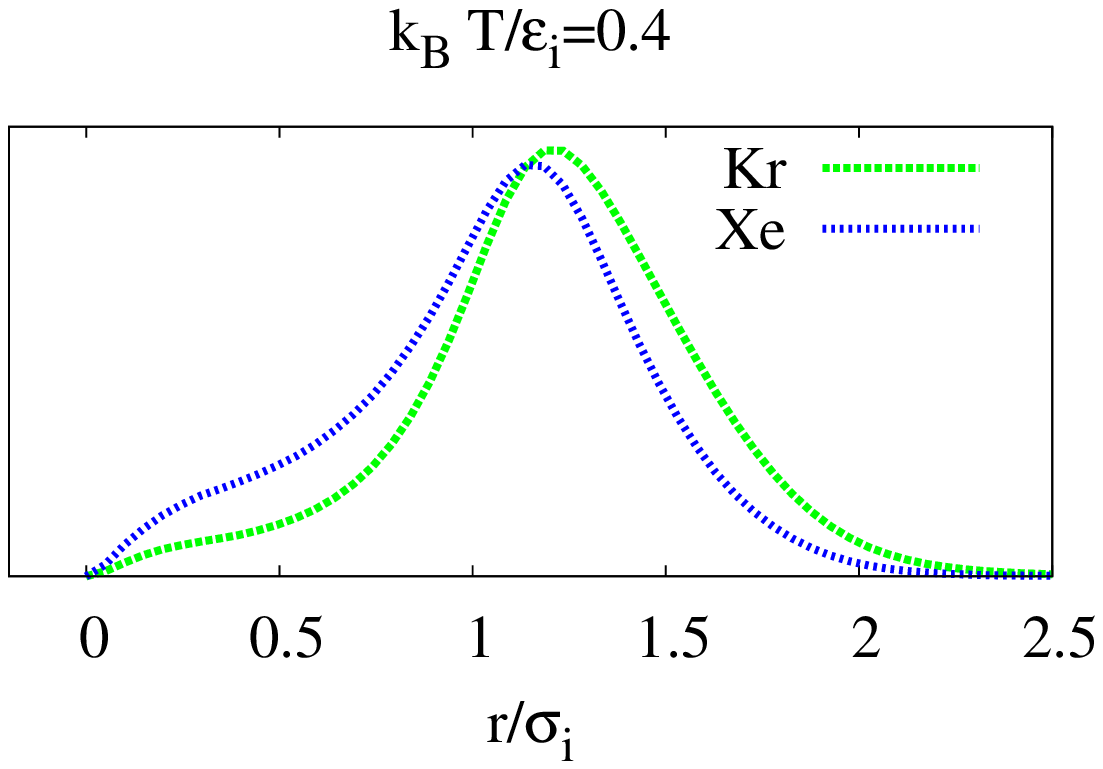}
  \end{minipage}
  \\
\caption{\label{fig:gs13} Radial distribution function for 13 atom clusters. The distribution of the figures is as follows: The columns corresponds to the compositions ArXe$_{N-1}$, KrXe$_{N-1}$, Ar$_{N-1}$Xe and Kr$_{N-1}$Xe. The top row of the panel shows the RDFs of the lowest energy structures. The bottom row contains the RDFs of the clusters once they have melted. The second and third rows correspond to intermediate temperatures of the melting range, illustrating the structural changes discussed in the text. This panel description applies also to figures  \ref{fig:gs55}, \ref{fig:gs147}, and \ref{fig:gs309}.}
\end{figure}

\begin{figure}[!ht]
  \begin{minipage}{0.23\textwidth}
    \centering ArXe$_{54}$
  \end{minipage}
  \ \hfill 
  \begin{minipage}{0.23\textwidth}
    \centering KrXe$_{54}$
  \end{minipage}
  \ \hfill
  \begin{minipage}{0.23\textwidth}
    \centering Ar$_{54}$Xe
  \end{minipage}
  \begin{minipage}{0.23\textwidth}
    \centering Kr$_{54}$Xe
  \end{minipage}
  \\
\bigskip
\bigskip
 \begin{minipage}{0.23\textwidth}
    \includegraphics[width=\textwidth]{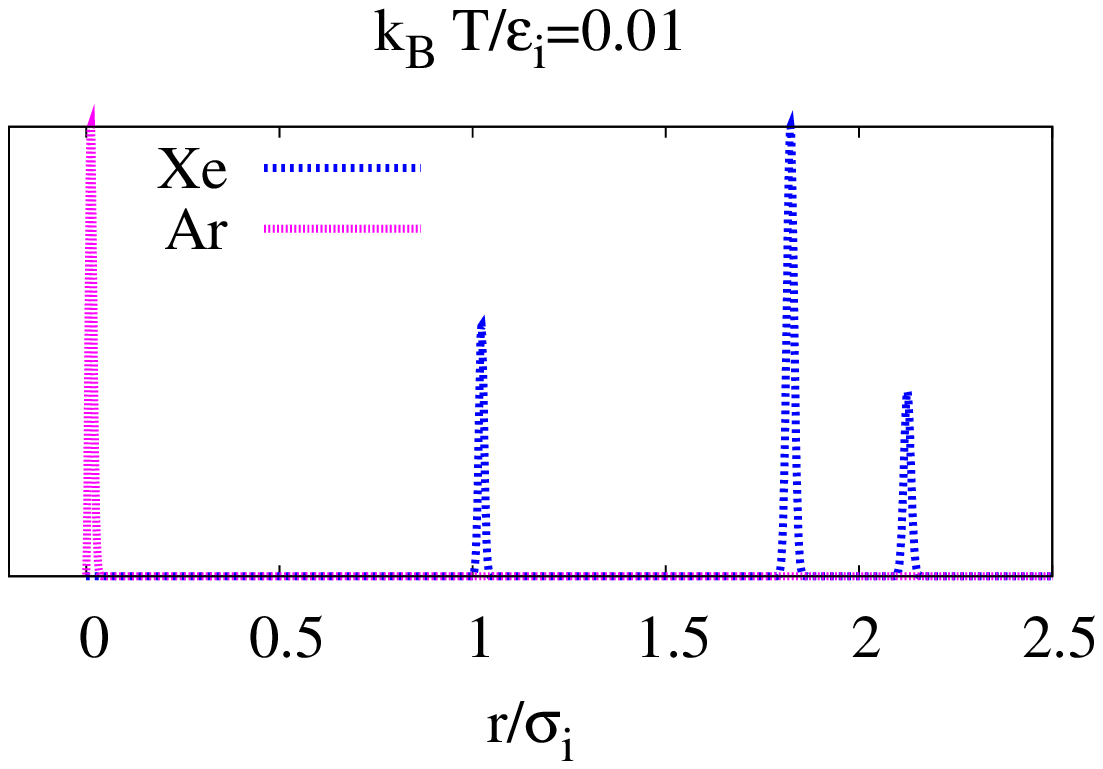}
  \end{minipage}
  \ \hfill 
 \begin{minipage}{0.23\textwidth}
    \includegraphics[width=\textwidth]{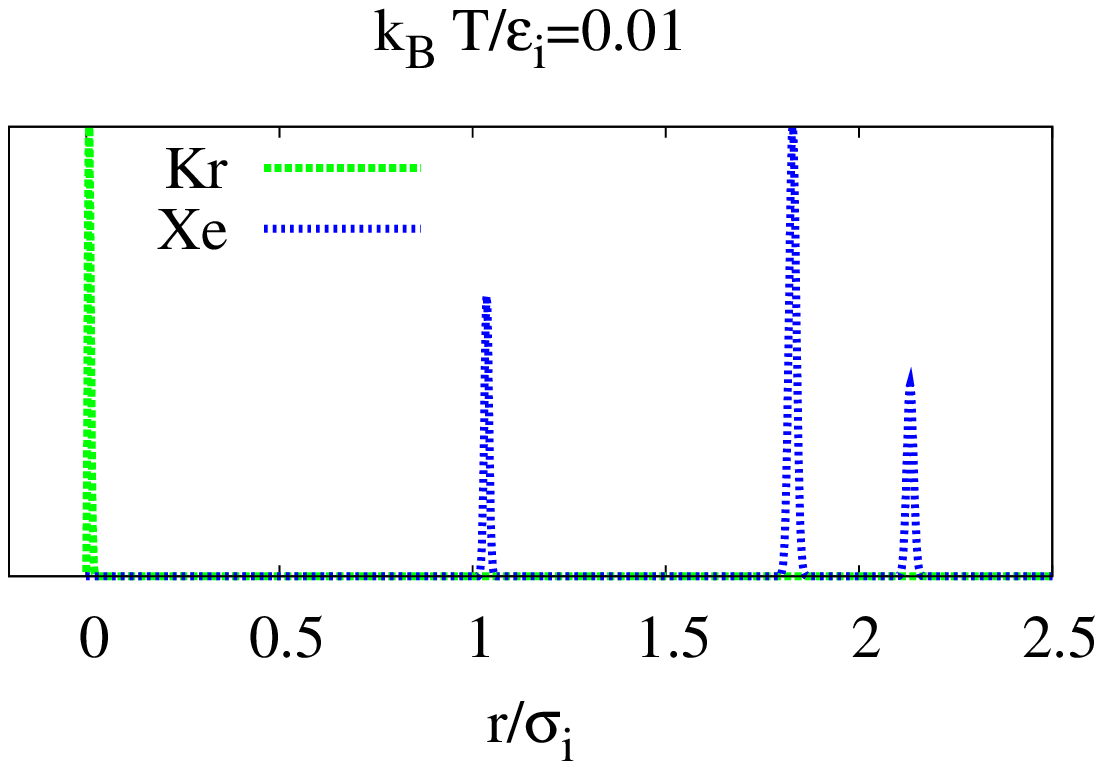}
  \end{minipage}
  \ \hfill 
  \begin{minipage}{0.23\textwidth}
    \includegraphics[width=\textwidth]{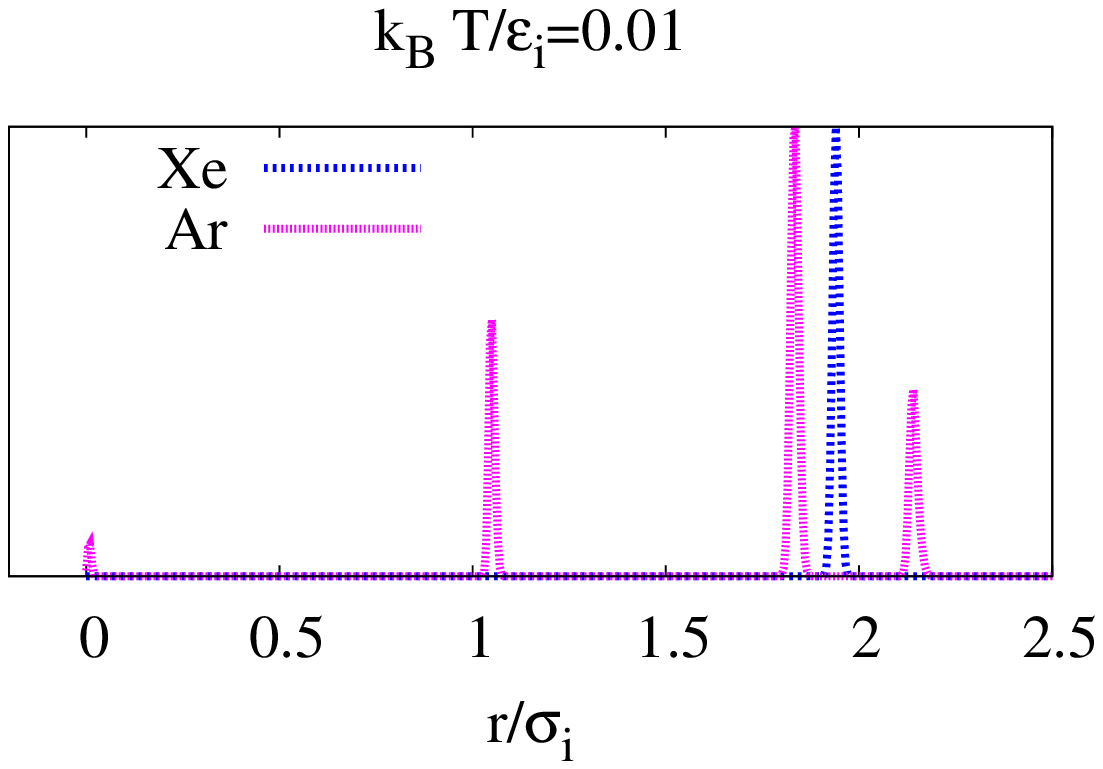}
  \end{minipage}
  \ \hfill
  \begin{minipage}{0.23\textwidth}
    \includegraphics[width=\textwidth]{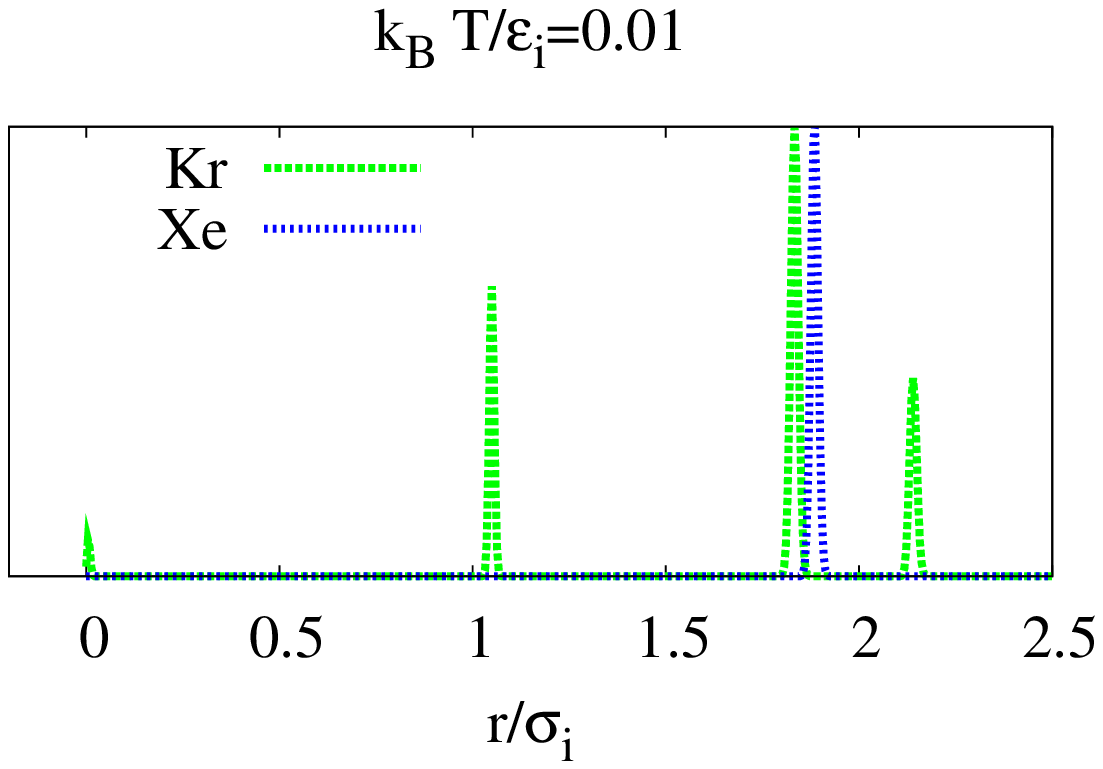}
  \end{minipage}
  \\
\bigskip
\bigskip
 \begin{minipage}{0.23\textwidth}
    \includegraphics[width=\textwidth]{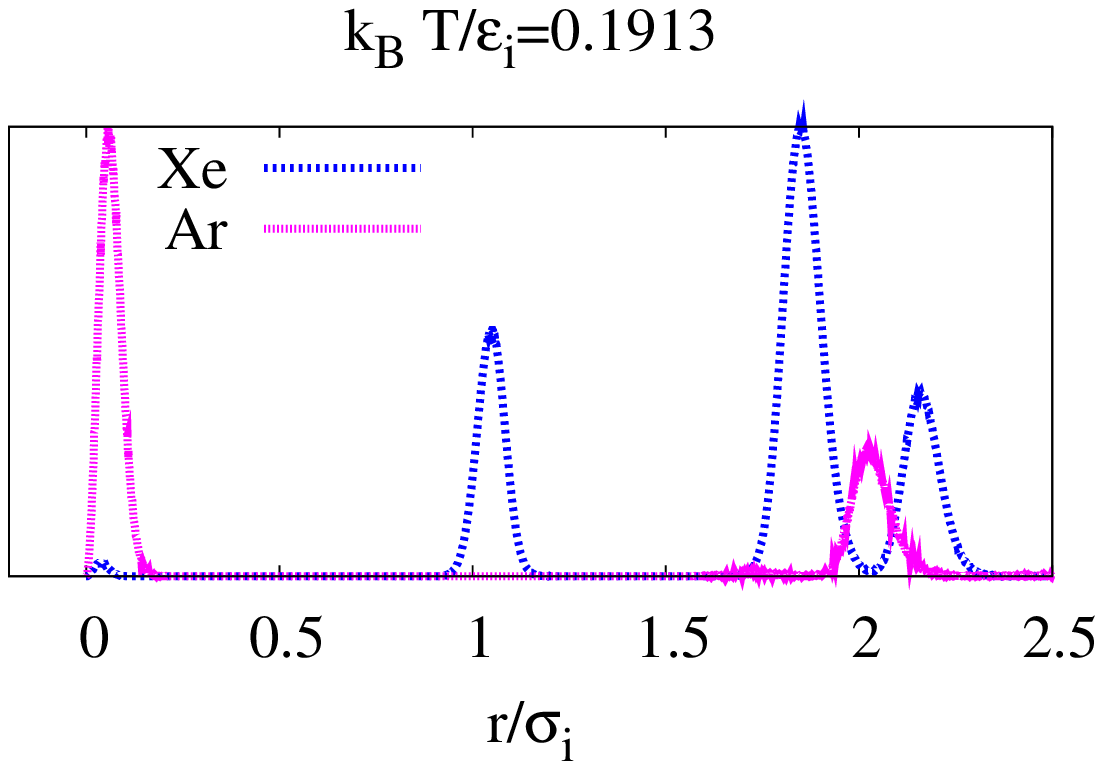}
  \end{minipage}
  \ \hfill 
 \begin{minipage}{0.23\textwidth}
    \includegraphics[width=\textwidth]{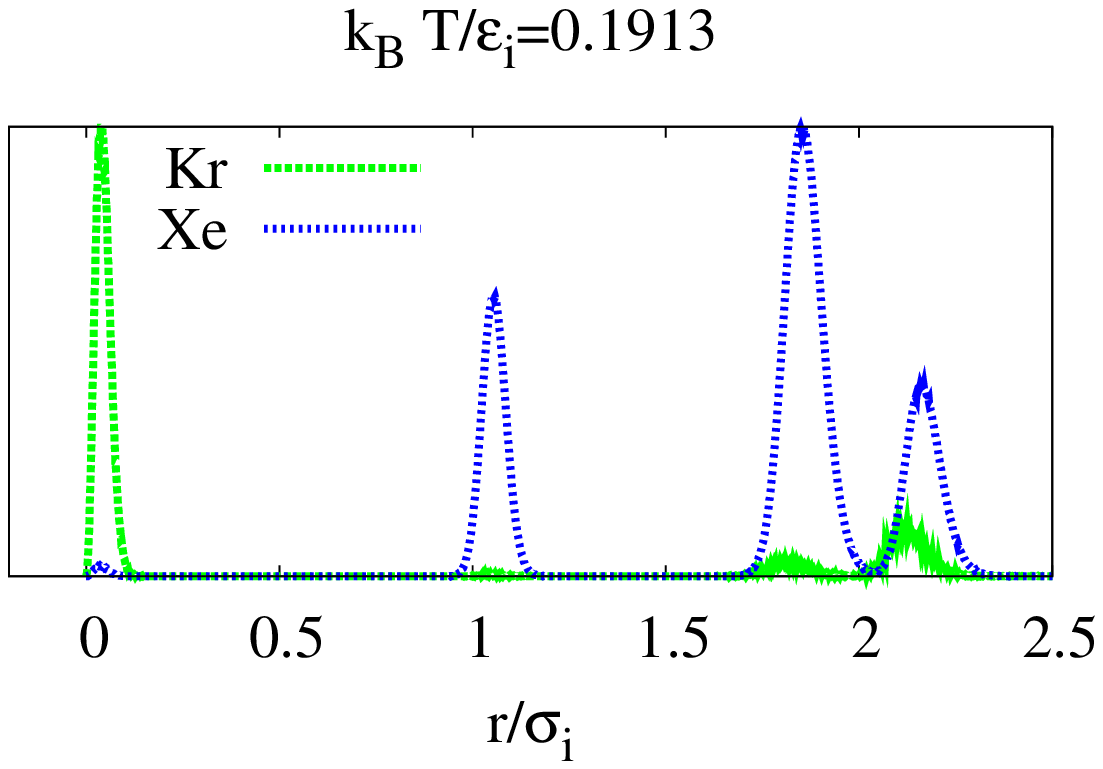}
  \end{minipage}
  \ \hfill 
  \begin{minipage}{0.23\textwidth}
    \includegraphics[width=\textwidth]{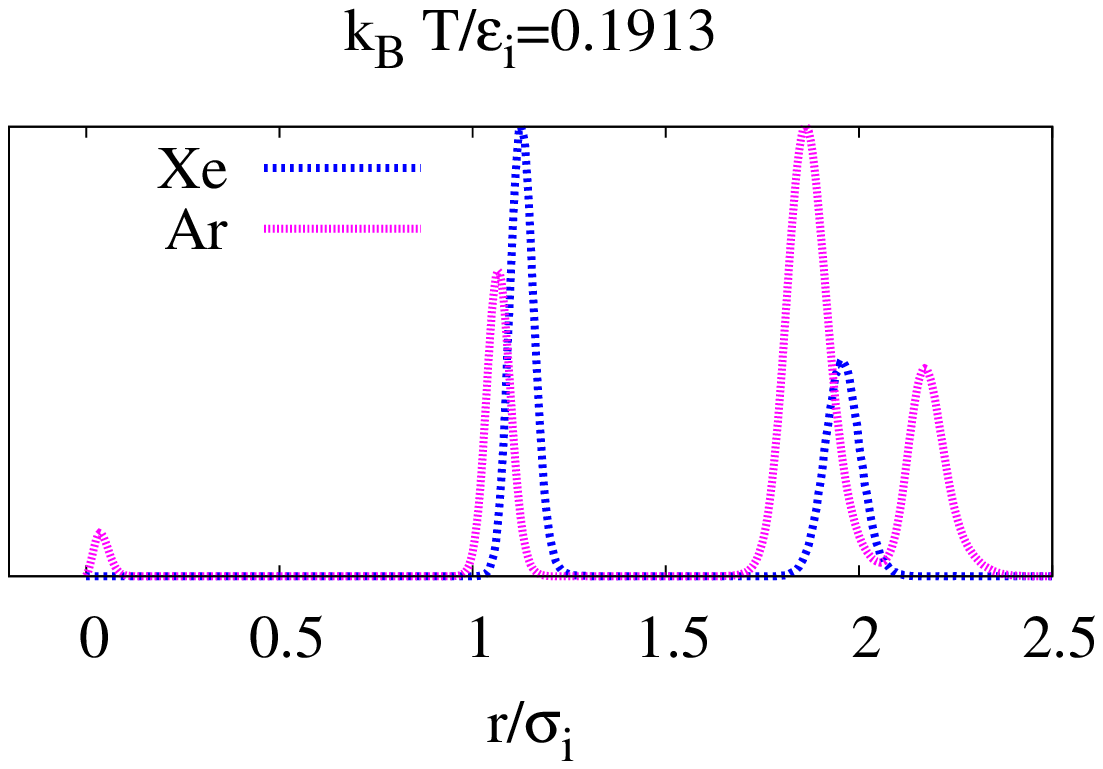}
  \end{minipage}
  \ \hfill
  \begin{minipage}{0.23\textwidth}
    \includegraphics[width=\textwidth]{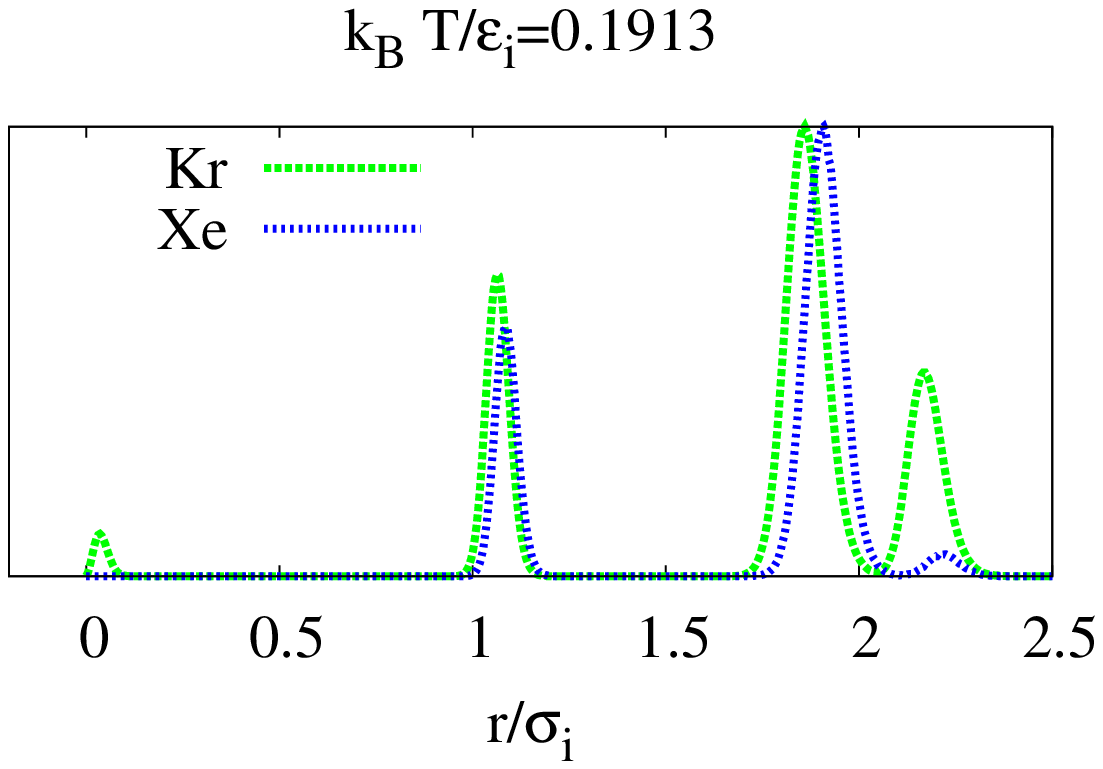}
  \end{minipage}
  \\
\bigskip
\bigskip
 \begin{minipage}{0.23\textwidth}
    \includegraphics[width=\textwidth]{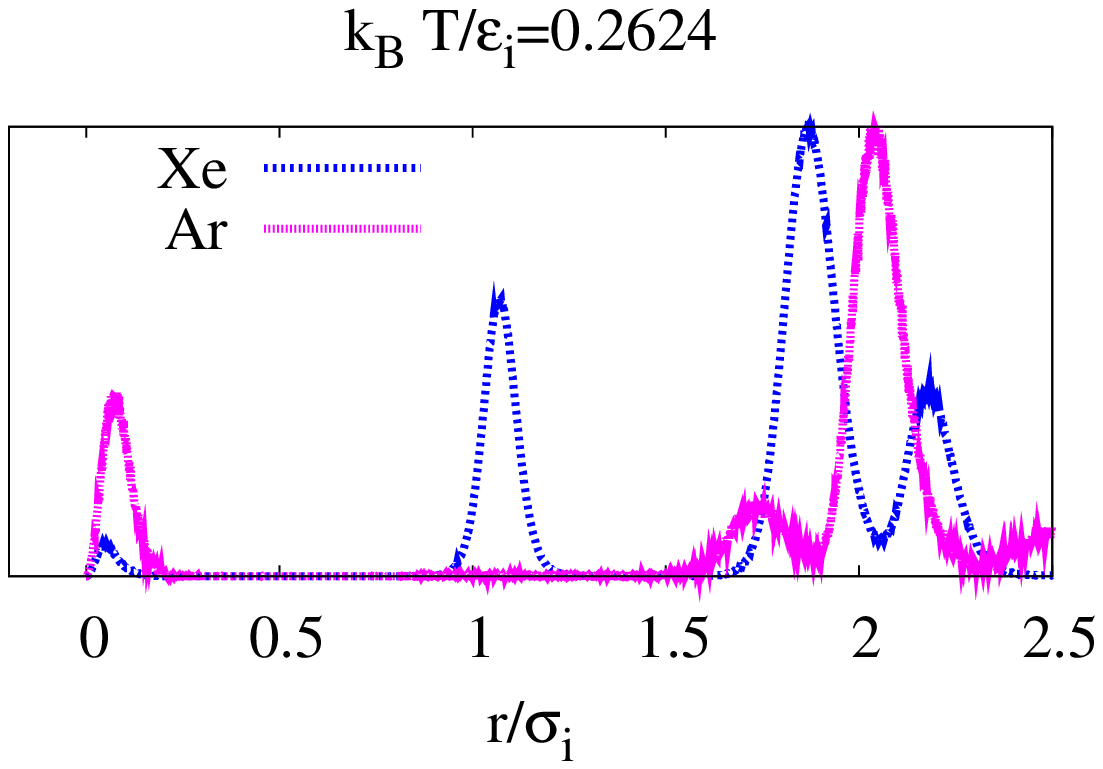}
  \end{minipage}
  \ \hfill  \begin{minipage}{0.23\textwidth}
    \includegraphics[width=\textwidth]{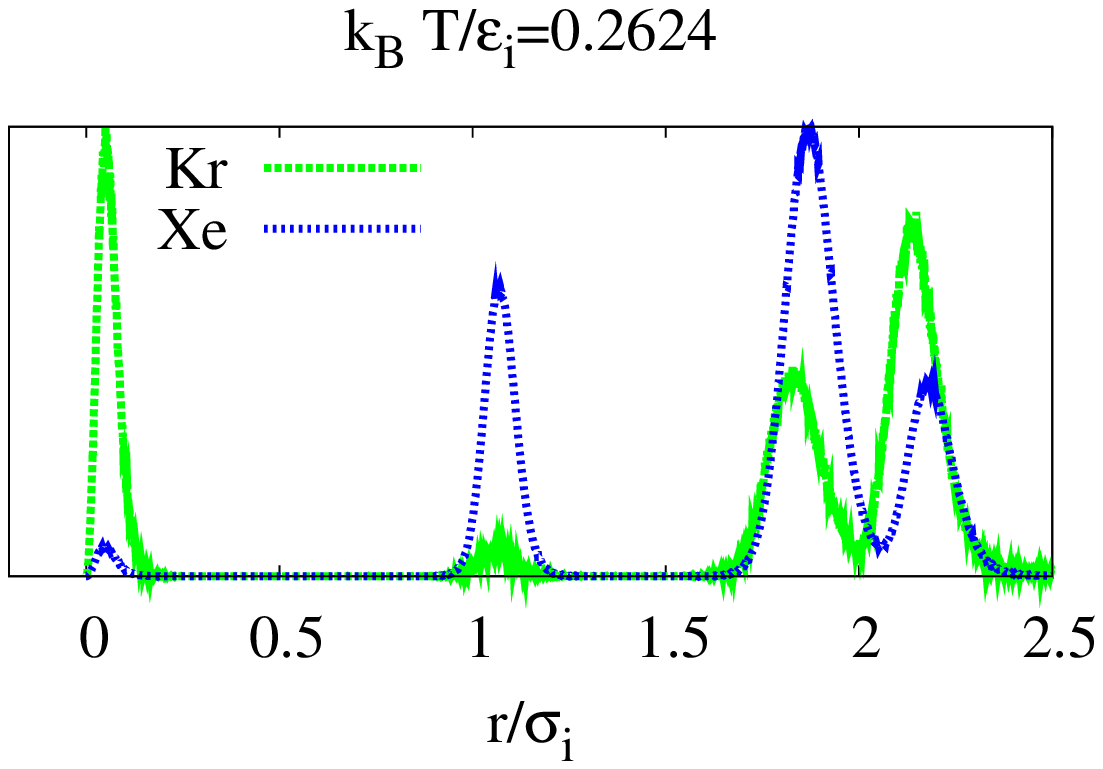}
  \end{minipage}
  \ \hfill 
  \begin{minipage}{0.23\textwidth}
    \includegraphics[width=\textwidth]{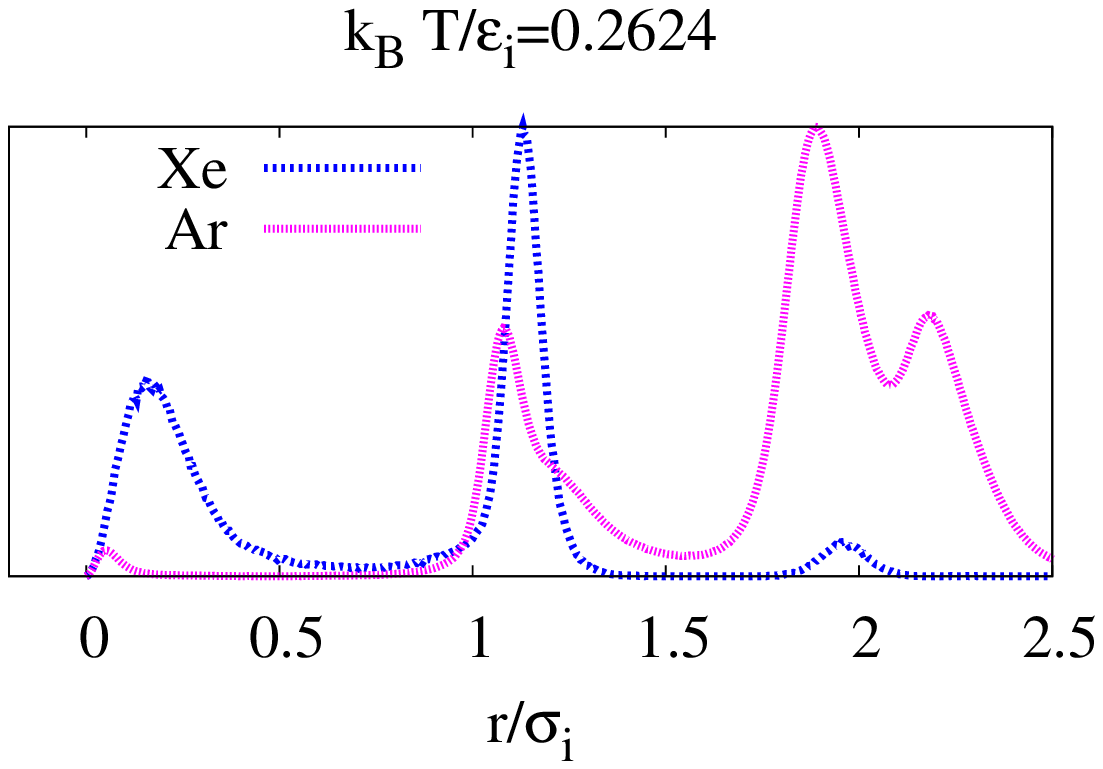}
  \end{minipage}
  \ \hfill
  \begin{minipage}{0.23\textwidth}
    \includegraphics[width=\textwidth]{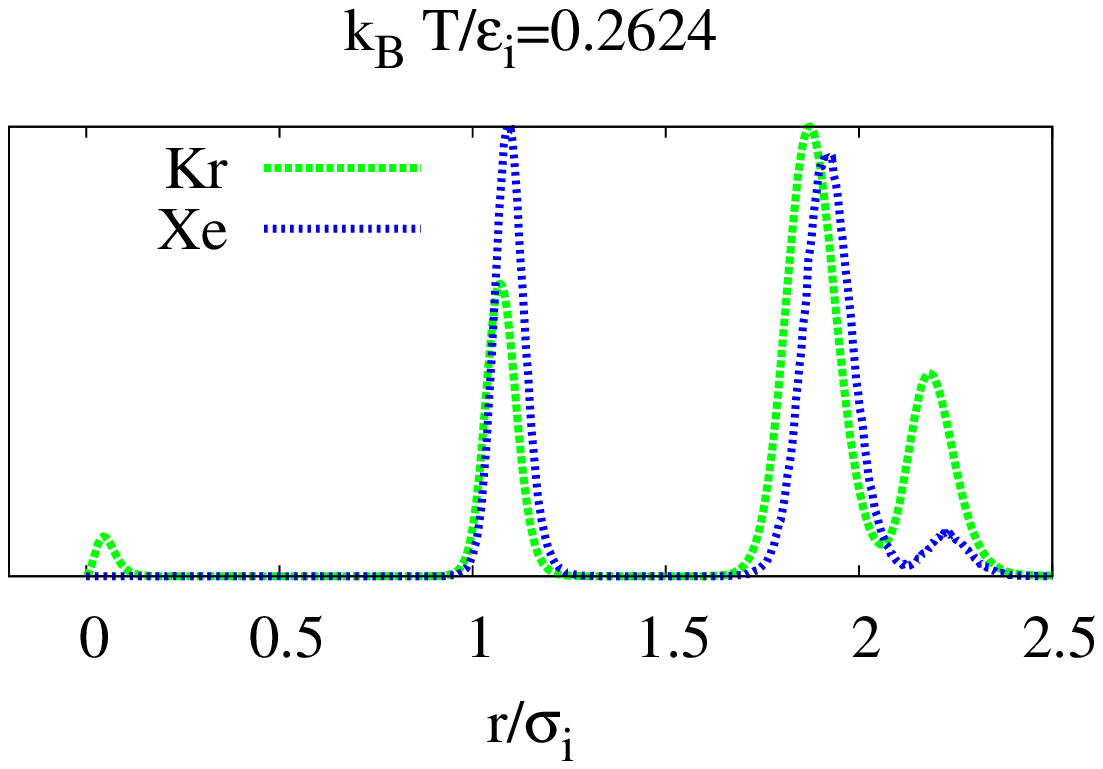}
  \end{minipage}
  \\
\bigskip
\bigskip
 \begin{minipage}{0.23\textwidth}
    \includegraphics[width=\textwidth]{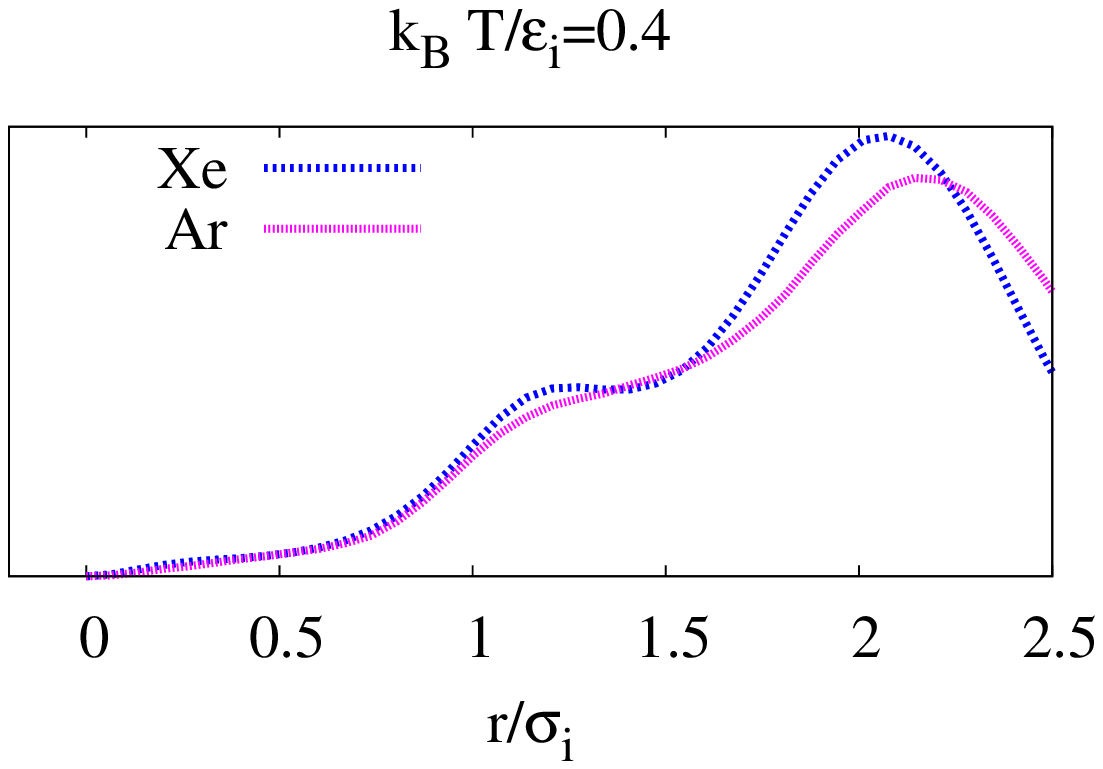}
  \end{minipage}
  \ \hfill  
   \begin{minipage}{0.23\textwidth}
    \includegraphics[width=\textwidth]{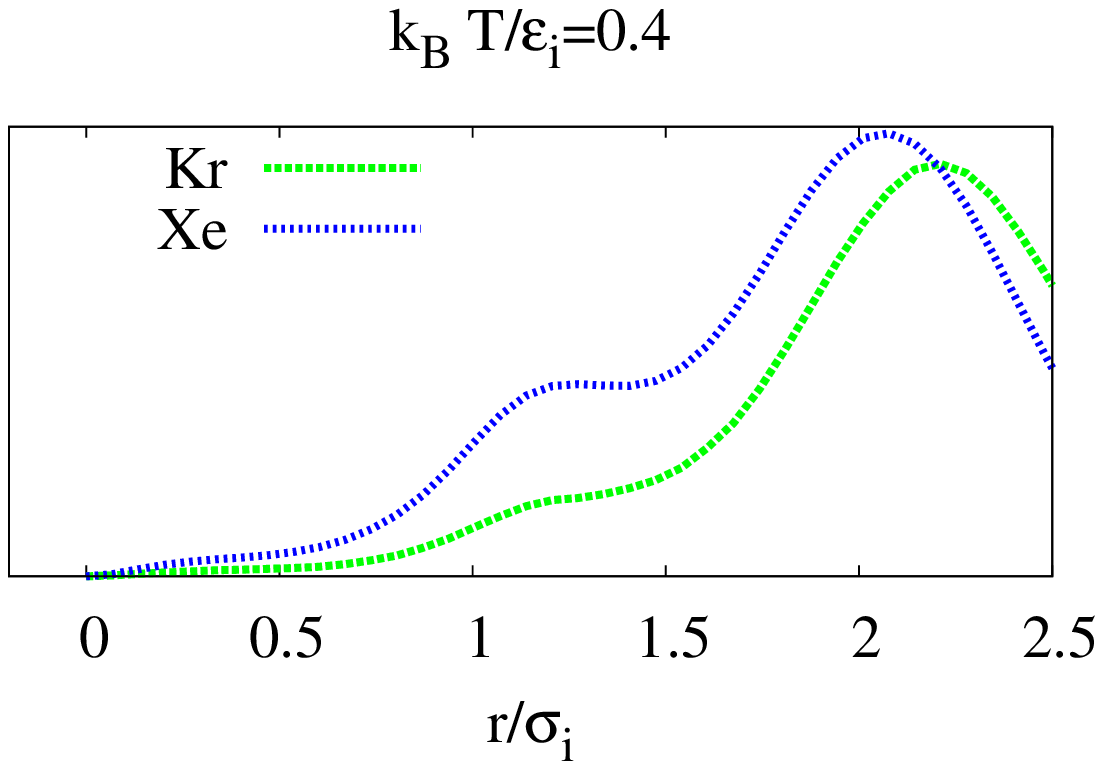}
  \end{minipage}
  \ \hfill 
  \begin{minipage}{0.23\textwidth}
    \includegraphics[width=\textwidth]{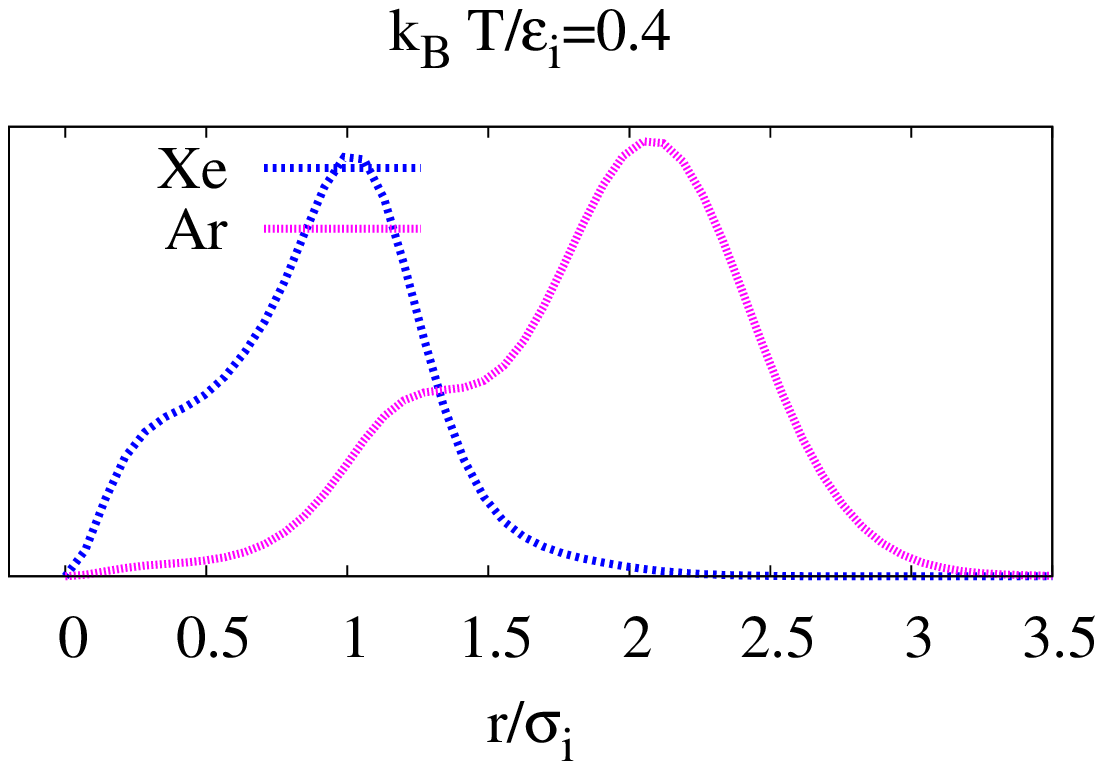}
  \end{minipage}
  \ \hfill
  \begin{minipage}{0.23\textwidth}
    \includegraphics[width=\textwidth]{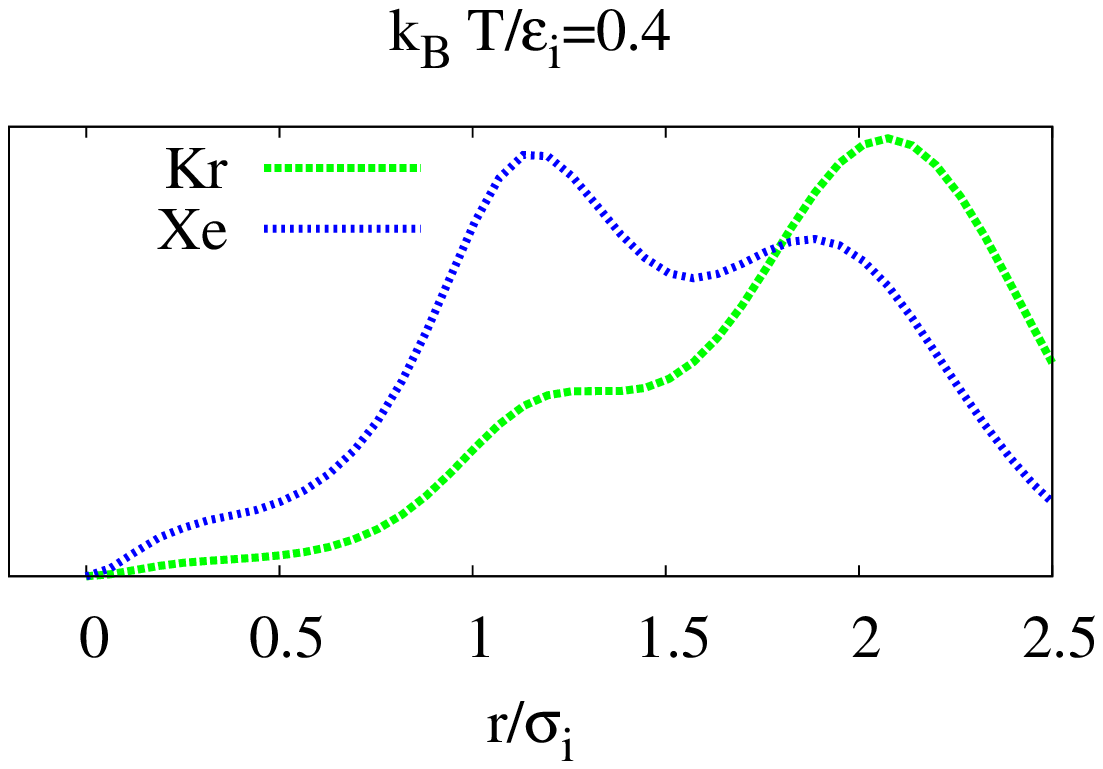}
  \end{minipage}
  \\

\caption{\label{fig:gs55} Radial distribution function for 55 atom clusters.}
\end{figure}

\begin{figure}[!ht]
  \begin{minipage}{0.23\textwidth}
    \centering ArXe$_{146}$
  \end{minipage}
  \ \hfill 
  \begin{minipage}{0.23\textwidth}
    \centering KrXe$_{146}$
  \end{minipage}
  \ \hfill
  \begin{minipage}{0.23\textwidth}
    \centering Ar$_{146}$Xe
  \end{minipage}
  \begin{minipage}{0.23\textwidth}
    \centering Kr$_{146}$Xe
  \end{minipage}
  \\
\bigskip
\bigskip
 \begin{minipage}{0.23\textwidth}
    \includegraphics[width=\textwidth]{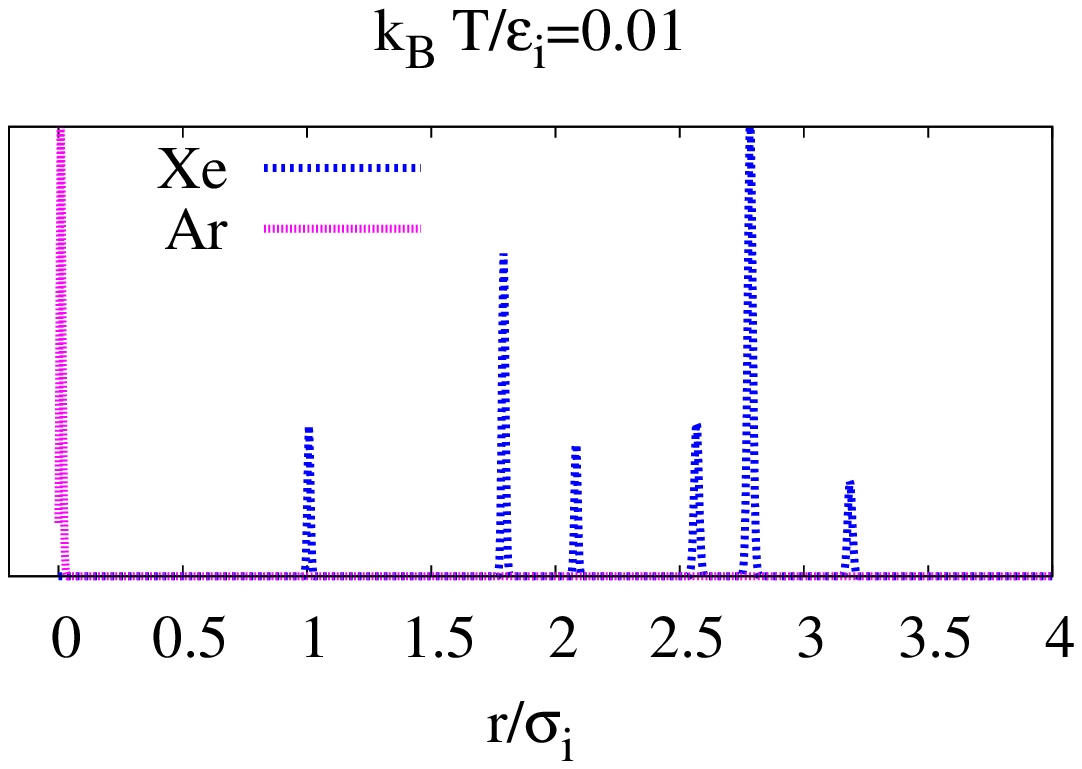}
  \end{minipage}
  \ \hfill 
 \begin{minipage}{0.23\textwidth}
    \includegraphics[width=\textwidth]{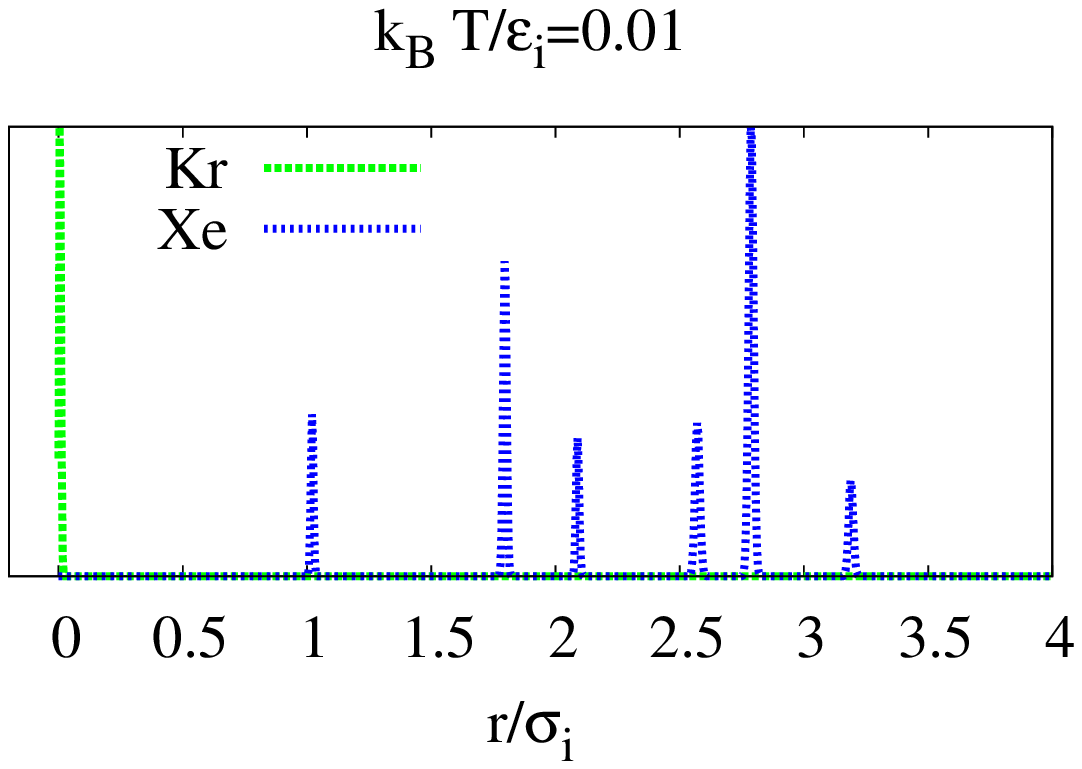}
  \end{minipage}
  \ \hfill 
  \begin{minipage}{0.23\textwidth}
    \includegraphics[width=\textwidth]{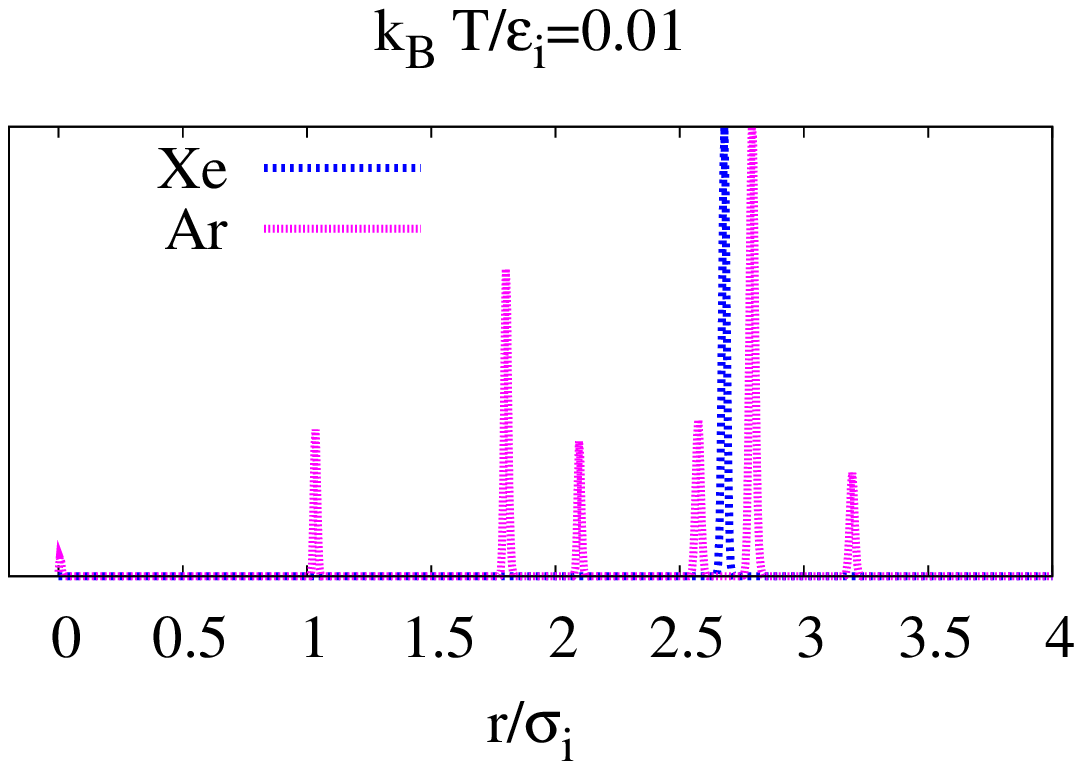}
  \end{minipage}
  \ \hfill
  \begin{minipage}{0.23\textwidth}
    \includegraphics[width=\textwidth]{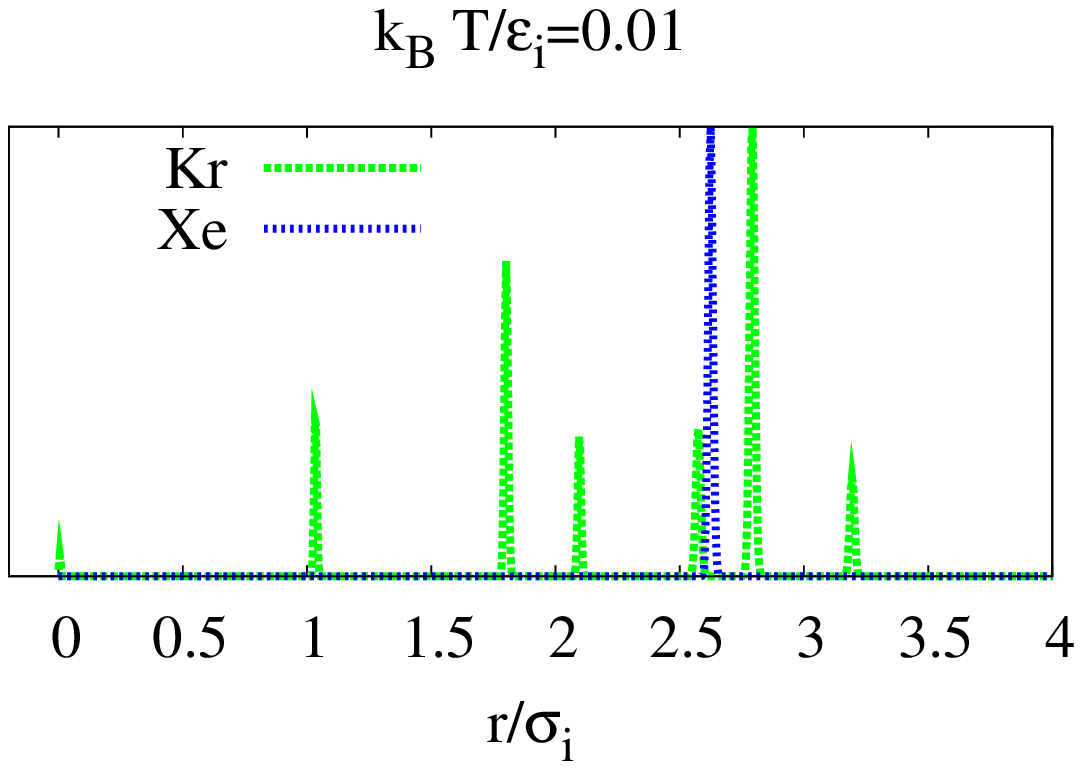}
  \end{minipage}
  \\
\bigskip
\bigskip
 \begin{minipage}{0.23\textwidth}
    \includegraphics[width=\textwidth]{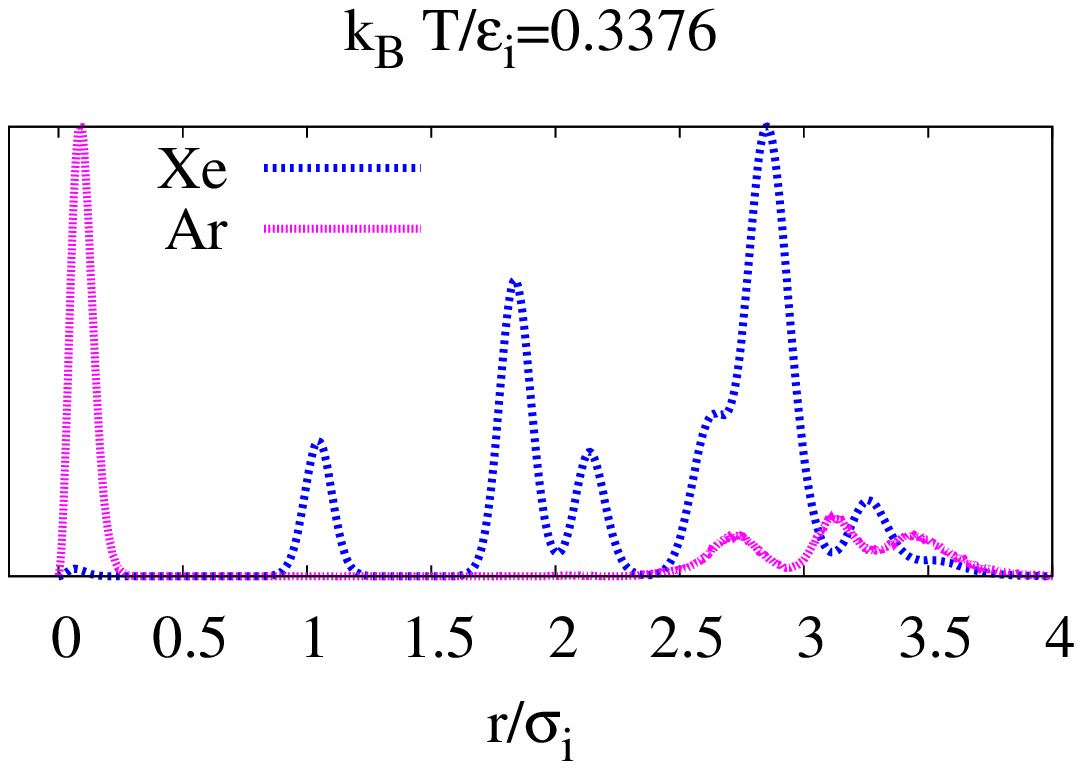}
  \end{minipage}
  \ \hfill 
 \begin{minipage}{0.23\textwidth}
    \includegraphics[width=\textwidth]{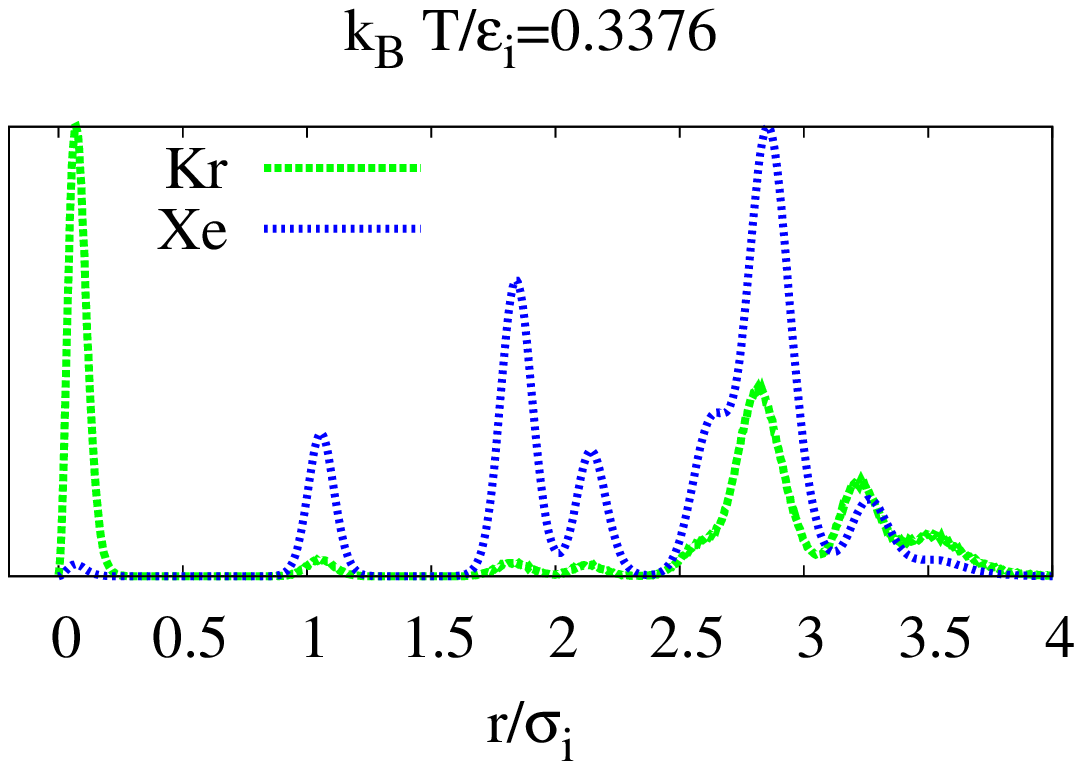}
  \end{minipage}
  \ \hfill 
  \begin{minipage}{0.23\textwidth}
    \includegraphics[width=\textwidth]{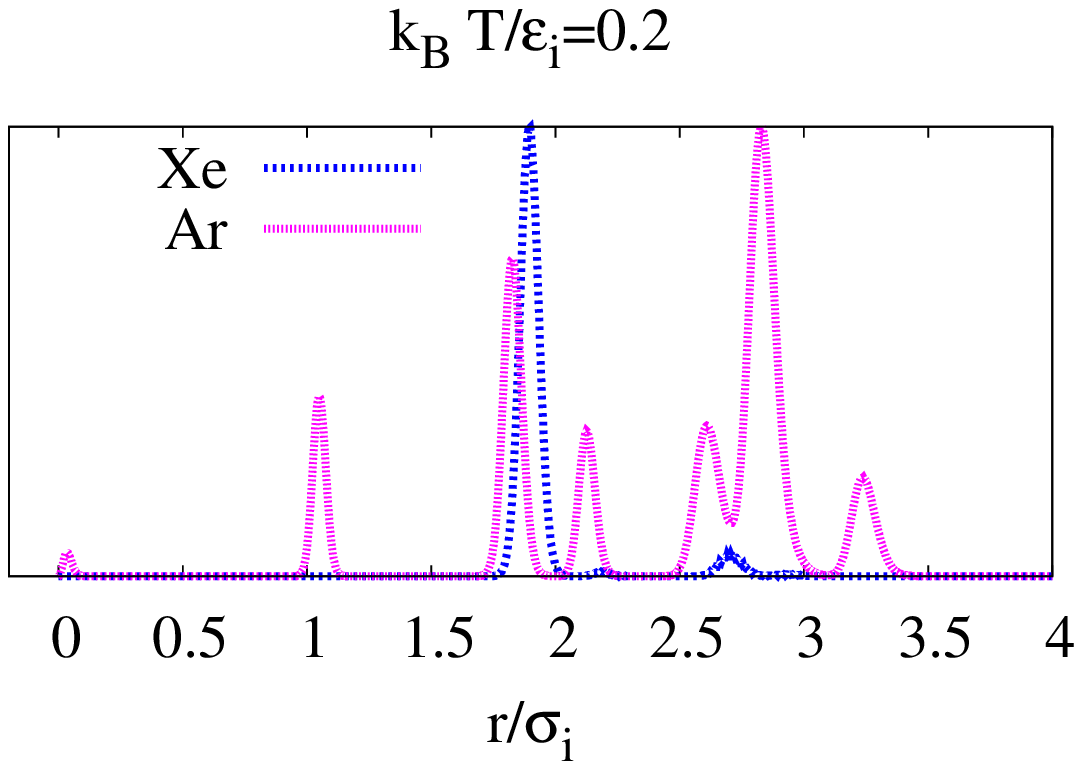}
  \end{minipage}
  \ \hfill
  \begin{minipage}{0.23\textwidth}
    \includegraphics[width=\textwidth]{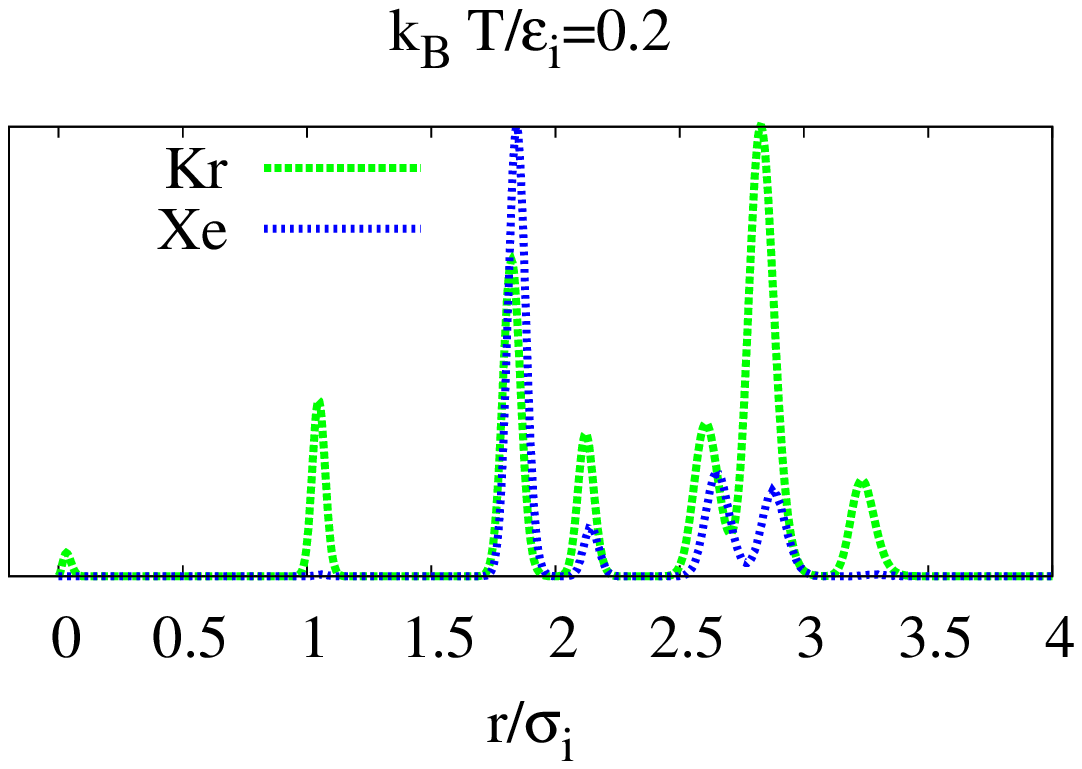}
  \end{minipage}
  \\
\bigskip
\bigskip
 \begin{minipage}{0.23\textwidth}
    \includegraphics[width=\textwidth]{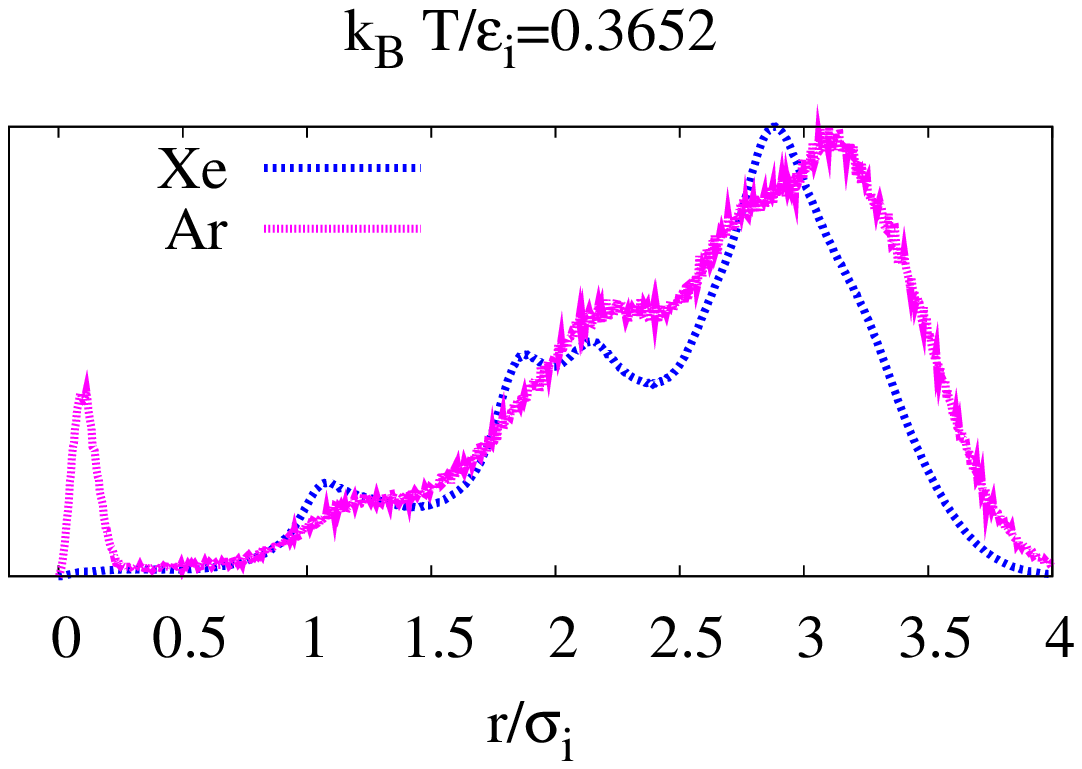}
  \end{minipage}
  \ \hfill  \begin{minipage}{0.23\textwidth}
    \includegraphics[width=\textwidth]{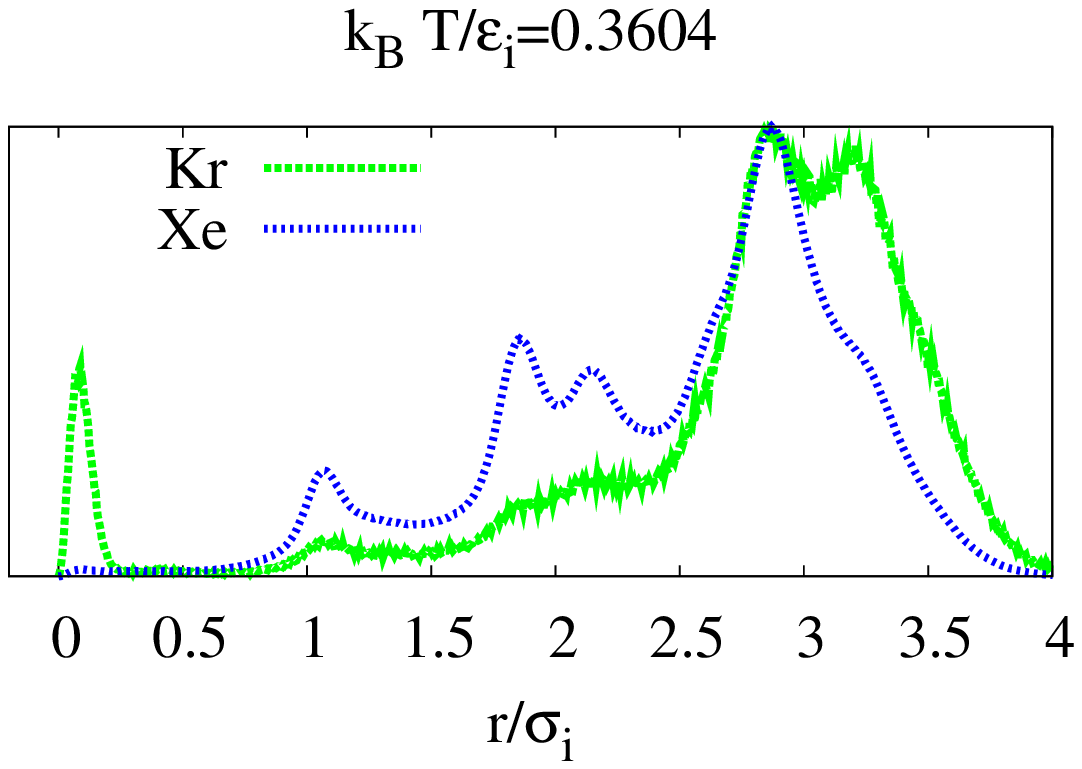}
  \end{minipage}
  \ \hfill 
  \begin{minipage}{0.23\textwidth}
    \includegraphics[width=\textwidth]{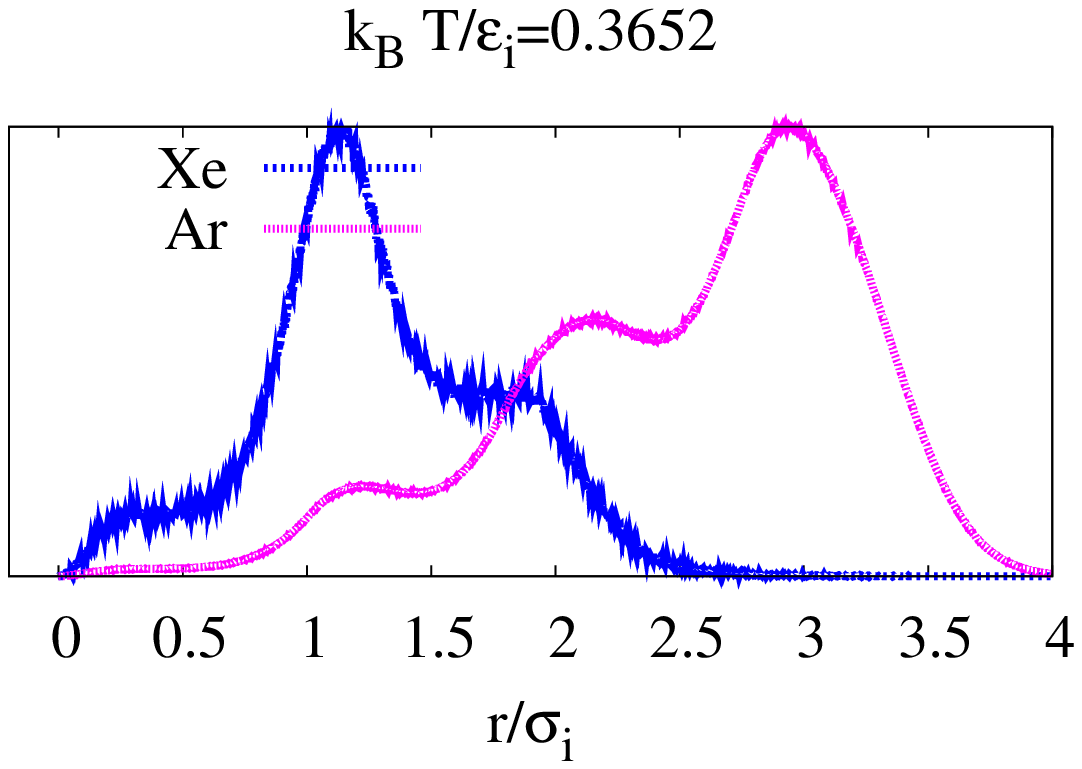}
  \end{minipage}
  \ \hfill
  \begin{minipage}{0.23\textwidth}
    \includegraphics[width=\textwidth]{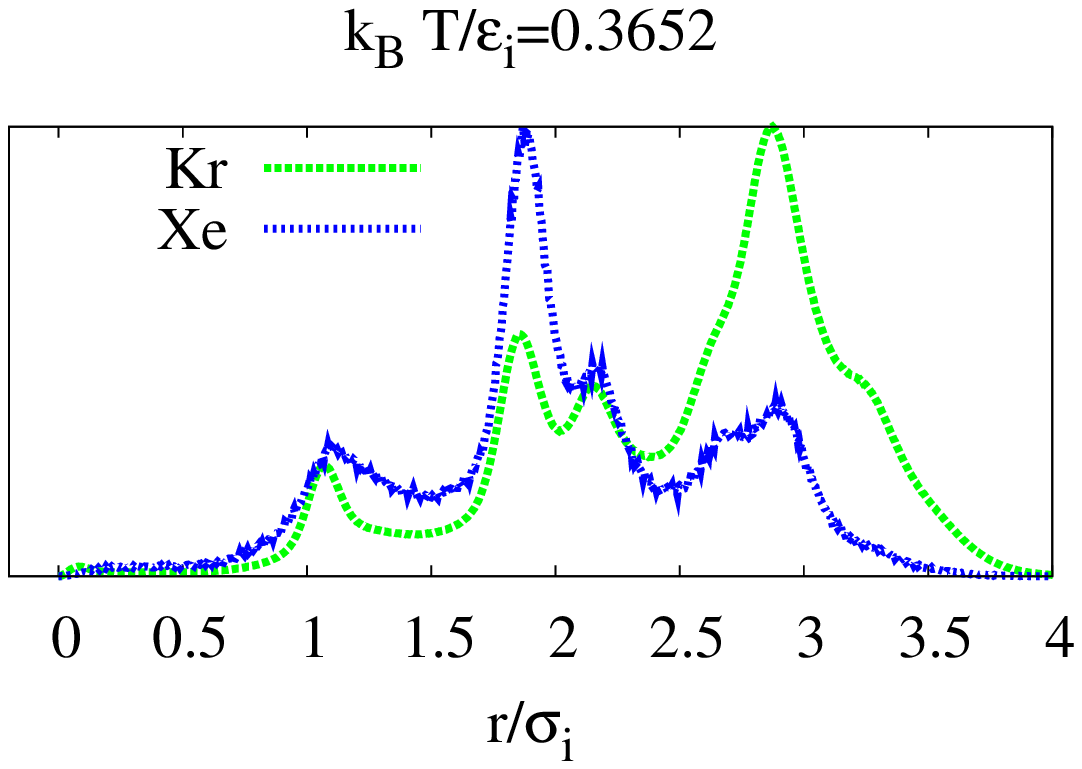}
  \end{minipage}
  \\
\bigskip
\bigskip
 \begin{minipage}{0.23\textwidth}
    \includegraphics[width=\textwidth]{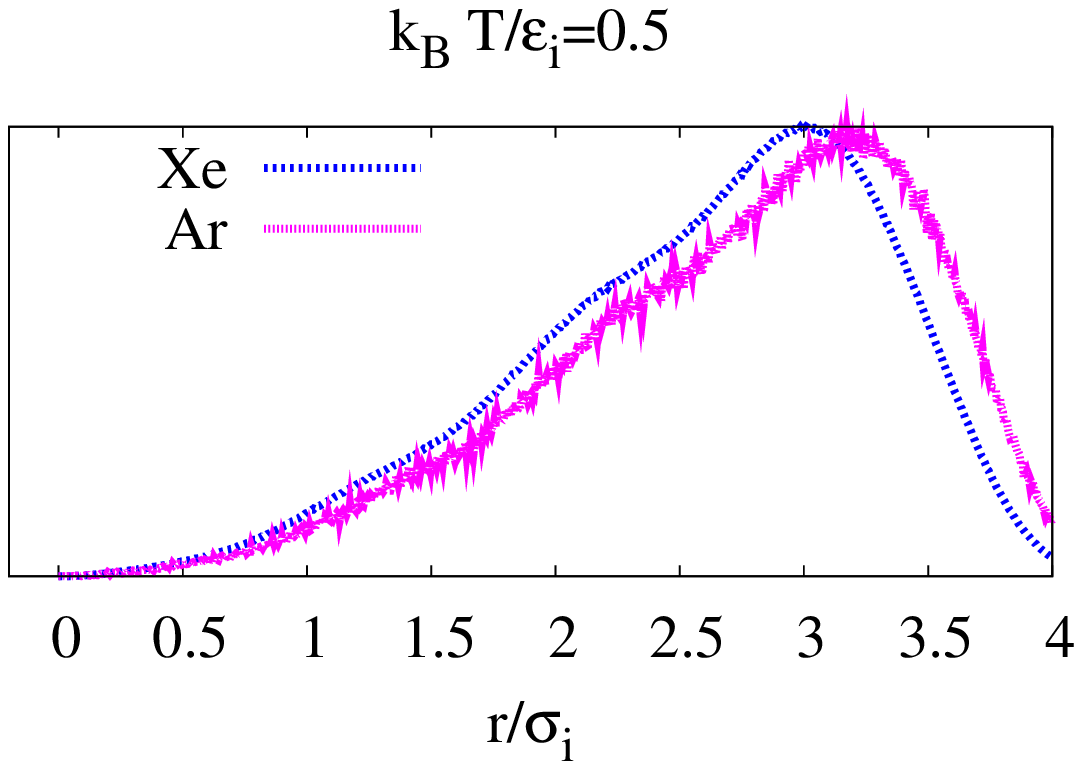}
  \end{minipage}
  \ \hfill  
   \begin{minipage}{0.23\textwidth}
    \includegraphics[width=\textwidth]{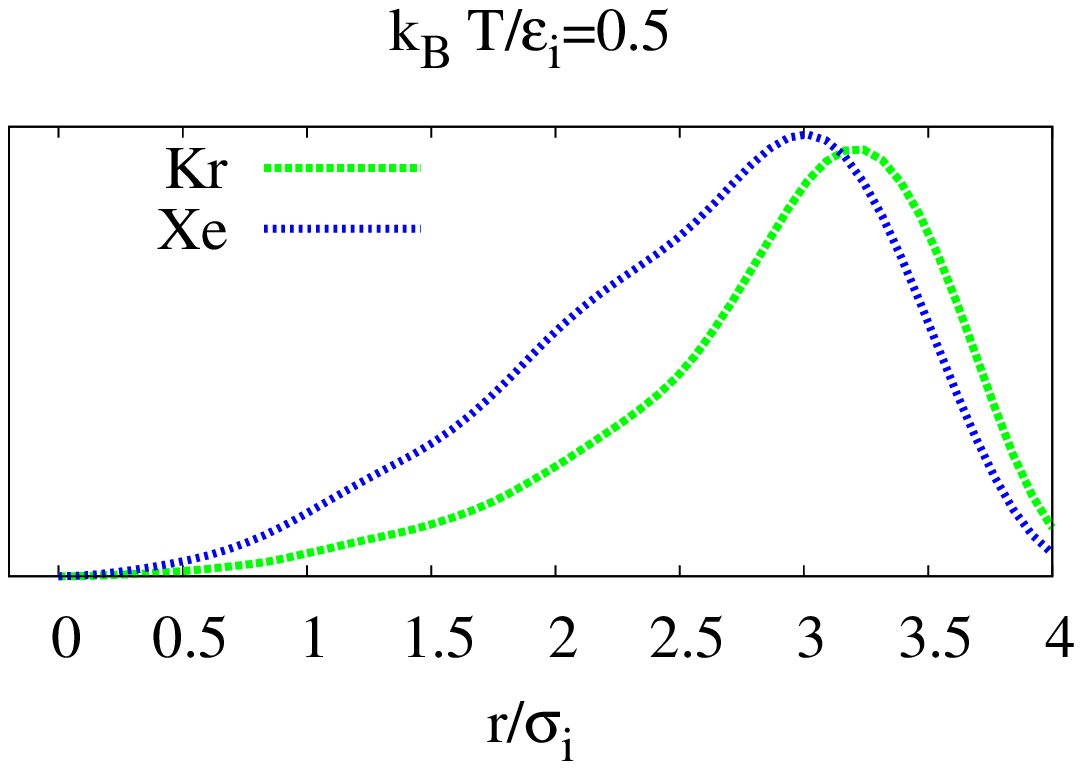}
  \end{minipage}
  \ \hfill 
  \begin{minipage}{0.23\textwidth}
    \includegraphics[width=\textwidth]{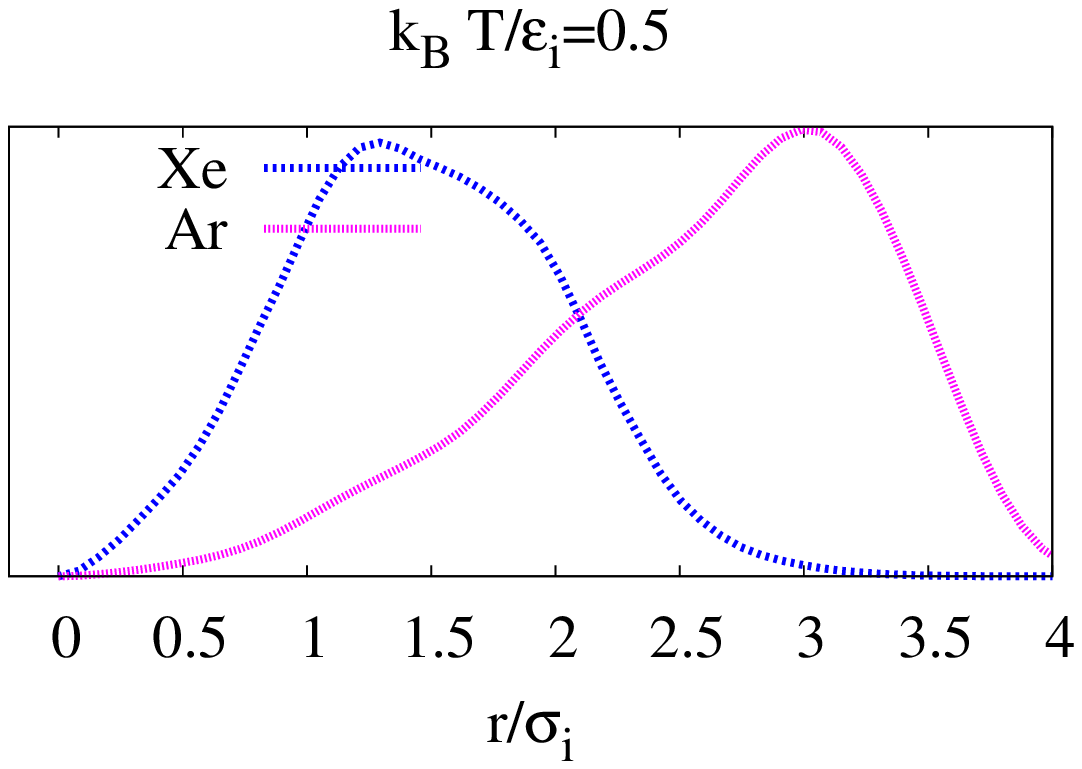}
  \end{minipage}
  \ \hfill
  \begin{minipage}{0.23\textwidth}
    \includegraphics[width=\textwidth]{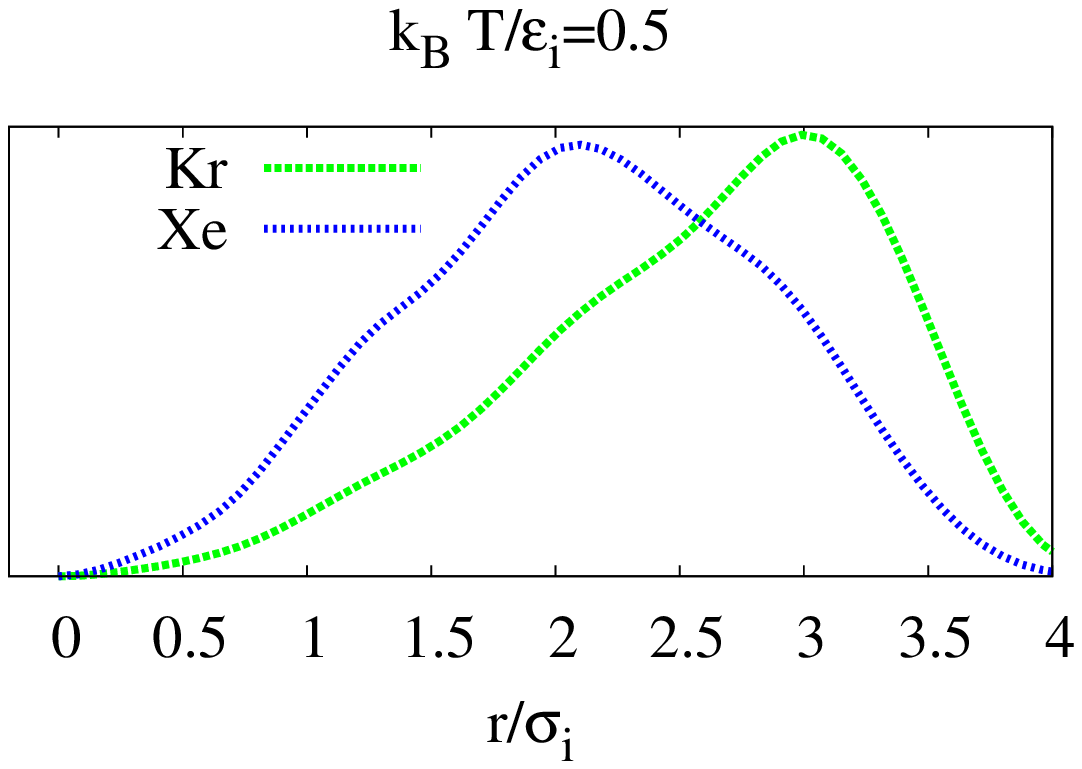}
  \end{minipage}
  \\

\caption{\label{fig:gs147} Radial distribution function for 147 atom clusters.}
\end{figure}

\begin{figure}[!ht]
  \begin{minipage}{0.23\textwidth}
    \centering ArXe$_{308}$
  \end{minipage}
  \ \hfill 
  \begin{minipage}{0.23\textwidth}
    \centering KrXe$_{308}$
  \end{minipage}
  \ \hfill
  \begin{minipage}{0.23\textwidth}
    \centering Ar$_{308}$Xe
  \end{minipage}
  \begin{minipage}{0.23\textwidth}
    \centering Kr$_{308}$Xe
  \end{minipage}
  \\
\bigskip
\bigskip
 \begin{minipage}{0.23\textwidth}
    \includegraphics[width=\textwidth]{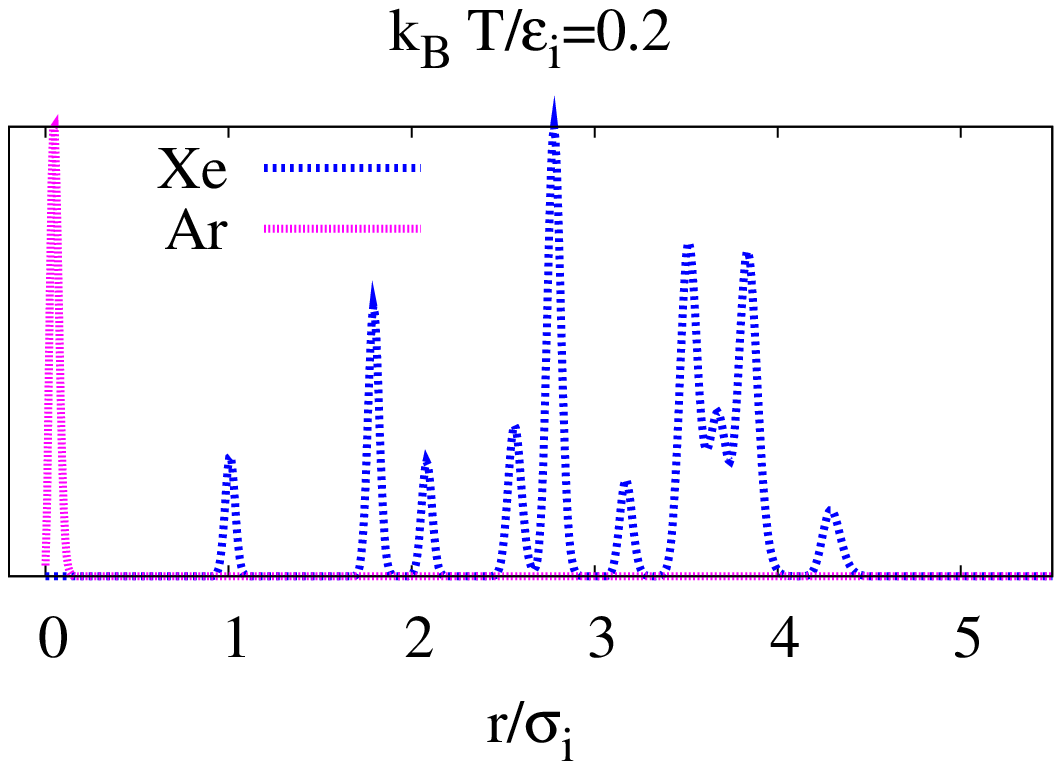}
  \end{minipage}
  \ \hfill 
 \begin{minipage}{0.23\textwidth}
    \includegraphics[width=\textwidth]{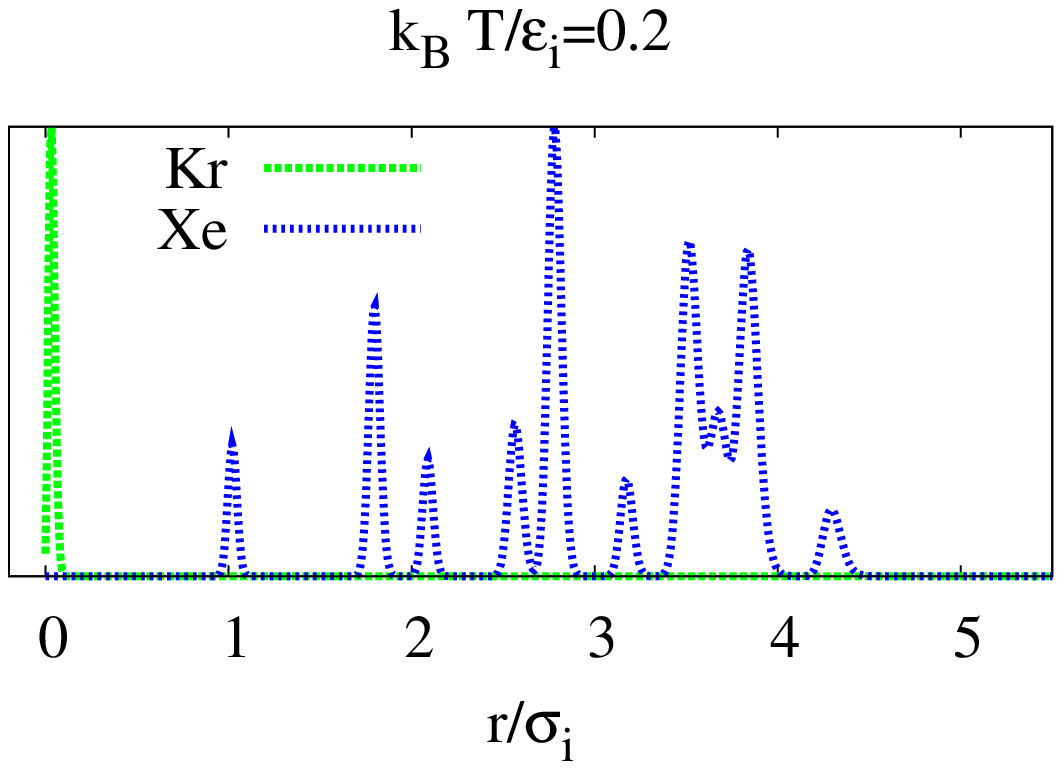}
  \end{minipage}
  \ \hfill 
  \begin{minipage}{0.23\textwidth}
    \includegraphics[width=\textwidth]{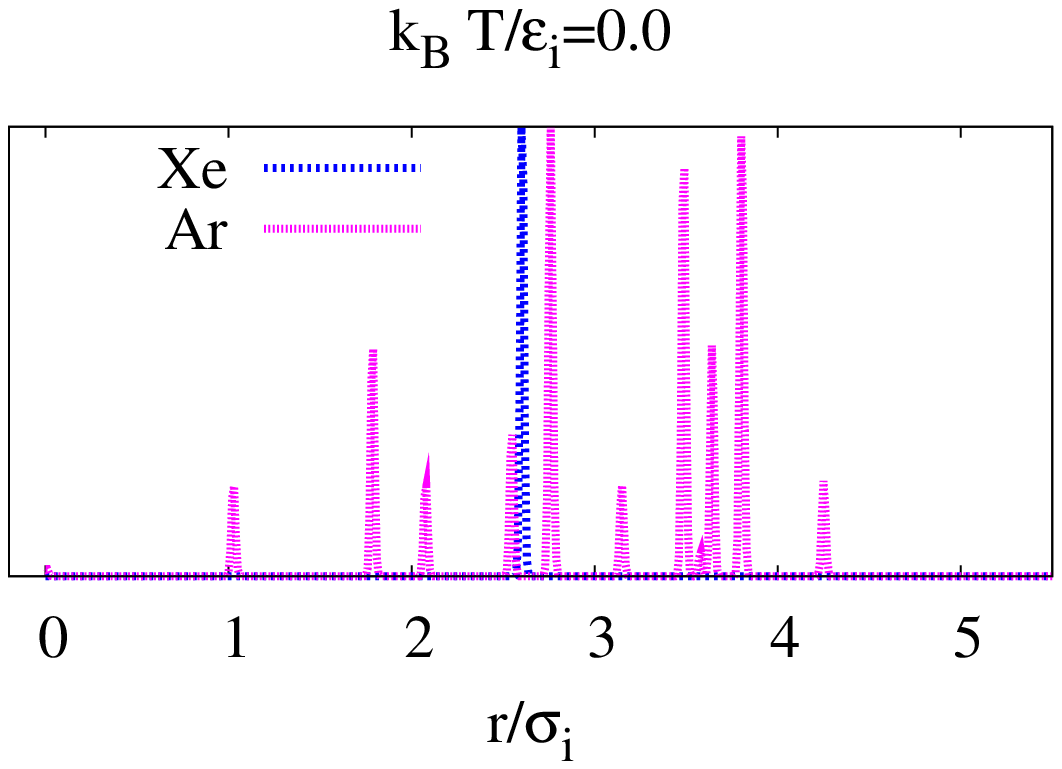}
  \end{minipage}
  \ \hfill
  \begin{minipage}{0.23\textwidth}
    \includegraphics[width=\textwidth]{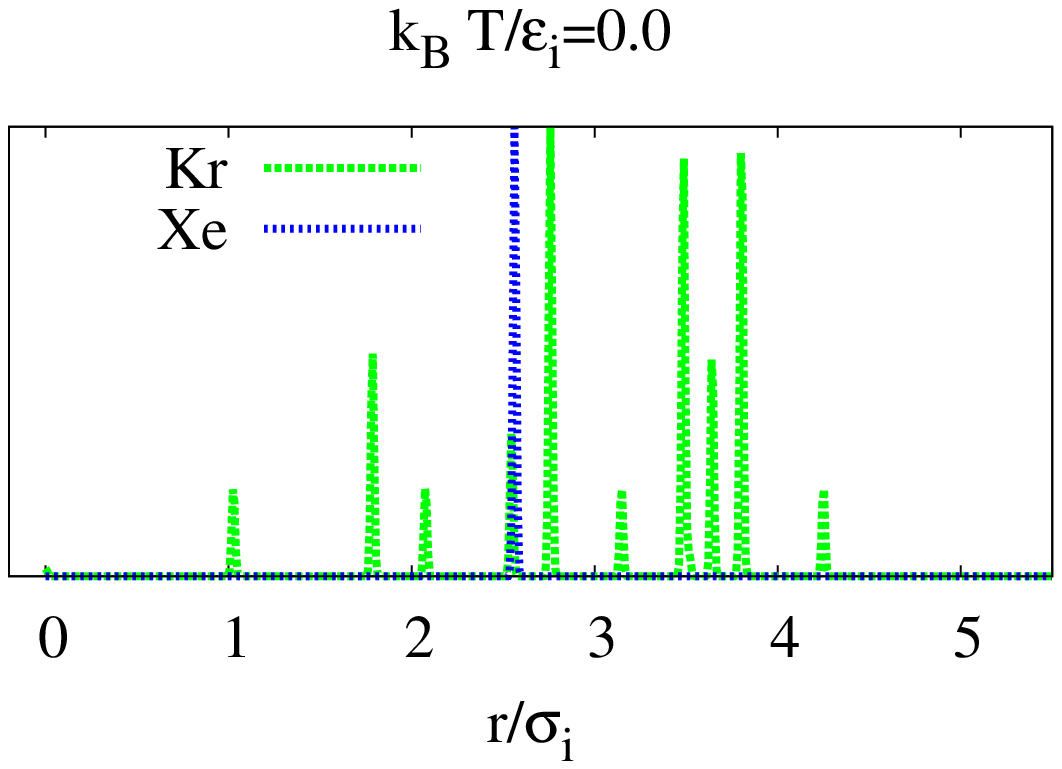}
  \end{minipage}
  \\
\bigskip
\bigskip
 \begin{minipage}{0.23\textwidth}
    \includegraphics[width=\textwidth]{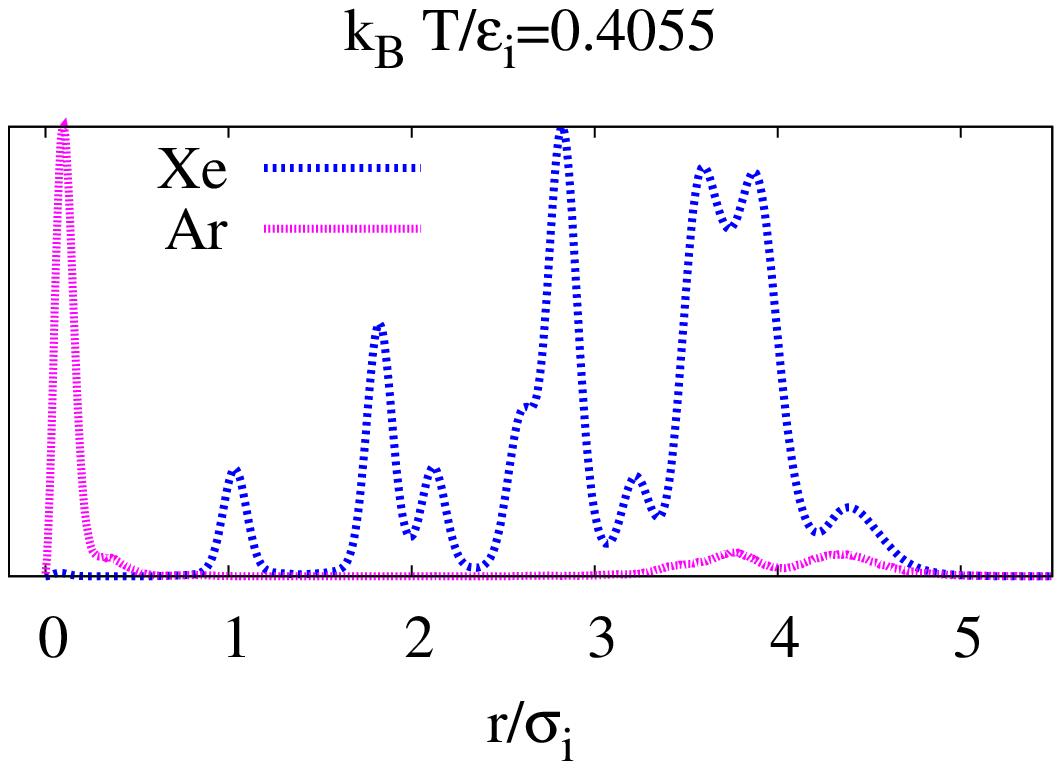}
  \end{minipage}
  \ \hfill 
 \begin{minipage}{0.23\textwidth}
    \includegraphics[width=\textwidth]{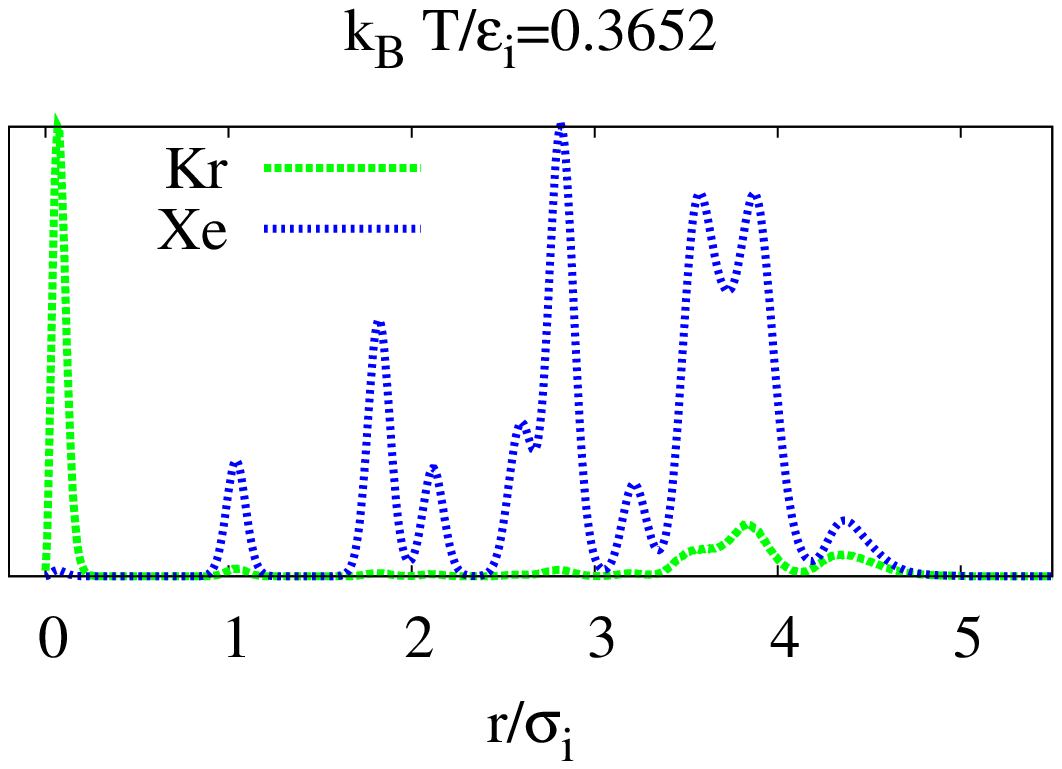}
  \end{minipage}
  \ \hfill 
  \begin{minipage}{0.23\textwidth}
    \includegraphics[width=\textwidth]{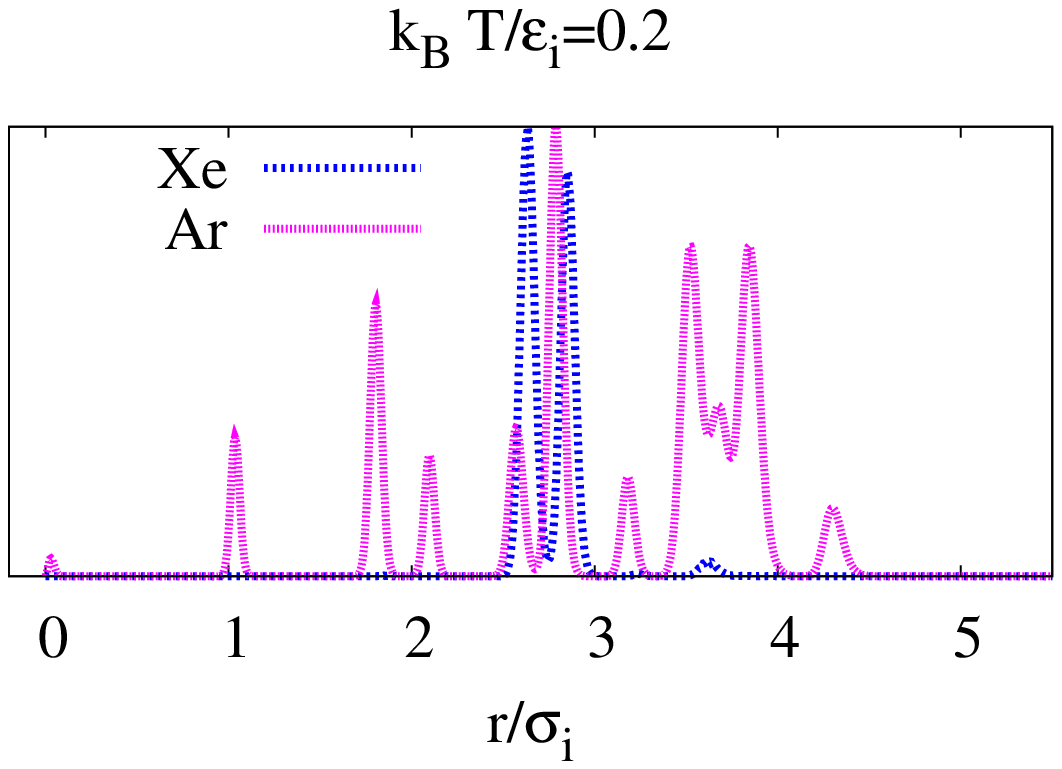}
  \end{minipage}
  \ \hfill
  \begin{minipage}{0.23\textwidth}
    \includegraphics[width=\textwidth]{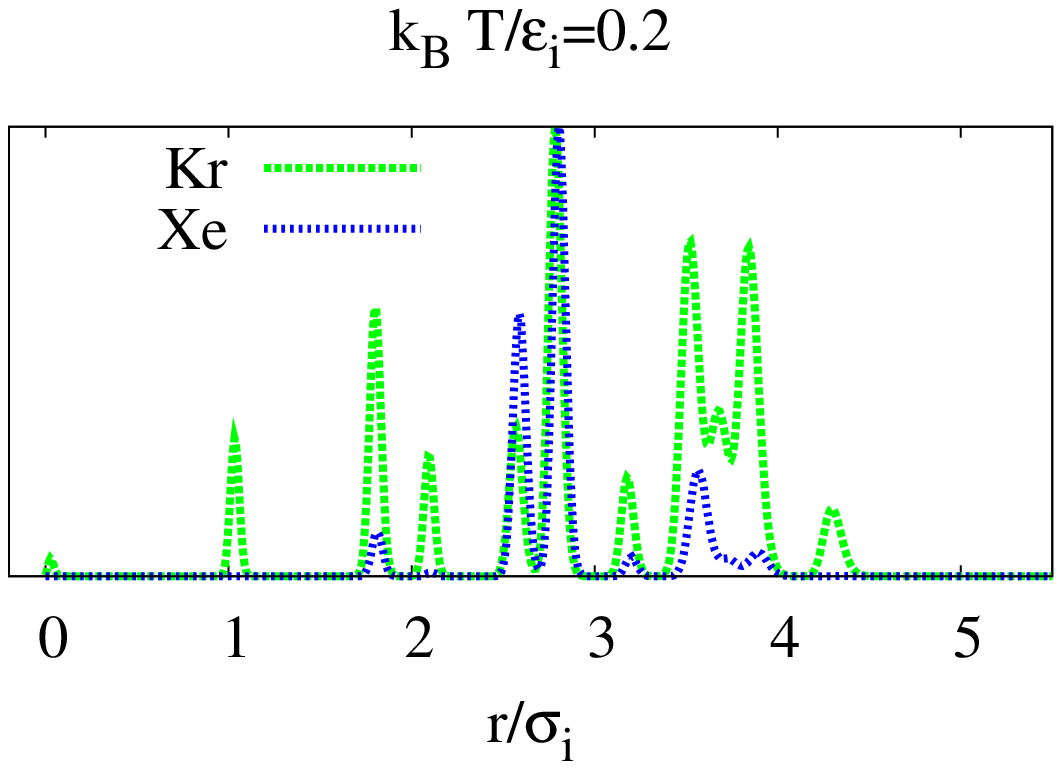}
  \end{minipage}
  \\
\bigskip
\bigskip
 \begin{minipage}{0.23\textwidth}
    \includegraphics[width=\textwidth]{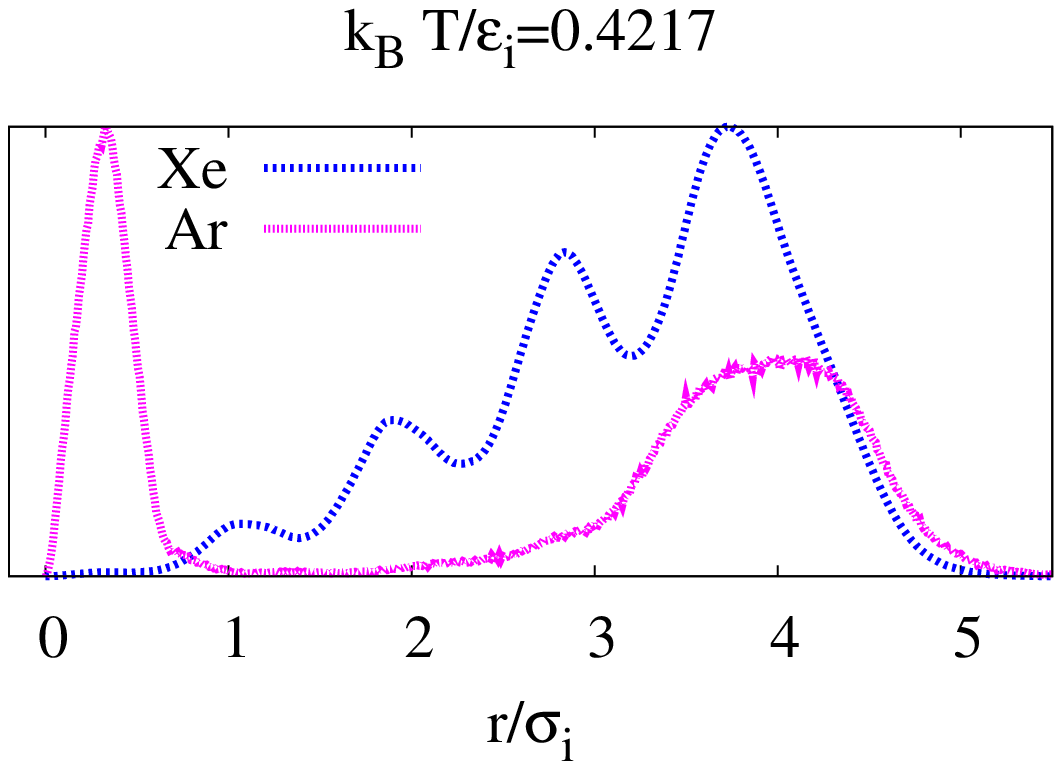}
  \end{minipage}
  \ \hfill  \begin{minipage}{0.23\textwidth}
    \includegraphics[width=\textwidth]{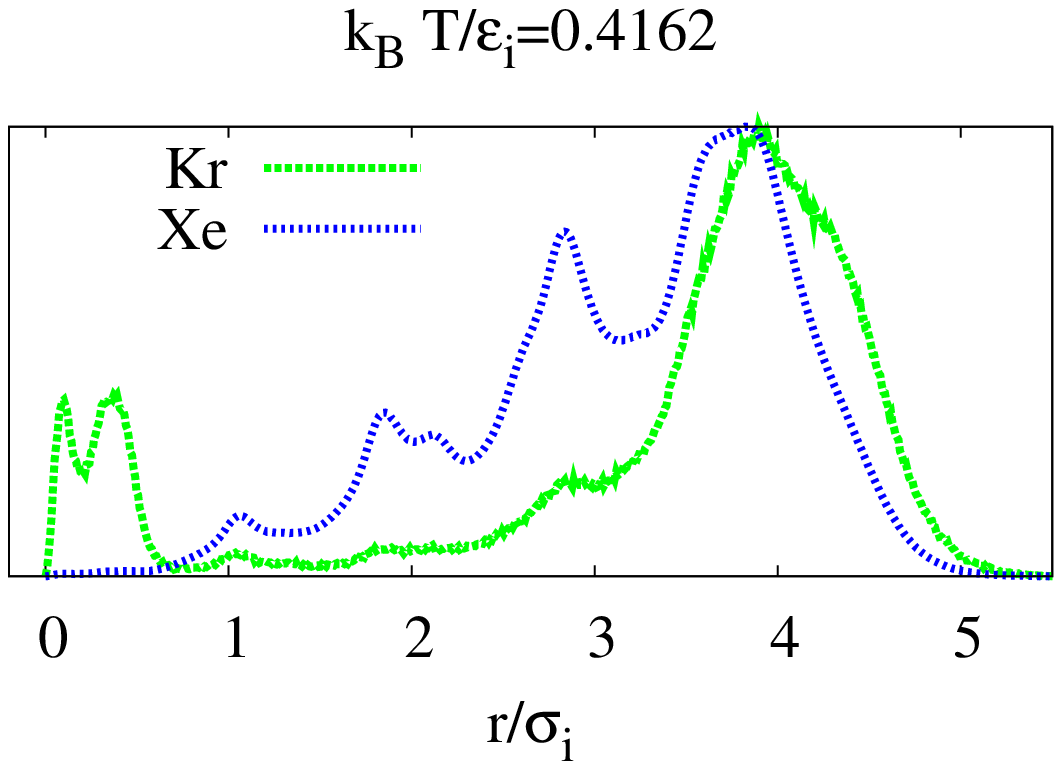}
  \end{minipage}
  \ \hfill 
  \begin{minipage}{0.23\textwidth}
    \includegraphics[width=\textwidth]{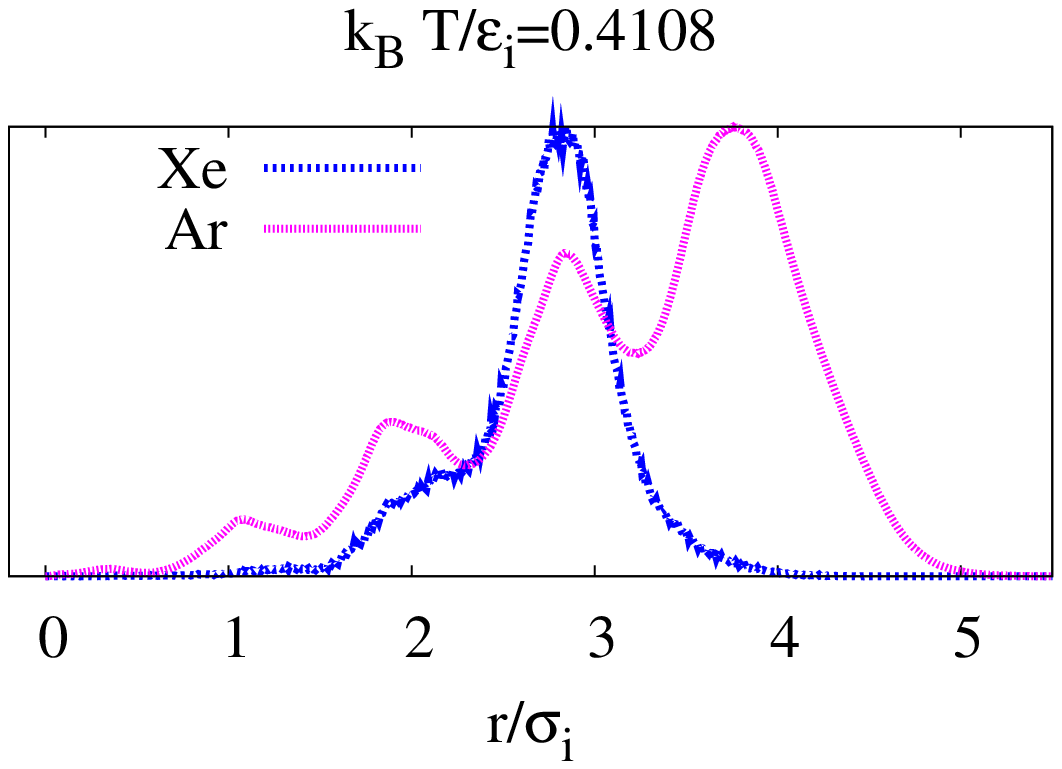}
  \end{minipage}
  \ \hfill
  \begin{minipage}{0.23\textwidth}
    \includegraphics[width=\textwidth]{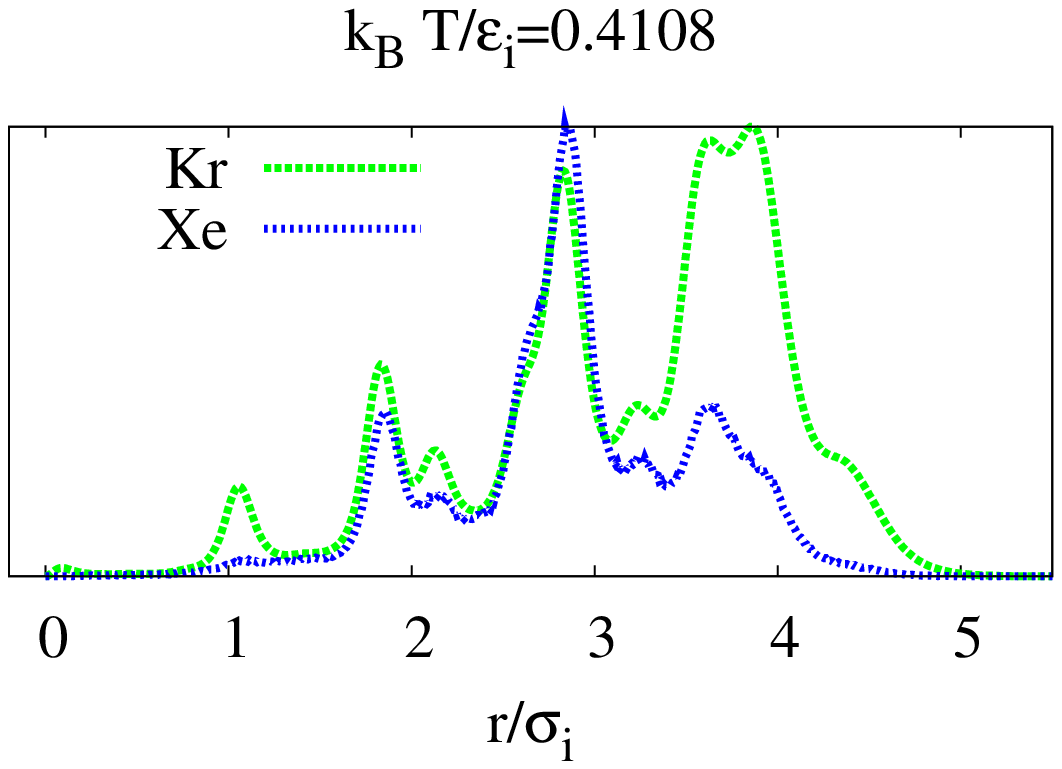}
  \end{minipage}
  \\
\bigskip
\bigskip
 \begin{minipage}{0.23\textwidth}
    \includegraphics[width=\textwidth]{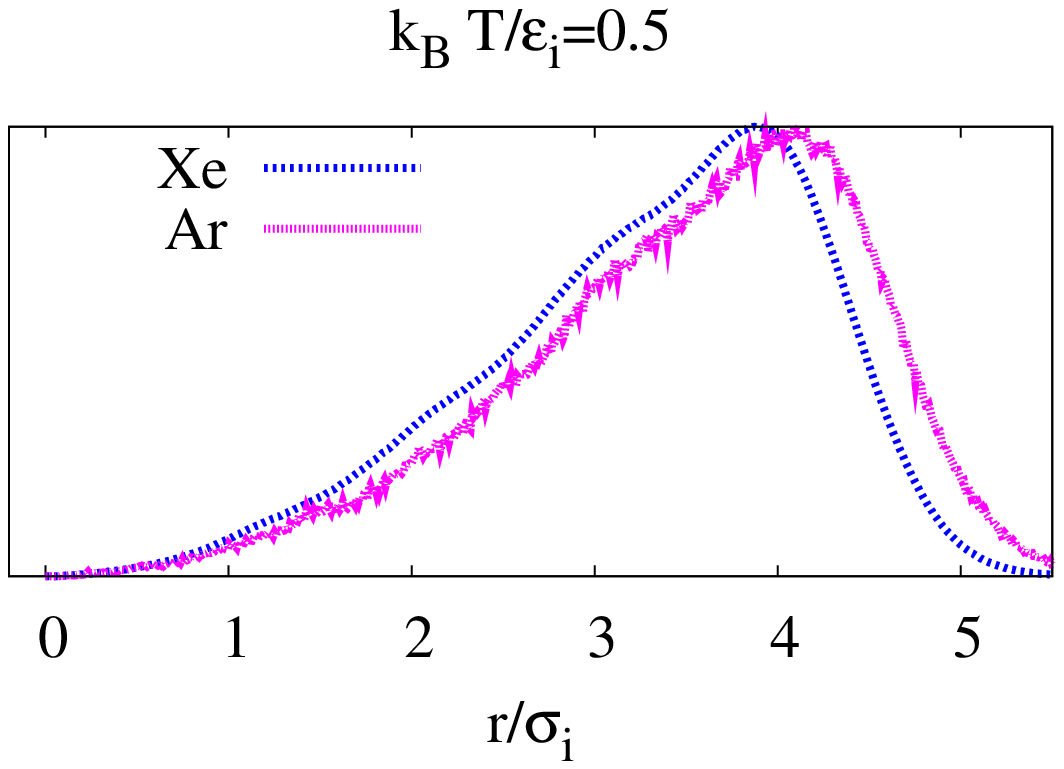}
  \end{minipage}
  \ \hfill  
   \begin{minipage}{0.23\textwidth}
    \includegraphics[width=\textwidth]{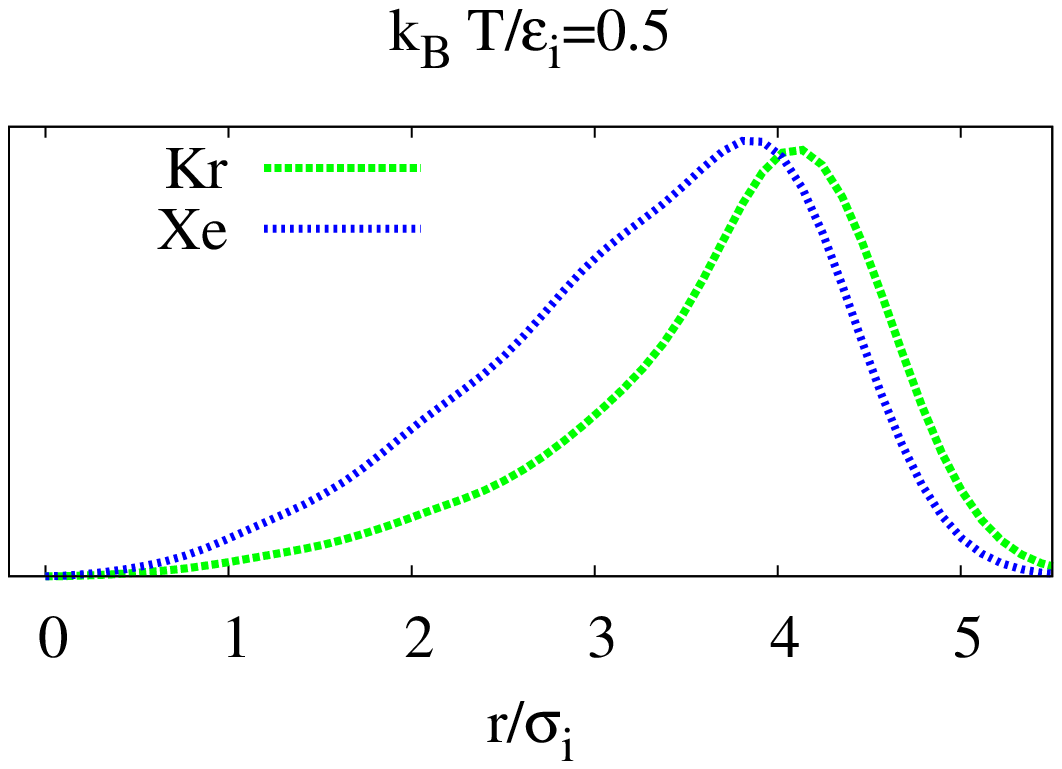}
  \end{minipage}
  \ \hfill 
  \begin{minipage}{0.23\textwidth}
    \includegraphics[width=\textwidth]{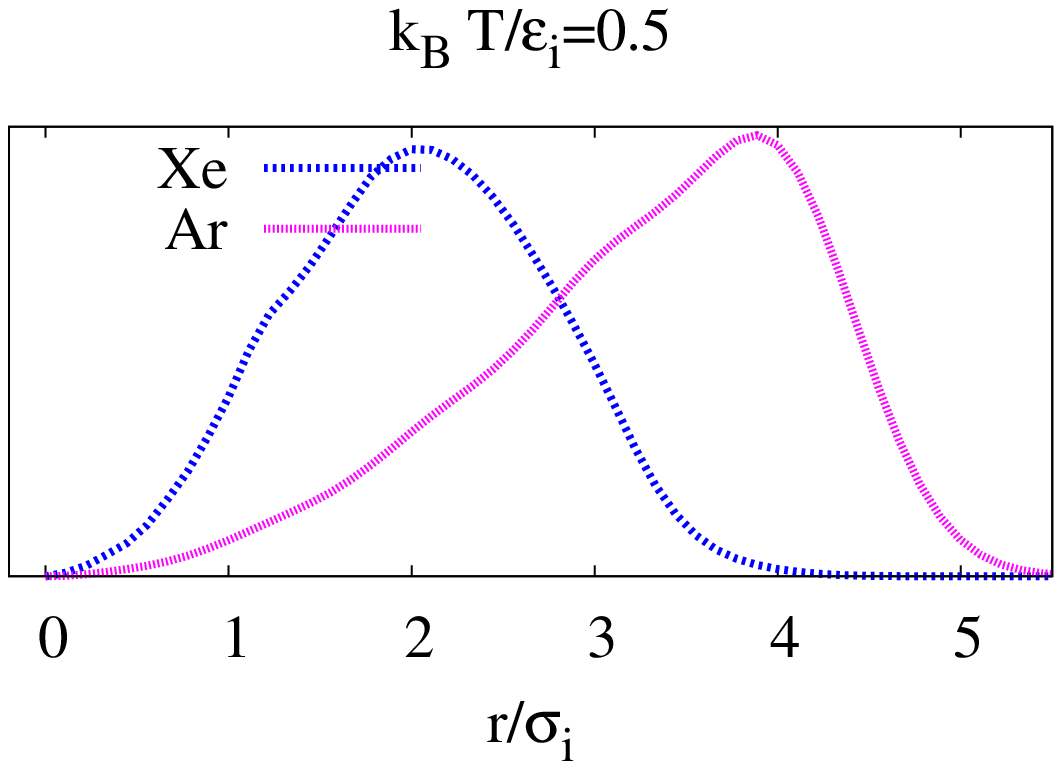}
  \end{minipage}
  \ \hfill
  \begin{minipage}{0.23\textwidth}
    \includegraphics[width=\textwidth]{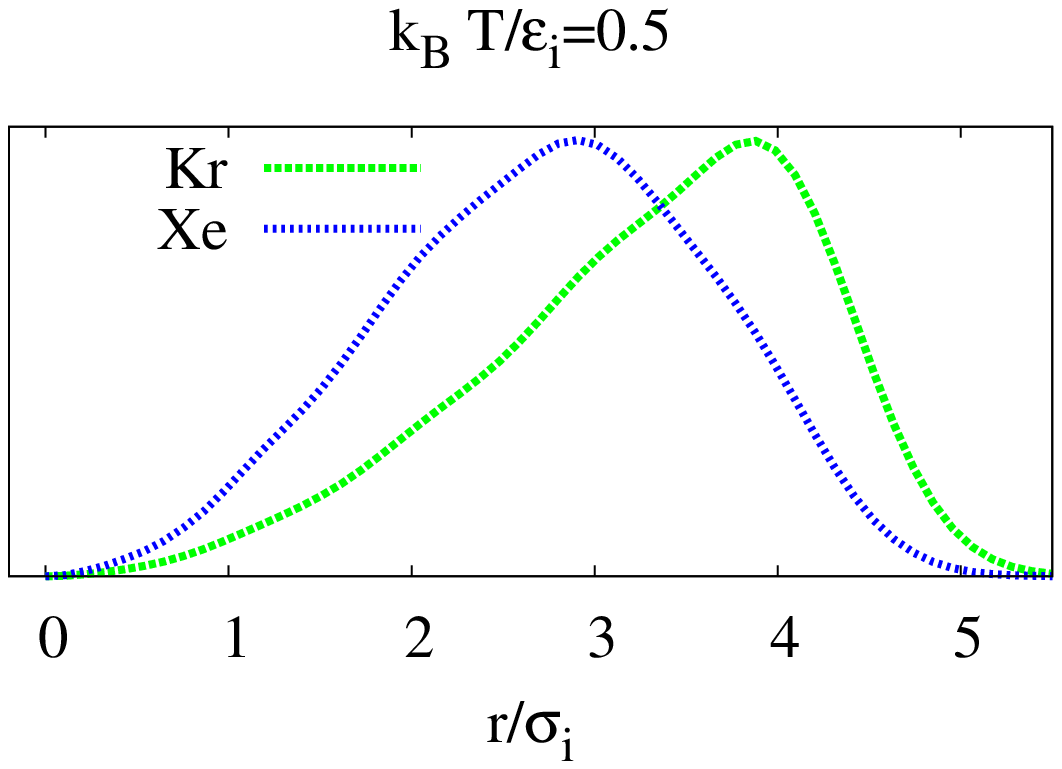}
  \end{minipage}
  \\

\caption{\label{fig:gs309} Radial distribution function for 309 atom clusters.}
\end{figure}

\begin{figure}[!th]
 \begin{minipage}{0.48\textwidth}
  \centering a)
\includegraphics[width=1.0\textwidth]{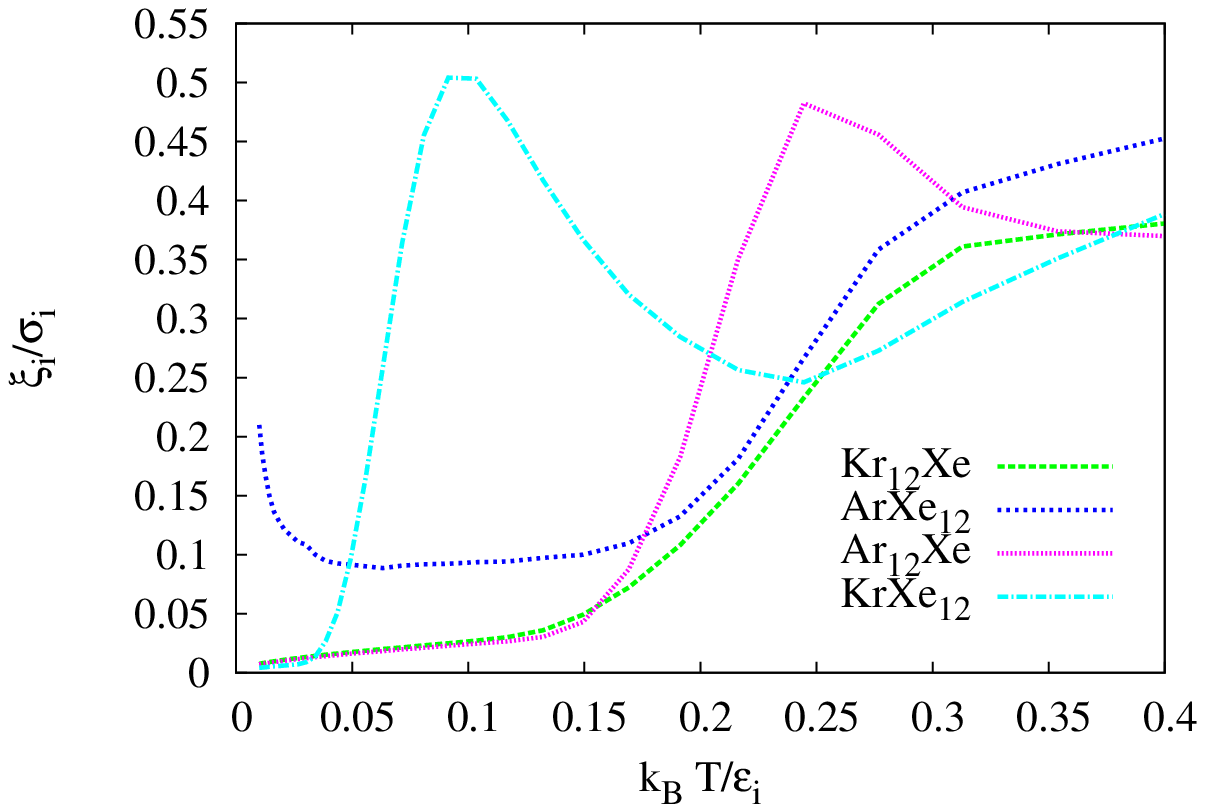}%
 \end{minipage}
\ \hfill
 \begin{minipage}{0.48\textwidth}
  \centering b)
\includegraphics[width=1.0\textwidth]{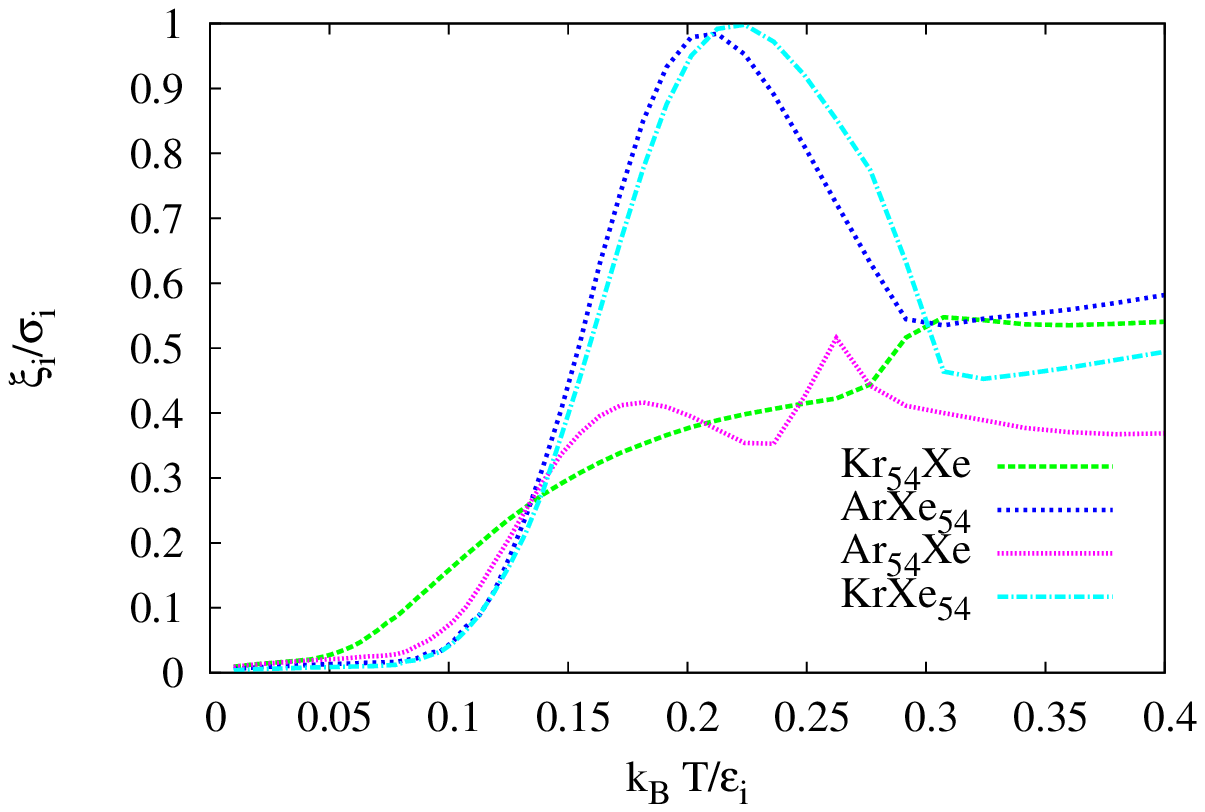}%
 \end{minipage}
\\
\bigskip
 \begin{minipage}{0.48\textwidth}
  \centering c)
\includegraphics[width=1.0\textwidth]{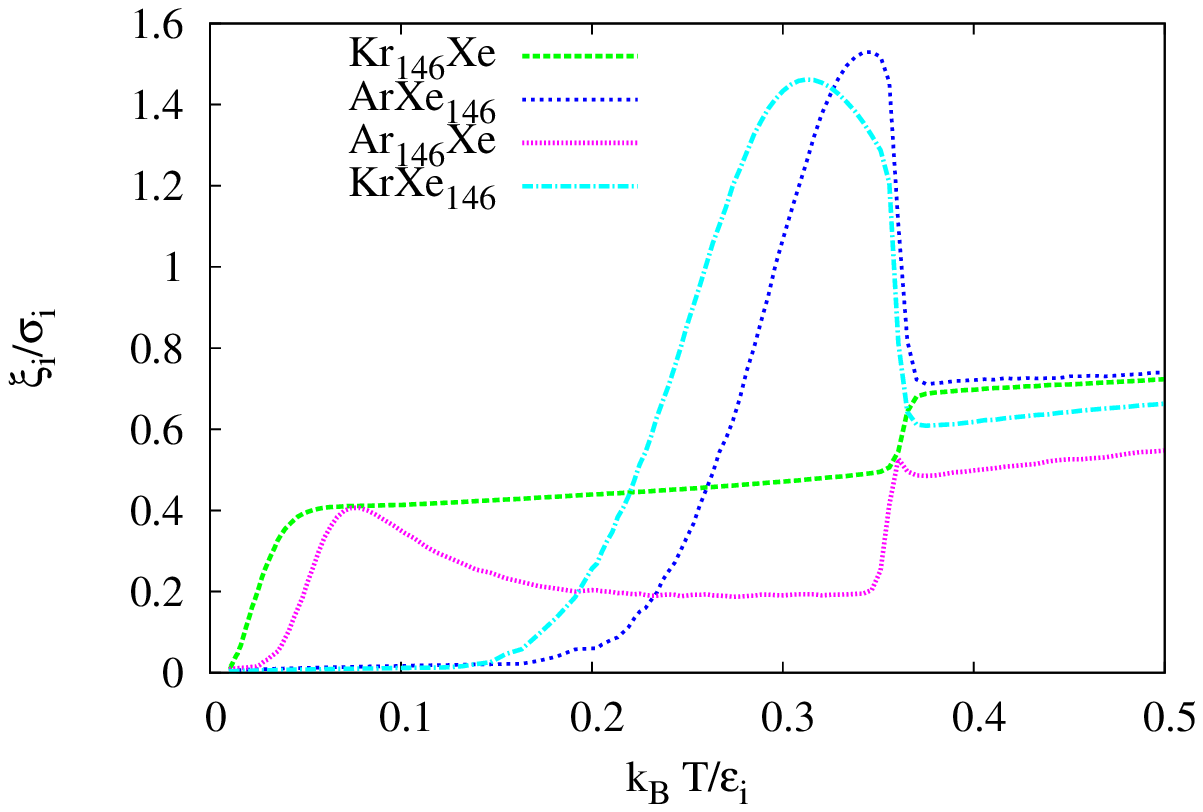}%
 \end{minipage}
\ \hfill
 \begin{minipage}{0.48\textwidth}
  \centering d)
\includegraphics[width=1.0\textwidth]{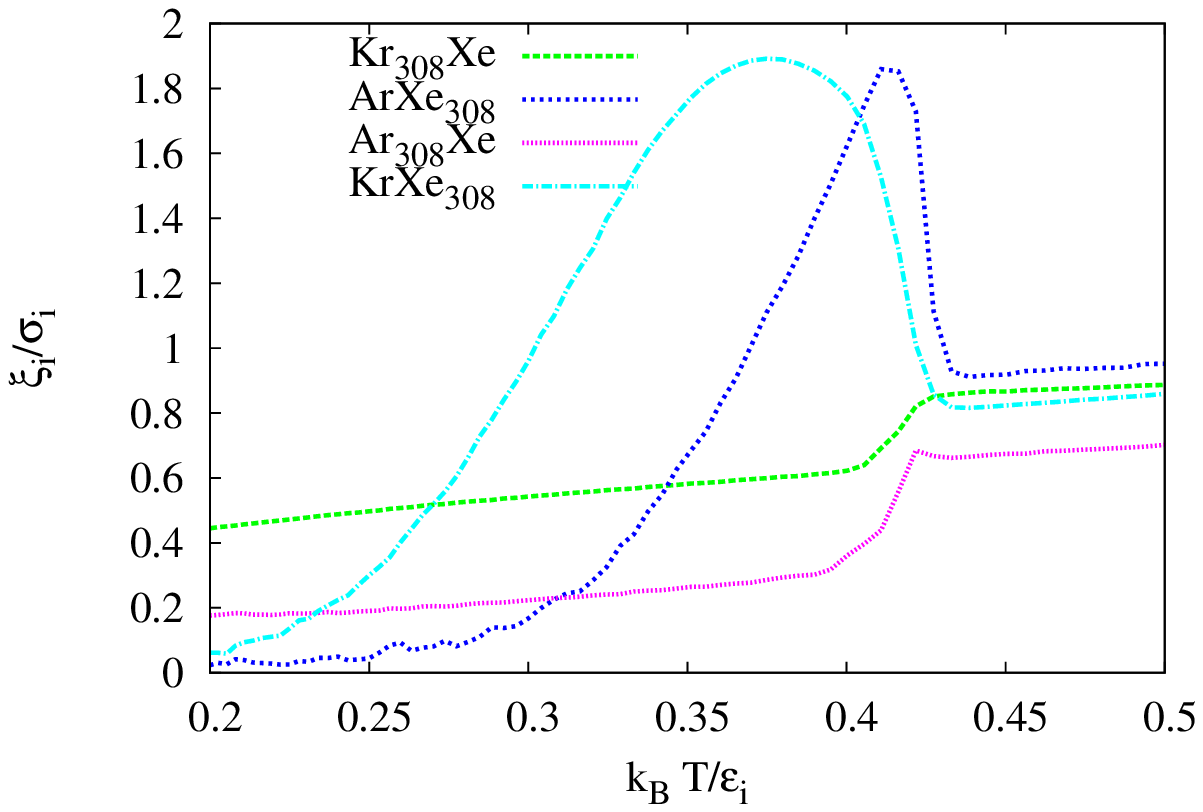}%
 \end{minipage}
\caption{\label{fig:var} Standard deviation of the position of the dopant atom for cluster sizes a)  $N$=13, b) $N$=55, c) $N$=147 d) $N$=309.}
\end{figure}

\begin{figure}[!ht]
  \begin{minipage}{0.3\textwidth}
%\centering
\begin{center}
 ArXe$_{54}$ (A)
    \includegraphics[width=\textwidth]{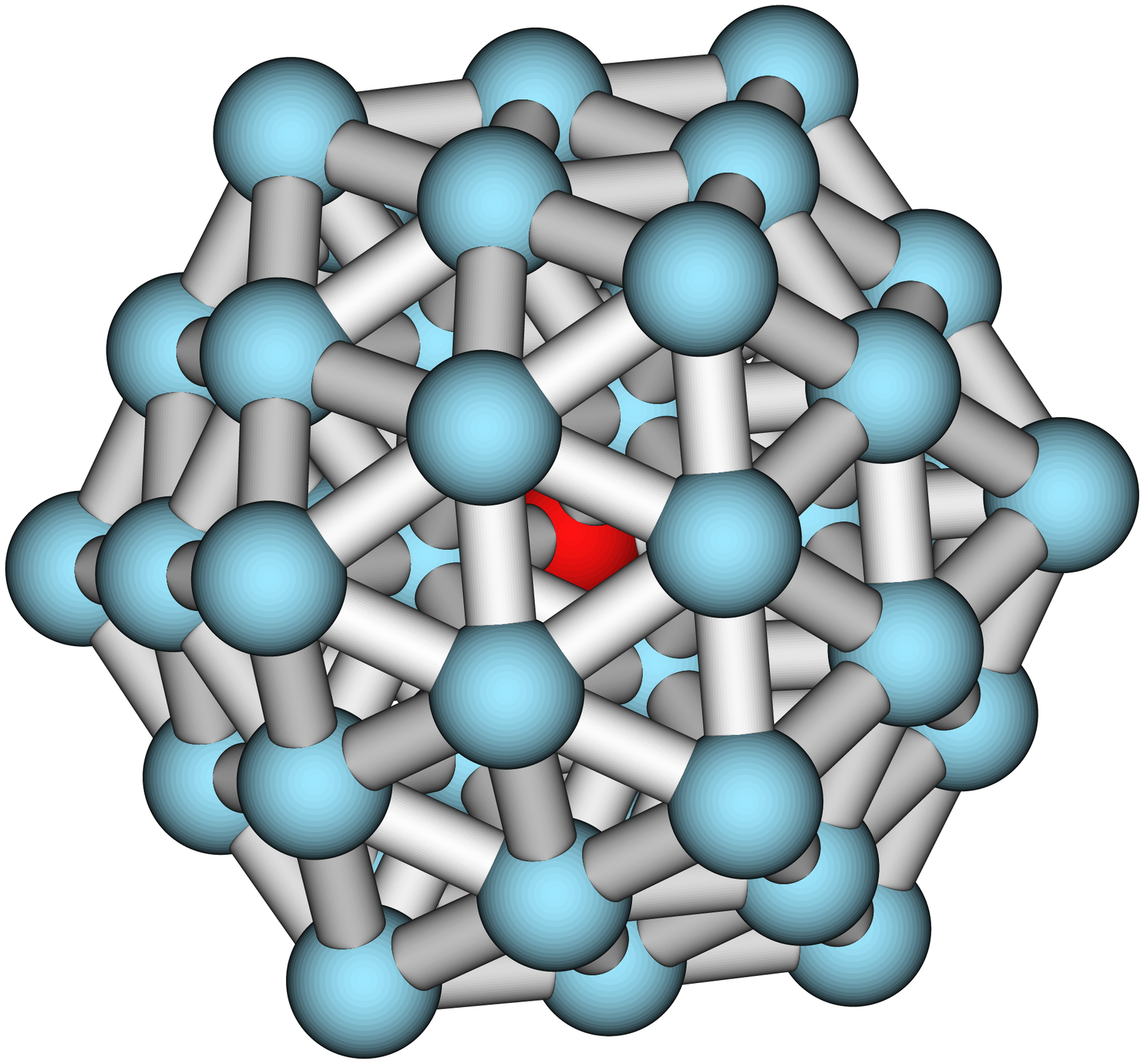}\end{center}
  \end{minipage}
%  \ \hfill 
  \begin{minipage}{0.3\textwidth}
%    \centering 
\begin{center}
ArXe$_{54}$ (B)
    \includegraphics[width=\textwidth]{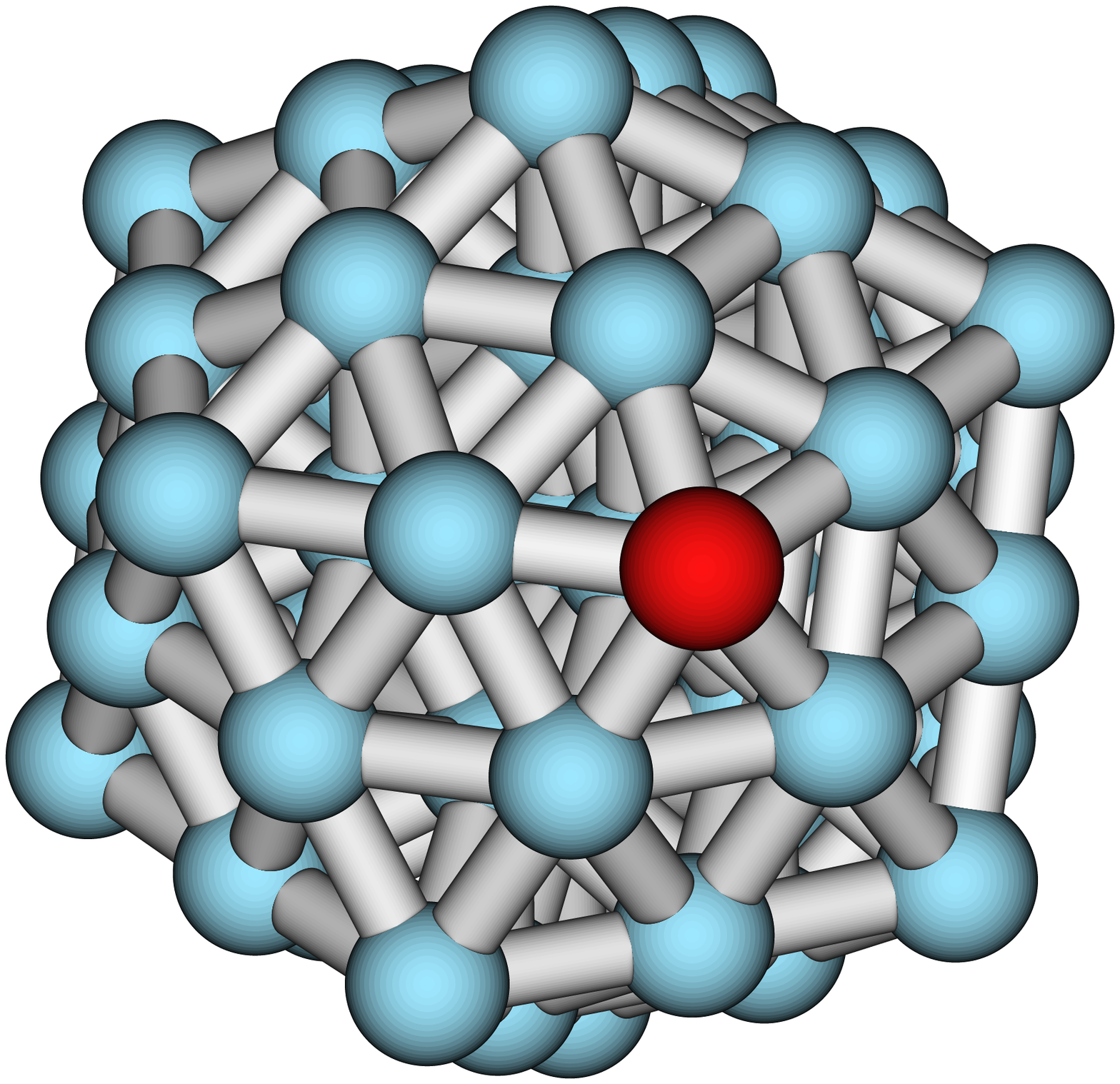}\end{center}
  \end{minipage}
\caption{\label{bolitas} The two lowest energy configurations of ArXe$_{54}$. These configurations correspond to the lines labeled (A) and (B) in figure \ref{fig:55quench}.}
\end{figure}

\begin{figure}[!th]
 \begin{minipage}{0.48\textwidth}
  \centering Kr$_{12}$Xe
\includegraphics[width=1.0\textwidth]{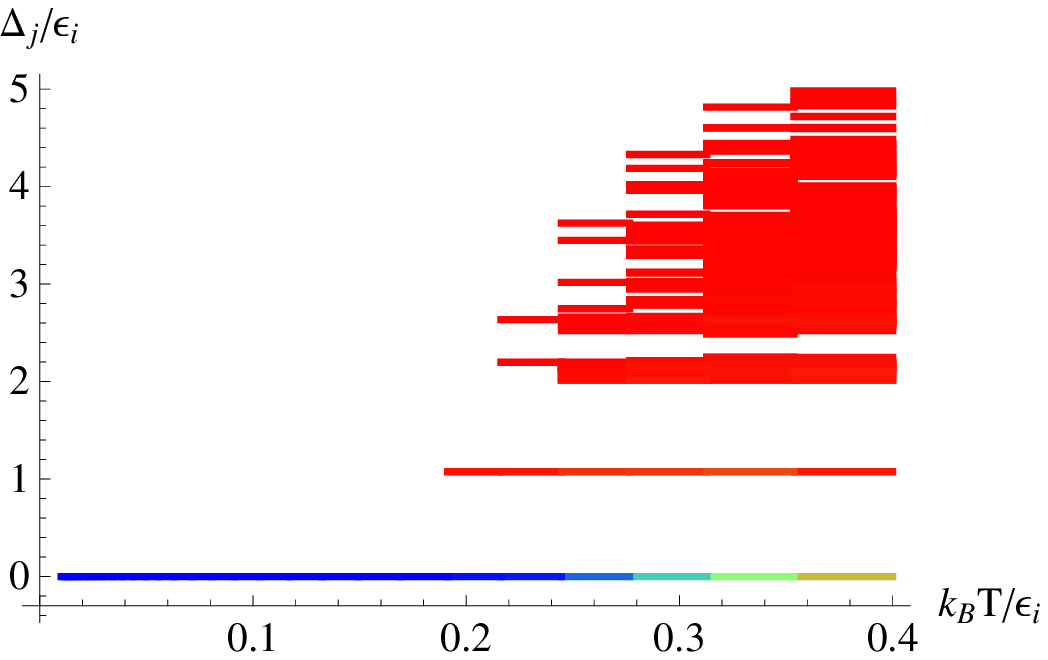}%
 \end{minipage}
\ \hfill
 \begin{minipage}{0.48\textwidth}
  \centering ArXe$_{12}$
\includegraphics[width=1.0\textwidth]{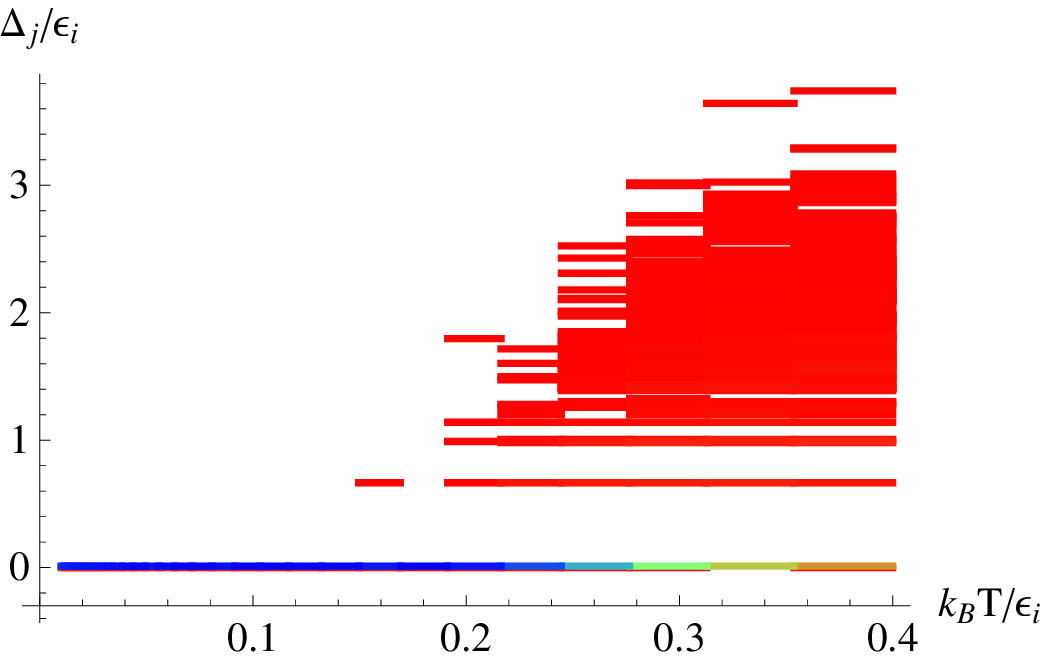}%
 \end{minipage}
\\
\bigskip
 \begin{minipage}{0.48\textwidth}
  \centering Ar$_{12}$Xe
\includegraphics[width=1.0\textwidth]{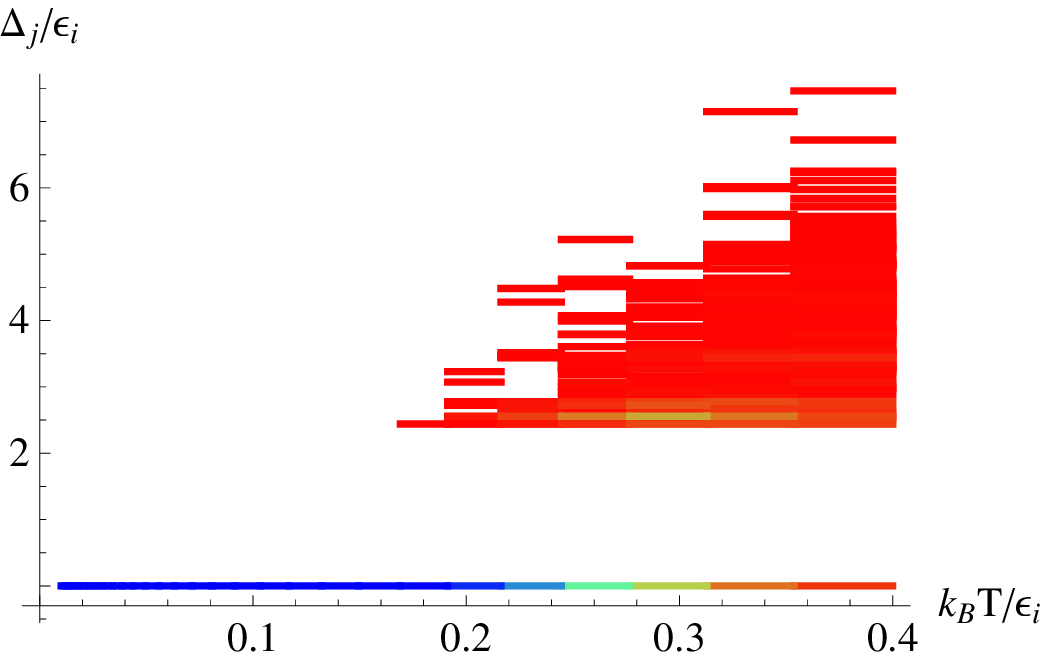}%
 \end{minipage}
\ \hfill
 \begin{minipage}{0.48\textwidth}
  \centering KrXe$_{12}$
\includegraphics[width=1.0\textwidth]{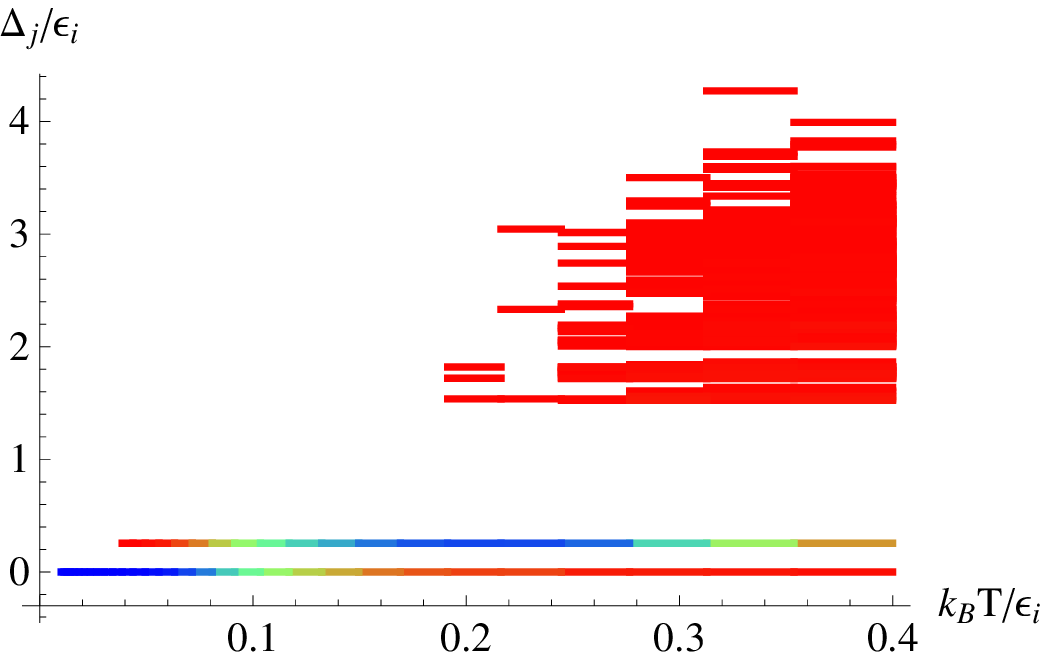}%
 \end{minipage}
\\
\bigskip
 \begin{minipage}{0.48\textwidth}
  \centering LJ$_{13}$
\includegraphics[width=1.0\textwidth]{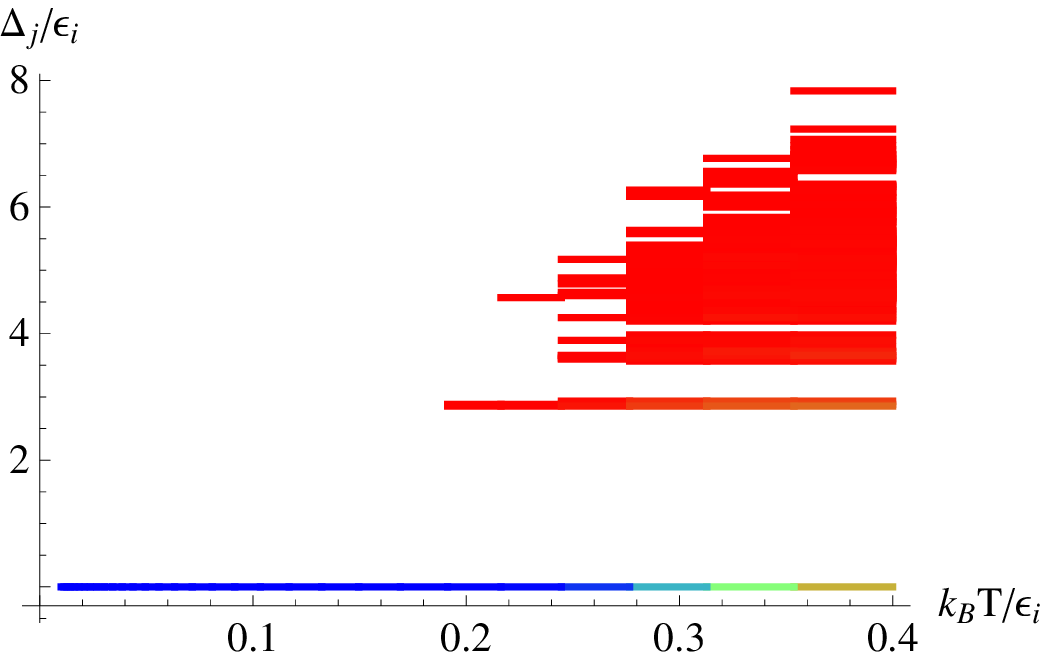}%
 \end{minipage}
\ \hfill
 \begin{minipage}{0.48\textwidth}
\includegraphics[width=0.2\textwidth]{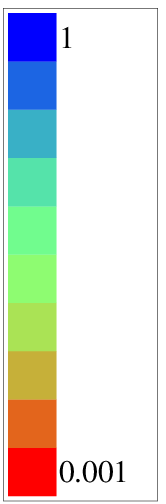}%
 \end{minipage}
\caption{\label{fig:13quench} Spectra of quenched energies ($\Delta_j=E_j-E_0$) for 13 atom pure and doped clusters. The color indicates the relative sampling frequency of each minimum at a given temperature. Notice that the two lowest states of ArXe$_{12}$ overlap almost completely due to their small energetic difference.}
\end{figure}

\begin{figure}[!th]
 \begin{minipage}{0.48\textwidth}
  \centering Kr$_{54}$Xe
\includegraphics[width=1.0\textwidth]{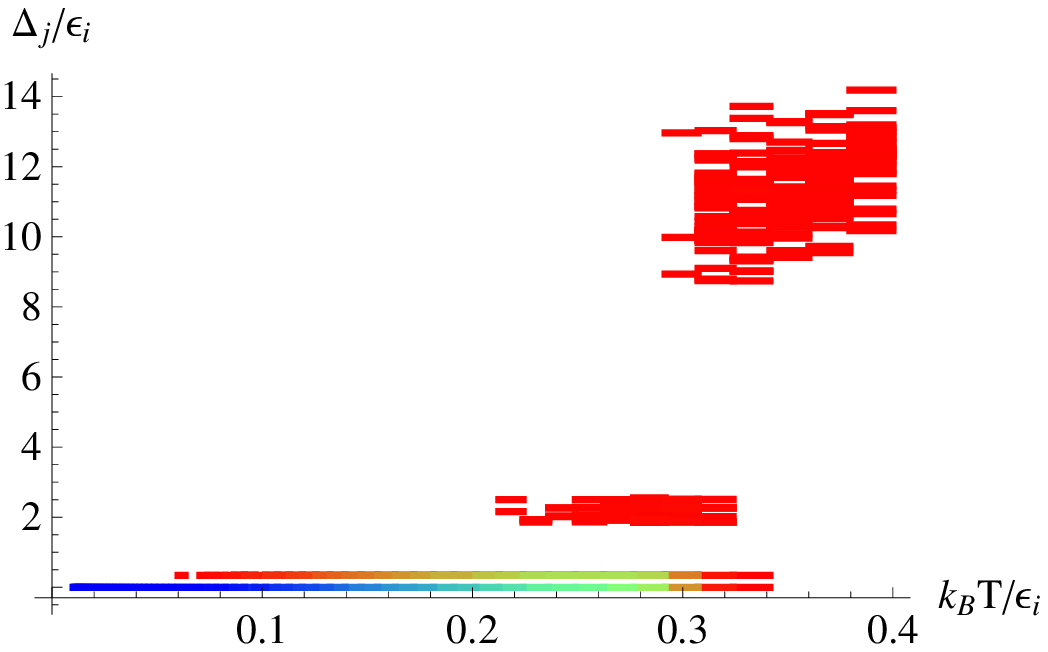}%
 \end{minipage}
\ \hfill
 \begin{minipage}{0.48\textwidth}
  \centering ArXe$_{54}$
\includegraphics[width=1.0\textwidth]{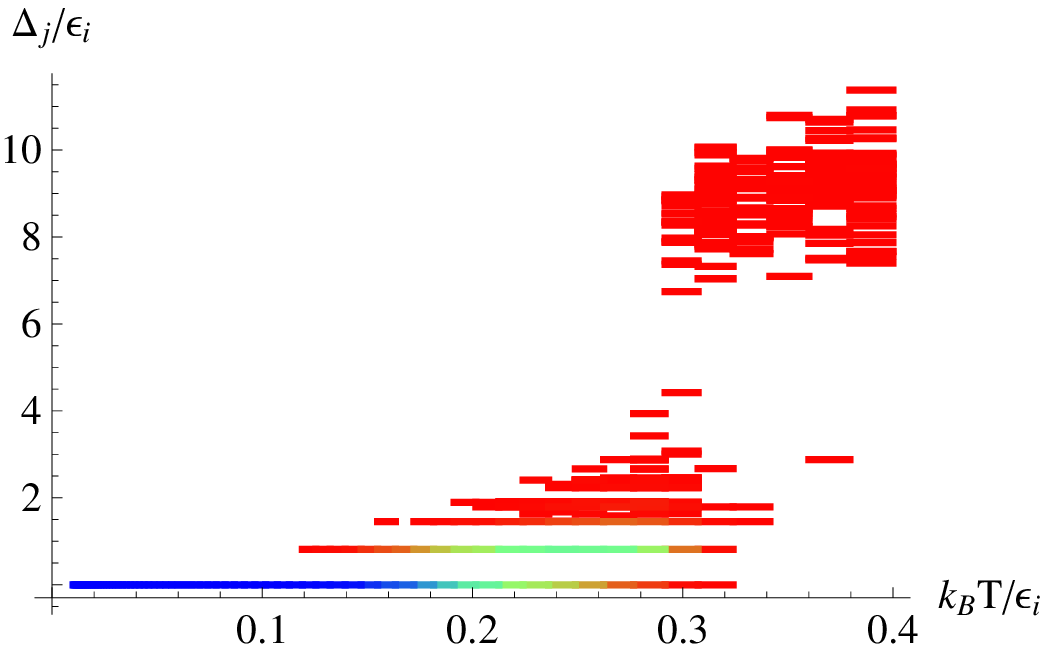}%
\put(-230,20){(A)}
\put(-180,30){(B)}

 \end{minipage}
\\
\bigskip
 \begin{minipage}{0.48\textwidth}
  \centering Ar$_{54}$Xe
\includegraphics[width=1.0\textwidth]{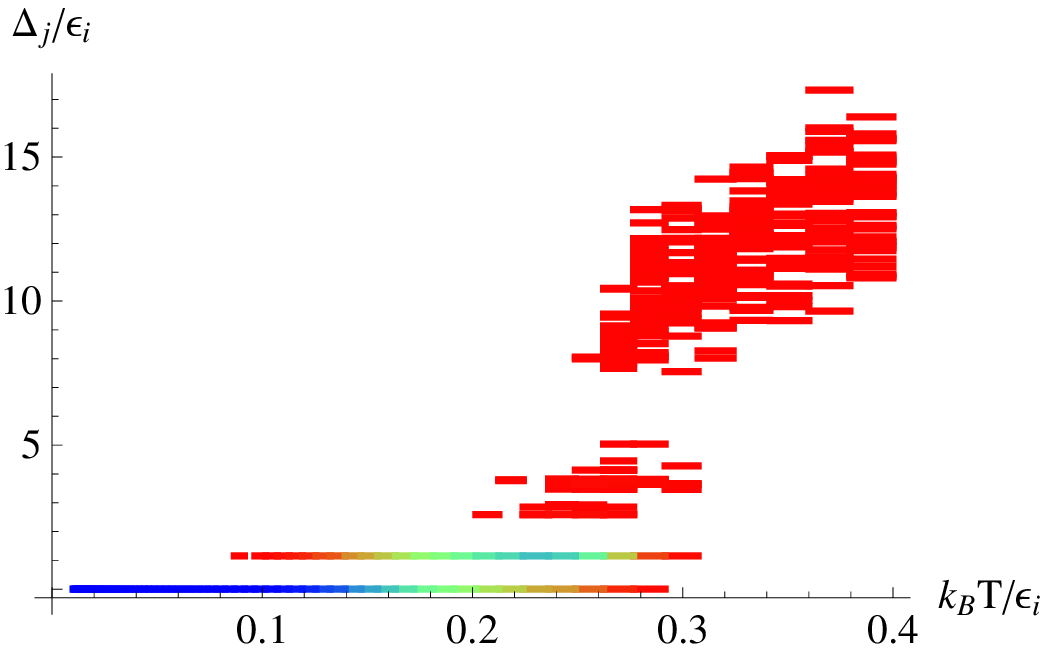}%
 \end{minipage}
\ \hfill
 \begin{minipage}{0.48\textwidth}
  \centering KrXe$_{54}$
\includegraphics[width=1.0\textwidth]{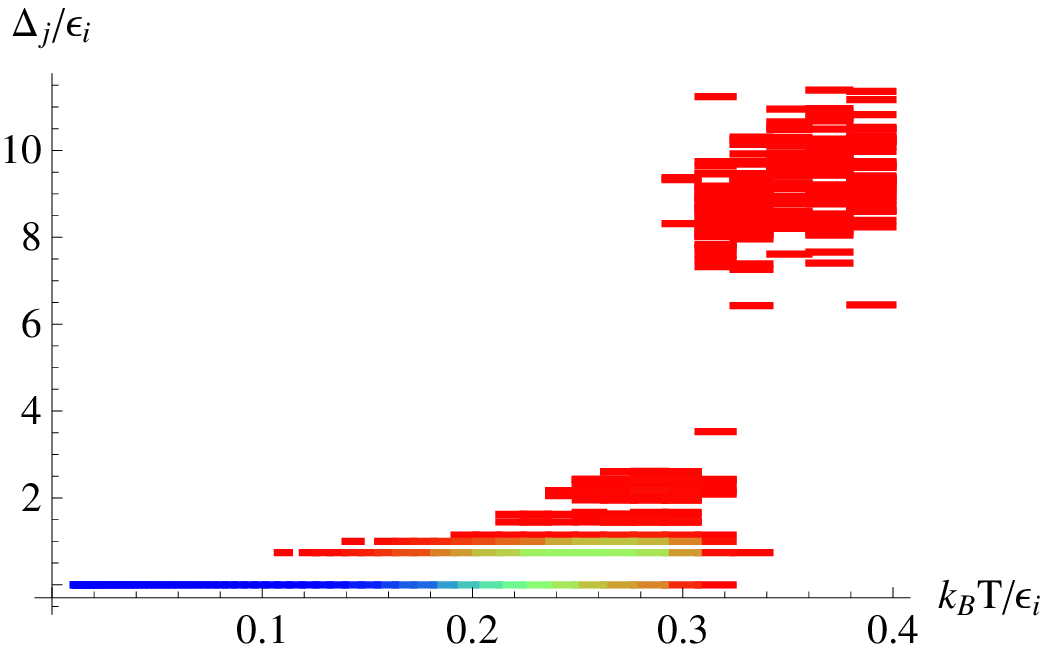}%
 \end{minipage}
\\
\bigskip
 \begin{minipage}{0.48\textwidth}
  \centering LJ$_{55}$
\includegraphics[width=1.0\textwidth]{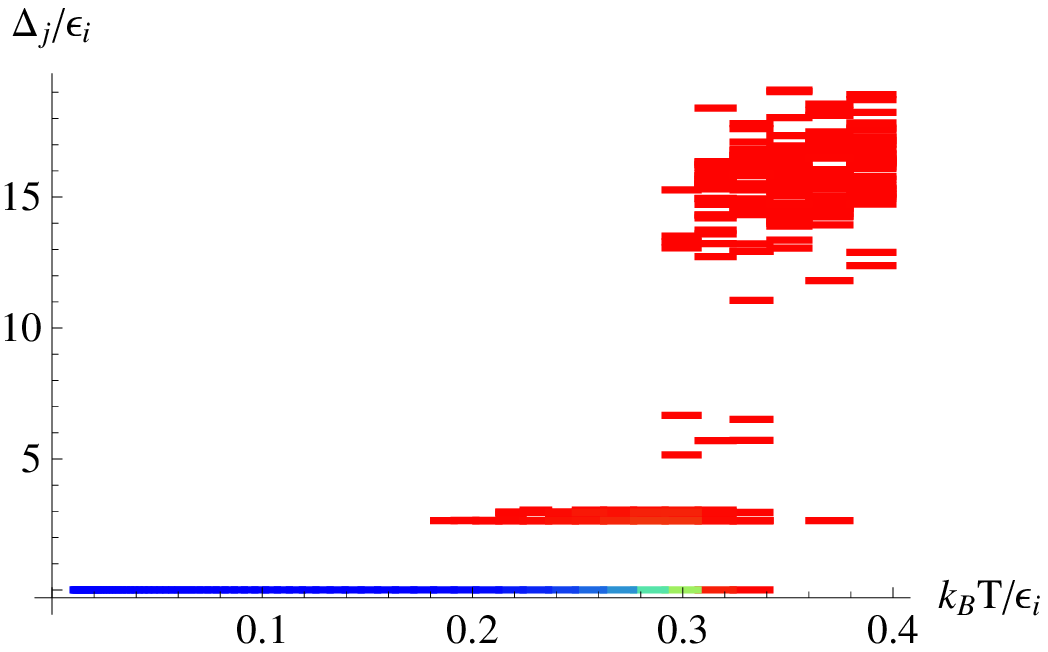}%
 \end{minipage}
\ \hfill
 \begin{minipage}{0.48\textwidth}
\includegraphics[width=0.2\textwidth]{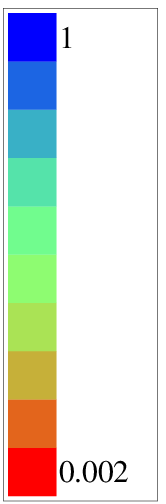}%
 \end{minipage}
\caption{\label{fig:55quench} Spectra of quenched energies ($\Delta_j=E_j-E_0$) for 55 atom pure and doped clusters. The color indicates the relative sampling frequency of each minimum at a given temperature.}
\end{figure}

\end{widetext}

\end{document}